\documentclass[aps,prb,twocolumn,showpacs,amsmath,amssymb,floatfix,footinbib]{revtex4-1}
\usepackage{subfigure}
\usepackage{makecell}
\usepackage[linktocpage=true]{hyperref}
\usepackage{graphicx}
\usepackage{booktabs}
\usepackage{epsfig}
\usepackage{color}
\usepackage{epstopdf}

\newcommand{\RNum}[1]{\uppercase\expandafter{\romannumeral #1\relax}}
\newcommand{\ket}[1]{\left|#1\right\rangle}

\begin{document}

\title{Finite-frequency noise in a quantum dot with normal and superconducting leads}

\author{Stephanie Droste$^{1}$, Janine Splettstoesser$^{2}$, and Michele Governale$^{1}$} 
\affiliation{$^{1}$School of Chemical and Physical Sciences and Mac Diarmid Institue for Advanced Materials and Nanotechnology, Victoria University of Wellington, PO Box 600, Wellington 6140, New Zealand\\
$^2$Department of Microtechnology and Nanoscience - MC2, Chalmers University of Technology, S-412 96 G\"oteborg, Sweden }
	
\date{\today}

\begin{abstract}

We consider a single-level quantum dot tunnel-coupled to one normal and one superconducting lead. 
We employ a diagrammatic real-time approach to calculate the finite-frequency current noise for subgap transport. The noise spectrum gives  direct access to the internal dynamics of the dot. In particular the noise spectrum  shows sharp dips at the frequency of the coherent oscillations of Cooper pairs between dot and superconductor. 
This feature is most pronounced when the superconducting correlation is maximal. 
Furthermore, in the quantum-noise regime, $\omega> k_\text{B}T,\mu_\text{N}$, the noise spectrum exhibits steps at frequencies equal to the Andreev addition energies. The height of these steps is related to the effective coupling strength of the excitations. The finite-frequency noise spectrum hence provides a full spectroscopy of the system. 
\end{abstract}

\maketitle

\section{Introduction}

Nanostructures with quantum dots in proximity to superconducting electrodes are an ideal playground to study superconducting correlations in systems with few degrees of freedom that exhibit strong Coulomb-interaction effects.~\cite{DeFranceschi10,Martin-Rodero11} Intriguingly, these types of structures are at the heart of recent  proposals to generate Majorana-fermion excitations in quantum dots~\cite{Leijnse12,sothmann_2013,Wright13, Brunetti13} and to establish and detect different symmetries of superconducting  pairing in a controllable way.~\cite{sothmann_2014} Another line of research has focussed on the possibility to use double quantum dots, tunnel coupled to superconductors as a source of entangled electron pairs.~\cite{hofstetter09,hermann10}  It is therefore of vital importance to get access to the properties of such hybrid quantum-dot systems.

Here we are interested in a setup, in which the tunnel-coupling rate between the superconductor and the quantum dot is strong (larger than the coupling to other normal-conducting leads eventually present in the device), such that it is possible to establish a BCS-like state in a single-level quantum dot even in the presence of strong Coulomb repulsion.~\cite{governale08} Such a state is characterised by a coherent exchange of Cooper pairs between the dot and the superconducting lead. 
If one considers the two-terminal case of a quantum dot, tunnel-coupled to one normal and one superconducting lead, the proximity effect is established by generating a non-equilibrium situation by means of an applied transport voltage. The presence of the proximity effect can be detected by measuring the Andreev current and its zero-frequency noise.~\cite{Braggio11} In various experimental studies the subgap spectrum of hybrid superconductor-quantum dot devices has been analyzed by Andreev level spectroscopy
\cite{Jespersen07,Eichler07,Deacon10,Pillet10,Dirks11,Lee12,Pillet13,Chang13,Lee14,Kumar14,Schindele14}, which allows to measure the Andreev addition energies and the total line width of the resonances via the differential conductance.

The coherent dynamics underlying the proximity effect in the dot shows up for example in the waiting time distribution of transport events~\cite{Brandes08,Albert12,Thomas13} in the normal lead, which exhibits an oscillatory behaviour due to the coherent exchange of Cooper pairs between dot and superconductor.~\cite{Rajabi13}
However, a direct measurement of the waiting time distribution might be challenging and it is therefore interesting to look at alternative possibilities to reveal the coherent tunnelling of Cooper pairs between dot and superconductors. 

A quantity of high interest to look at  is the finite-frequency noise of the Andreev current. Current noise spectroscopy in mesoscopic systems has become a standard tool to gain information on the transport processes and internal time scales of mesoscopic conductors.~\cite{Koch82,Schoelkopf97,Deblock03,Onac06,Billangeon06,Billangeon07,Zakka07,Gabelli08,Ubbelohde12,Maisi14} 
Indeed, the non-equilibrium finite-frequency noise of quantum dots in different regimes and setups has previously been at the focus of various theoretical studies.~\cite{Averin93,Ding97,Aguado00,Blanter00,Engel04,Braun06,Wohlman07,Rothstein09,Gabdank11,Orth12,Brandschaedel11,Marcos11,Joho12,Sothmann10,Jin12,Mueller13,Dong13,Kirton12,Soller14,Moca14}

This manuscript focuses on the finite-frequency current noise for subgap transport through a single-level quantum dot tunnel-coupled to one normal and one superconducting lead. We consider strong coupling between quantum dot and superconductor, while the dot is only weakly coupled to the normal conducting lead. We employ a non-equilibrium real-time diagrammatic perturbation expansion in the tunnel-coupling to the normal lead.~\cite{Koenig96,Koenig96B,Thielmann03,Braun06}
We find that the coherent oscillations between dot states with different  particle numbers leads to a resonant feature at the oscillation frequency in the finite-frequency noise spectrum. The magnitude of this feature (a sharp dip in the spectrum) is directly related to the pair amplitude in the dot.   
In the quantum noise regime, $\omega> k_\text{B}T,\mu_\text{N}$, it is possible to extract information on the relative coupling of different BCS-like states to the normal lead. 
Beyond the knowledge of the Andreev addition energies, which can also be obtained from Andreev level spectroscopy by means of a differential conductance measurement,~\cite{Jespersen07,Eichler07,Deacon10,Pillet10,Dirks11,Lee12,Pillet13,Chang13,Lee14,Kumar14,Schindele14} the finite-frequency noise additionally provides information on the coherent dynamics of the system and its characteristic time scale, and the effective coupling strengths of the different Andreev levels.
 
The remainder of this paper is organised as follows. In Sec.~\ref{modelandformalism}, we introduce the mathematical model used to describe the hybrid quantum-dot system and present the diagrammatic approach employed to calculate its noise spectrum. After a brief overview over the properties of the Andreev current in Sec.~\ref{sec:current}, we show the results for the finite-frequency current noise in Sec.~\ref{results}  organised by biasing and frequency regimes. 
The main conclusions of the paper are summarised in Sec.~\ref{conclusions}.

\section{Model and Formalism}\label{modelandformalism}

In this section we present the model for the interacting single-level quantum dot attached to one normal and one superconducting lead and the real-time diagrammatic approach to obtain the finite-frequency current noise. 

\subsection{Quantum-dot Hamiltonian}

In this paper, we study subgap transport through a single-level quantum dot tunnel-coupled to one normal and one superconducting lead. We consider strong coupling between quantum dot and superconductor, while the dot is only weakly coupled to the normal conducting lead.  
We restrict ourself to the case when the the temperature is larger than the tunnel-coupling strength  of the normal-conducting lead ($k_\text{B}T \gg \Gamma_\text{N}$,  where $\Gamma_\text{N}$ will be defined in terms of tunnel amplitude and  the density of states of the lead later in this section). In this particular regime the Kondo correlations due to the coupling with the normal lead are negligible and we can treat the tunnelling with the normal lead to lowest non-vanishing order. A large body of theoretical work regards the interplay of superconductivity and Kondo physics\cite{Cuevas01,Tanaka07,Yamada10,Yamada11,Zapalska14,zitko}.
For the subgap transport characteristics of the system, the superconductor can be described by means of an effective Hamiltonian which becomes exact in the regime of infinite superconducting gap. However the effective Hamiltonian still describes well the subgap transport features even for finite values of the gap as long as the temperature is larger than the Kondo temperature related to the Kondo screening by the quasiparticle excitations in the superconductor.
A detailed study of the reliability of this approximation can be found in Ref. \onlinecite{Futterer13} .  

\begin{figure}[b]
\subfigure{\includegraphics[width=0.25\textwidth]{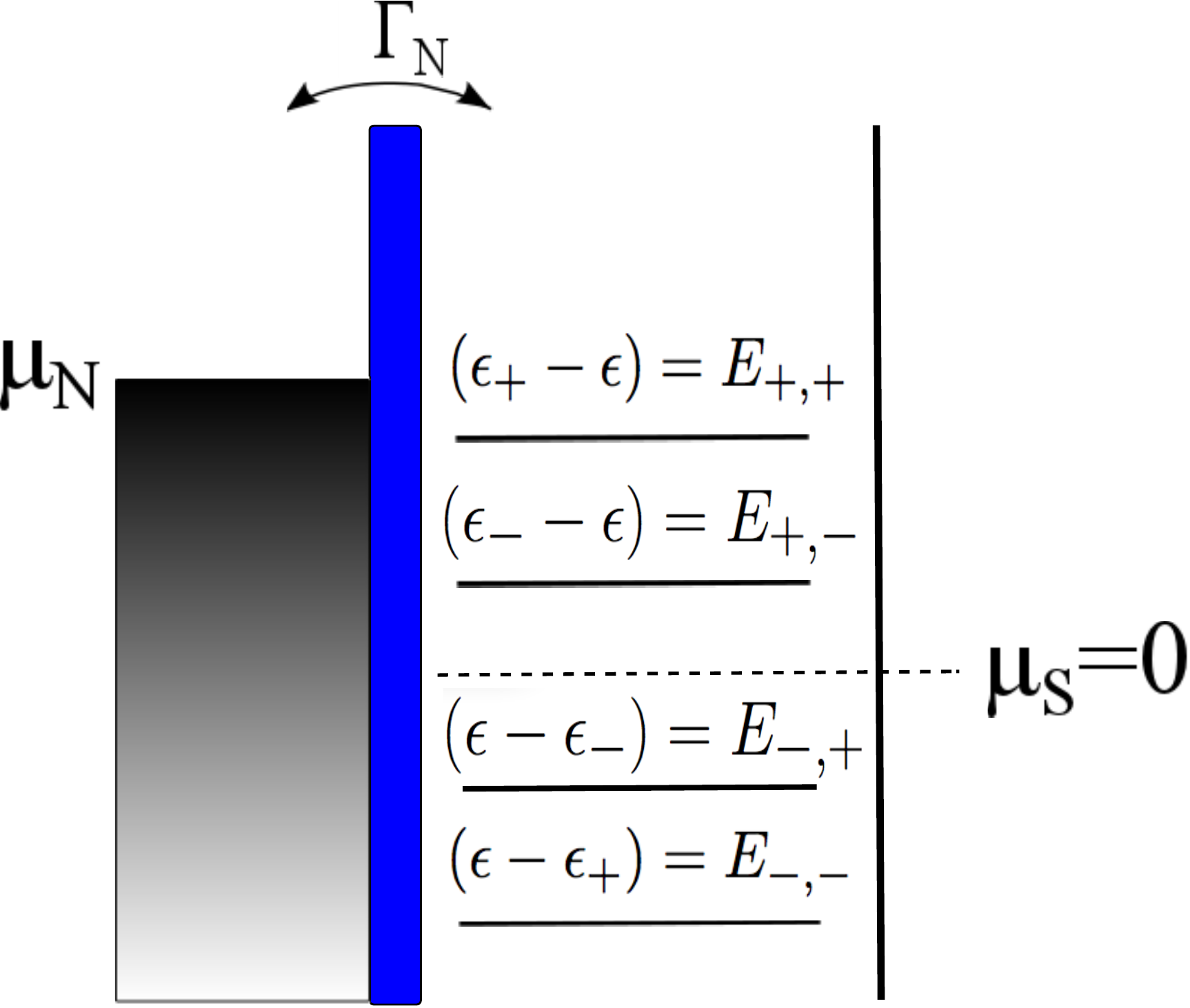}}
\caption{Sketch of the energy landscape of the effective dot-superconductor subsystem coupled to a normal-conducting lead which acts as a bath.  }
\label{fig:model}
\end{figure}

The Hamiltonian of the system can be written as the sum of three terms:  $H=H_{\text{eff}}+ H_{\text{N}}+H_{\text{tunn}}$. 
The first term, $H_\text{eff}$, is the Hamiltonian for the hybrid system composed of the dot and the superconductor in the limit of large superconducting gap, the second term, $H_\text{N}$, describes the non-interacting normal lead and the third one, $H_\text{tunn}$, the tunnel-coupling between the proximized dot and the normal lead. 

We model the quantum dot tunnel-coupled to the superconducting lead by means of the following effective Hamiltonian, which becomes exact in the limit of very large superconducting gap~\cite{Rozhkov00}
\begin{equation}
H_\mathrm{eff}=
\sum_{\sigma}\epsilon \hat{n}_{\sigma}+U\hat{n}_{\uparrow}\hat{n}_{\downarrow}
-\frac{\Gamma_\text{S}}{2}\left(d^{\dagger}_{\uparrow}d^{\dagger}_{\downarrow}+d_{\downarrow}^{}d_{\uparrow}^{}\right)\ .
\label{eq:H_eff}
\end{equation}
Here, $\epsilon$ is the dot level energy, $U$ the on-site Coulomb repulsion and $\Gamma_{\text{S}}$ the tunnel-coupling strength between dot and superconducting lead. All energies are measured with respect to the chemical potential of the superconductor, i.e. $\mu_{\text{S}}=0$. 
Here, $d^{}_{\sigma}(d^{\dagger}_{\sigma})$ is the annihilation (creation) operator for an electron on the dot with spin $\sigma=\uparrow,\downarrow$  and $\hat{n}_{\sigma}=d^{\dagger}_{\sigma}d^{}_{\sigma}$ the corresponding number operator.
The Hilbert space of the proximized dot is spanned by the states: $|0\rangle$ (empty), $|\sigma\rangle=d^{\dagger}_{\sigma}|0\rangle$ (singly occupied) and $|d\rangle=d^{\dagger}_{\uparrow}d^{\dagger}_{\downarrow}|0\rangle$ (doubly occupied). 

The normal lead is described by the non-interacting Hamiltonian $H_\text{N}=\sum_{k,\sigma}\epsilon_{k} c^{\dagger}_{ k \sigma} c^{}_{ k \sigma}$, where $c^{}_{ k \sigma}(c^{\dagger}_{k \sigma})$ is the annihilation (creation) operator for an electron with spin $\sigma$ in the single-particle state of the lead characterised by the momentum quantum number $k$ with energy $\epsilon_k$. The normal-conducting lead has an electrochemical potential $\mu_\mathrm{N}$, which in general differs from zero. 

The tunnel-coupling between the dot and the normal lead is modelled by means of the tunnelling Hamiltonian, $H_\mathrm{tunn}=\sum_{k, \sigma} t_{\text{N}}   c^{\dagger}_{ k \sigma}d^{}_{\sigma}+\mathrm{H.c.} $ with the tunnelling amplitude $t_\text{N}$.
We assume the density of states, $\rho_{\text{N}}$, of the normal lead to be constant and spin- and momentum independent; we define the tunnel-coupling strength as $\Gamma_{\text{N}}=2\pi\rho_{\text{N}} |t_{\text{N}}|^2$. The effective dot-superconductor subsystem coupled to a normal-conducting lead is sketched in Fig.~\ref{fig:model}. When discussing the results of this paper, we will always assume $\Gamma_\text{N}\ll\Gamma_\text{S}$.

We proceed to discuss the eigenstates of the effective Hamiltonian in the absence of coupling to the normal lead. The singly-occupied states  $|\sigma\rangle$ are not affected by the proximity effect and are eigenstates of $H_{\text{eff}}$ with eigenenergy $\epsilon$.
Due to the tunnel-coupling to the superconductor the states $\ket{0}$ and $\ket{d}$ form  Andreev bound states (ABS) 
 \begin{equation}
 \ket{\pm}=\frac{1}{\sqrt{2}}\sqrt{1\mp\frac{\delta}{2\epsilon_\text{A}}}\ket{0}\mp\frac{1}{\sqrt{2}}\sqrt{1\pm\frac{\delta}{2\epsilon_\text{A}}}\ket{d}\ ,
 \label{eq:new_eigenstates}
 \end{equation}
with the eigenenergies of the effective Hamiltonian, $\epsilon_\pm=\delta/2\pm\epsilon_\text{A}$. Here,  $\delta=2\epsilon+U$  is the detuning between the empty and the doubly-occupied states and  $2\epsilon_\text{A}=\sqrt{\delta^{2}+\Gamma^{2}_\text{S}}$ is the energy splitting between the  $\ket{+}$ and $\ket{-}$ states. 
The excitation energies of the dot are the so called Andreev addition energies, which are given by the differences of the eigenenergies of those states which  have occupation numbers differing by one
\begin{equation}
E_{\gamma',\gamma}=\pm(\epsilon_{\pm}-\epsilon)=\gamma'\frac{U}{2}+\gamma\epsilon_\text{A}
\label{eq:Andreev_energies}
\end{equation}
with $\gamma',\ \gamma=\pm1$. 
When an electron leaves or enters the normal lead, its energy must account for the energy difference between the initial and final state of the dot-superconductor subsystem, which are the Andreev addition energies, represented by the energy levels in the sketch, Fig.~\ref{fig:model}.

At this stage it is useful to introduce effective coupling strengths which describe the coupling between the electronic reservoir and the dot resonances, namely the Andreev levels.
These effective coupling strengths turn out to be essential to understand the form of the finite-frequency noise spectrum. The effective tunnel-coupling strengths of the Andreev levels to the normal-conducting lead $\Gamma_\text{N}$ corresponding to the transition  $|\sigma\rangle$ to $|\pm\rangle$ are given by
\begin{equation}
\Gamma_{\sigma\rightarrow\pm}=\frac{\Gamma_\text{N}}{2}\left(1\pm\frac{\delta}{2\epsilon_\text{A}}\right)\ ,
\label{eq:Gamma_effective1}
\end{equation}
while for the opposite transition they read
\begin{equation}
\Gamma_{\pm\rightarrow\sigma}=\frac{\Gamma_\text{N}}{2}\left(1\mp\frac{\delta}{2\epsilon_\text{A}}\right)\ .
\label{eq:Gamma_effective2}
\end{equation}
In the following, we use the convention $\hbar=e=1$.

\subsection{Diagrammatic real-time approach for noise}\label{sec:diagrammaticApproach}
We aim to study the finite-frequency noise for transport through a quantum dot coupled to a normal- and a superconducting lead as described by the above introduced model. We take into account on-site Coulomb interaction of arbitrary magnitude and non-equilibrium conditions without resorting to the linear-response regime. While we are interested in a strong coupling of the quantum dot to the superconducting lead, leading to strong superconducting correlations, we assume the coupling to the normal-conducting lead to be weak ($\Gamma_\text{N}\ll k_\text{B} T$, where $k_\text{B}$ is the Boltzmann constant and $T$ is the absolute temperature). Considering these conditions, we make use of a diagrammatic real-time perturbation theory in the tunnel-coupling with the normal lead~\cite{Koenig96,Koenig96B} and its extension to a system with superconducting electrodes,~\cite{governale08} in order to derive the current and the finite-frequency current noise. The formalism to obtain the finite-frequency noise using this real-time diagrammatic approach has been introduced previously, where it was applied to the case of normal-conducting electrodes~\cite{Thielmann03} as well as for the ferromagnetic case.~\cite{Braun06} 

In this section we review the formalism to obtain the finite-frequency current noise by  relating it directly to the system of a quantum dot coupled to one normal and one superconducting lead. 
The aim is to formulate a method which allows to calculate the reduced density matrix of the proximized quantum dot as well as the current through it and the finite-frequency current noise. 

The full system is represented by a density matrix describing the normal-conducting lead (which has many degrees of freedom but is non-interacting) coupled to  the interacting quantum dot proximized by the superconducting condensate (the latter having just a few degrees of freedom). Since we are not interested in the degrees of freedom of the normal-conducting reservoir we trace them out, making use of Wick's theorem. We are then left with an effective description by means of  the reduced density matrix of the quantum dot proximized by the superconductor. This reduced density matrix has the form
\begin{equation}
\boldsymbol{P}=
\begin{pmatrix}
P^{+}_{+} & P^{+}_{-} & 0 & 0 \\  
P^{-}_{+} & P^{-}_{-} & 0 & 0 \\  
0 & 0 & P^{\uparrow}_{\uparrow} & P^{\uparrow}_{\downarrow} \\
0 & 0 & P^{\downarrow}_{\uparrow} & P^{\downarrow}_{\downarrow} \\
\end{pmatrix}\ ,
\label{eq:red_matrix_coherences}
\end{equation}
where the diagonal elements are the probabilities to find the dot singly occupied, $P^\sigma_{\sigma}\equiv P_{\sigma}$, or in a BCS-like state, $P^+_{+}\equiv P_{+}$ or $P^-_{-}\equiv P_{-}$. The off-diagonal elements, which we refer to as the coherences, describe coherent superpositions of two eigenstates of the proximized dot. Importantly, the time evolution of the coherences between states of single occupation decouples from the one of the diagonal elements due to spin-conserving tunnelling and these coherences will hence be disregarded in the following. In contrast, in order to fully describe the short-time dynamics of the system, it is necessary to consider the off-diagonal elements in the reduced density matrix between the Andreev bound states, $P^+_-$ and $P^-_+$. It turns out that these become important for the finite-frequency noise at frequencies $\omega\sim \pm(\epsilon_{+}-\epsilon_{-})$. 

\begin{figure}[b]
 \subfigure{ \includegraphics[width=0.45\textwidth]{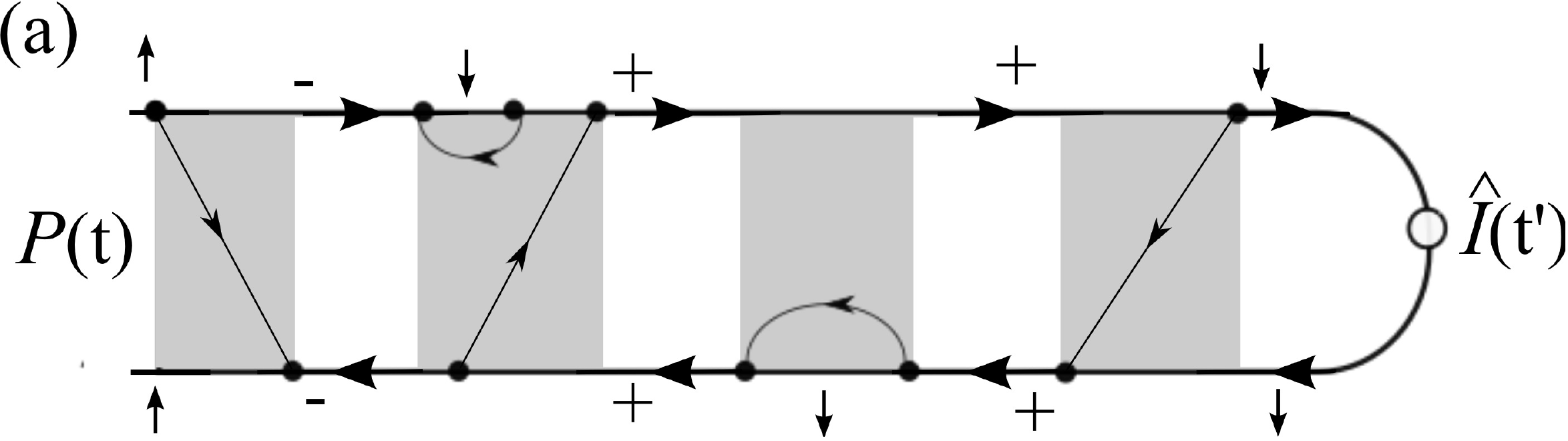}}
  \subfigure{ \includegraphics[width=0.45\textwidth]{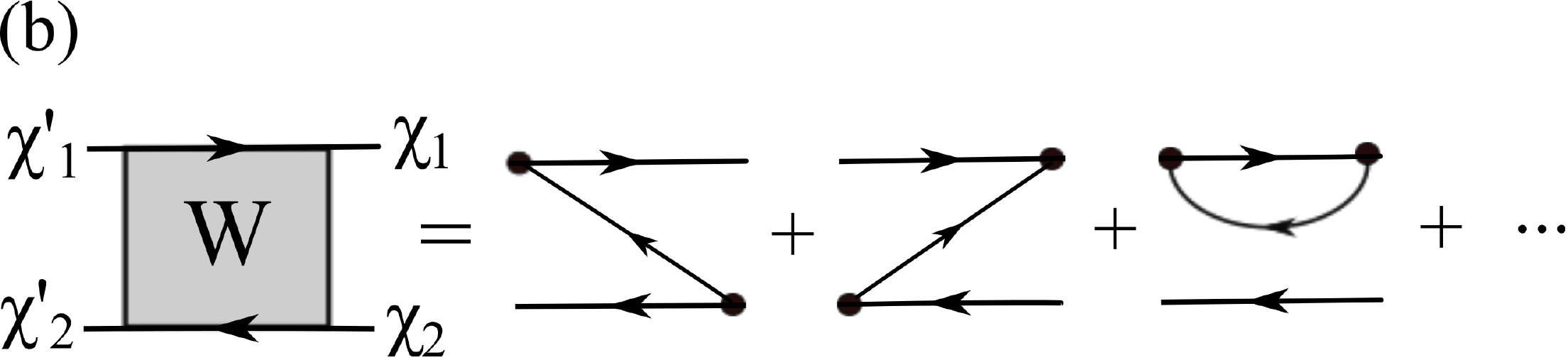}}
  \caption{(a) Example of the time evolution of the reduced density matrix $P$ for the evaluation of the expectation value of the current. The reduced system propagates forward in time along the top path from $t$ to $t'$ at which the observable $I$ is measured, then the system propagates back to time $t$. (b) Diagrammatic representation of the matrix element $W^{\chi_{1}\chi_{1}'}_{\chi_{2}\chi_{2}'}$.}
	\label{fig:contour}
\end{figure}

The non-equilibrium time evolution of the reduced density matrix can be depicted on the Keldysh contour and expressed in terms of a propagator $\boldsymbol{P}(t)=\boldsymbol{\Pi}(t,t')\boldsymbol{P}(t')$. An example of the Keldysh contour for the calculation of the current is shown in the sketch in  Fig.~\ref{fig:contour}~(a). The upper (lower) horizontal time line stands for the propagation of the individual dot state forward (backward) in real-time, indicated by arrows.
In frequency space, this full propagator  can be written in terms of a Dyson equation, 
\begin{equation}
\begin{split}
\boldsymbol{\Pi}(\omega) & =\boldsymbol{\Pi}_{0}(\omega)+\boldsymbol{\Pi}_{0}(\omega)\boldsymbol{W}(\omega) \boldsymbol{\Pi}(\omega)\\
& =\left[\boldsymbol{\Pi}_{0}(\omega)^{-1}-\boldsymbol{W}(\omega)\right]^{-1}\ ,
\end{split}
\label{eq:full_prop_finite}
\end{equation}
with the frequency-dependent free propagator on the Keldysh contour $\boldsymbol{\Pi}_{0}(\omega)$ and the kernel $\boldsymbol{W}(\omega)$, representing the self energy of the Dyson equation due to coupling to the normal-conducting reservoir. 
The full propagator is broken up in two types of blocks on the Keldysh contour,  irreducible self energy insertions and free propagation, as depicted in Fig.~\ref{fig:contour} (a). The matrix elements of the free propagator are given by
\begin{equation}
\Pi_{0}(\omega)^{\chi_{1}\chi'_{1}}_{\chi_{2}\chi'_{2}}=\frac{i\delta_{\chi_{1}\chi'_{1}}\delta_{\chi_{2}\chi'_{2}}}{\epsilon_{\chi_{2}}-\epsilon_{\chi_{1}}-\omega+i0^{+}}
\label{eq:free_prop}
\end{equation}
where $\chi^{}_{i},\chi'_{i}$ denote the different dot states at different times $t,t'$.  The kernel $\boldsymbol{W}$ describes transitions between different reduced-density-matrix elements due to tunnel events between quantum dot and normal-conducting lead. The kernel $\boldsymbol{W}$  is defined as the sum of all irreducible diagrams and can be obtained diagrammatically, based on a perturbation expansion in the tunnel-coupling to the normal-conducting lead as displayed in Fig.~\ref{fig:contour}~(b). A tunnel event, where an electron hops between dot and normal-conducting lead, is represented by an internal vertex (black dot) on the Keldysh contour.  A directed tunnel line between two vertices indicates the contribution due to the contraction 
 of two lead operators.  
The transformation into frequency space enters the diagrammatic representation by an additionaly horizontal bosonic line transporting the energy $\omega$. 

Finally, in the stationary limit, the reduced density matrix, $\boldsymbol{P}_\mathrm{stat}$, is found from the solution of a generalised master equation 
\begin{equation}
0=\left[\boldsymbol{\Pi}^{-1}_{0}(\omega=0)-\boldsymbol{W}(\omega=0)\right]\boldsymbol{P}_\mathrm{stat} 
\label{eq:master_equation}
\end{equation}
containing the coherent evolution of the reduced system described by the zero-frequency contribution to the free propagator and the dissipative coupling to the normal-conducting lead described by the zero-frequency contribution to the kernel. With the help of the solution for $\boldsymbol{P}_\mathrm{stat}$, we will in the following be able to determine the expectation values of the current and the current-current correlator yielding the finite-frequency noise. 

In the results part of this manuscript, we will restrict ourselves to the weak-coupling regime, performing a perturbation expansion with respect to the tunnel coupling to the normal lead.  The explicit expression for the kernel can be obtained by using diagrammatic rules, see Appendix~\ref{App:rules}.

\subsubsection{Current}
The current through the hybridised quantum dot is given by the operator representing the rate of change of the number of electrons in the normal lead:  $\hat{I}=\frac{i}{\hbar}\left[\hat{N},H\right]$, where $\hat{N}=\sum_{k,\sigma}c^{\dagger}_{ k \sigma} c^{}_{ k \sigma}$.
When calculating the time-dependent expectation value of the current operator, the latter acts as an external vertex on the Keldysh contour.

\begin{figure}[tb]
    \subfigure{ \includegraphics[width=0.45\textwidth]{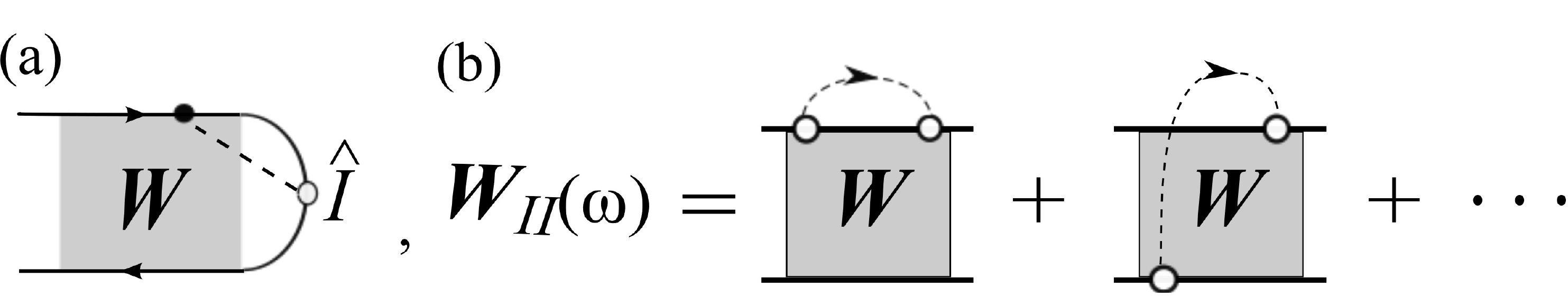}}
      \subfigure{ \includegraphics[width=0.45\textwidth]{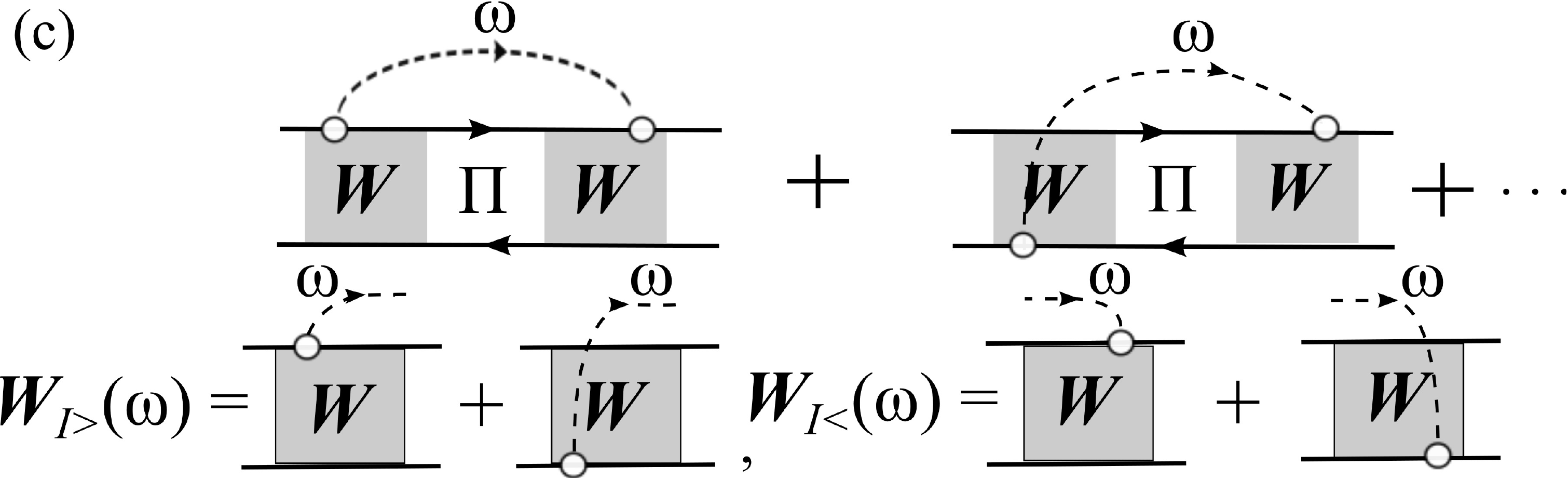}}
  \caption{Diagrammatic representation of (a) the current, (b) the contribution to noise with both current operators in one block and (c) in different ones separated by the full propagator. }
	\label{fig:noise}
\end{figure}

In order to calculate the expectation value of the charge current, $I$, the current operator is placed at the rightmost point  
of the Keldysh contour, see Fig.~\ref{fig:noise}~(a), and contracted to an internal tunnel vertex via a tunnelling line.  
It turns out that the current can be expressed as
\begin{equation}
I=\frac{1}{2}\mathrm{Tr}\left[\boldsymbol{W}_{\RNum{1}}\boldsymbol{P}_\mathrm{stat}\right]\quad.
\label{eq:current_average}
\end{equation}
The kernel $\boldsymbol{W}_{\RNum{1}}$ can be obtained from  $\boldsymbol{W}$ by replacing one of the internal tunnelling vertices (black dot)  by an external current vertex (open circle).~\cite{Thielmann03,Braun06}
The current kernel $\boldsymbol{W}_{\RNum{1}}$ takes into account whether an electron enters or leaves the dot through the normal lead. The diagrammatic rules to compute the kernel $\boldsymbol{W}_{\RNum{1}}$ are summarised in Appendix~\ref{App:rules} in lowest order in the tunnel coupling.

\subsubsection{Noise}
The symmetrized finite-frequency current noise is defined as the Fourier transform of $S(t)=\langle\hat{I}(t)\hat{I}(0)\rangle+\langle\hat{I}(0)\hat{I}(t)\rangle-2\langle\hat{I}\rangle^{2}$, namely of the current-current correlator at different times,
\begin{equation}
\begin{split}
S(\omega) & =\int^{0}_{-\infty}dt\left[\langle\hat{I}(t)\hat{I}(0)\rangle+\langle\hat{I}(0)\hat{I}(t)\rangle\right]\left(e^{-i\omega t}+e^{+i\omega t}\right) \\
& \quad-4\pi\delta(\omega)\langle\hat{I}\rangle^{2}\quad.
\end{split}
\label{eq:noise_FT}
\end{equation}
By construction, the finite-frequency current noise Eq.~(\ref{eq:noise_FT}), also referred to as the power spectral density, is symmetric with respect to frequency, $S(\omega)=S(-\omega)$. It represents a real quantity, which can be measured by a classical detector.\cite{Aguado00,Gavish00} 

Experimentally, the current noise can be measured in the normal lead. However, at finite-frequencies so-called displacement currents appear and the tunneling current $\hat{I}$ is not equal to the measured currents. The displacement current can be included in the calculation by means of the Ramo-Shockley theorem~\cite{Blanter00,Shockley38}. 
A derivation of the Ramo-Shockley theorem for the effective Hamiltonian Eq.~\eqref{eq:H_eff} can be found in Appendix~\ref{App:diss_current}. 
In the following we assume the capacitance of the superconducting junction to be much larger than the capacitance of the normal junction. In this case the displacement current in the normal-conducting lead can be neglected, as discussed in Appendix~\ref{App:diss_current}. This assumption is consistent with $\Gamma_\mathrm{S}\gg\Gamma_\mathrm{N}$. 

In order to calculate the current correlator, two current operators at different times have to be placed on the Keldysh contour. Diagrammatically, this means that two internal tunnelling vertices have to be replaced by external current vertices. The contributions to the current-current correlator can be grouped into two different classes, as shown in Fig.~\ref{fig:noise}~(b) and (c). Either both current vertices are placed in the same irreducible block or in two different ones separated by a propagator. 

These external operators are connected by additional bosonic (dashed) lines, carrying the frequency $\omega$ of the Fourier transform. 
The symmetrized finite-frequency noise can be written as~\cite{Braun06}
\begin{eqnarray}
S(\omega) & = & \frac{1}{2}\mathrm{Tr}\left[\boldsymbol{W}_{\RNum{2}}(\omega)\boldsymbol{P}_\mathrm{stat}+\right.\nonumber\\
&&\left.\boldsymbol{W}_{\RNum{1<}}(\omega)\boldsymbol{\Pi}(\omega)\boldsymbol{W}_{\RNum{1>}}(\omega)\boldsymbol{P}_\mathrm{stat}\right]\nonumber\\
&&-2\pi\delta(\omega)\langle\hat{I}\rangle^{2}+(\omega\rightarrow -\omega)\quad.
\label{eq:noise_diagram}
\end{eqnarray}
Here, the kernels $\boldsymbol{W}_{\RNum{1>}}(\omega)$  and $\boldsymbol{W}_{\RNum{1<}}(\omega)$ are the sum of all diagrams, where one tunnel vertex (black dot) is replaced by a current vertex (open circle) and a frequency line $\omega$ is attached to the current vertex. The indices $>$ and $<$ indicate whether the frequency line leaves the diagram to the right or enters it from the left as shown in Fig~\ref{fig:noise}~(c). The kernel $\boldsymbol{W}_{\RNum{2}}(\omega)$ contains diagrams with both current vertices in the same irreducible block (see Fig.~\ref{fig:noise}~(b)).

\section{Andreev current} \label{sec:current}
\begin{figure}[b]
\centering
  \includegraphics[width=0.4\textwidth]{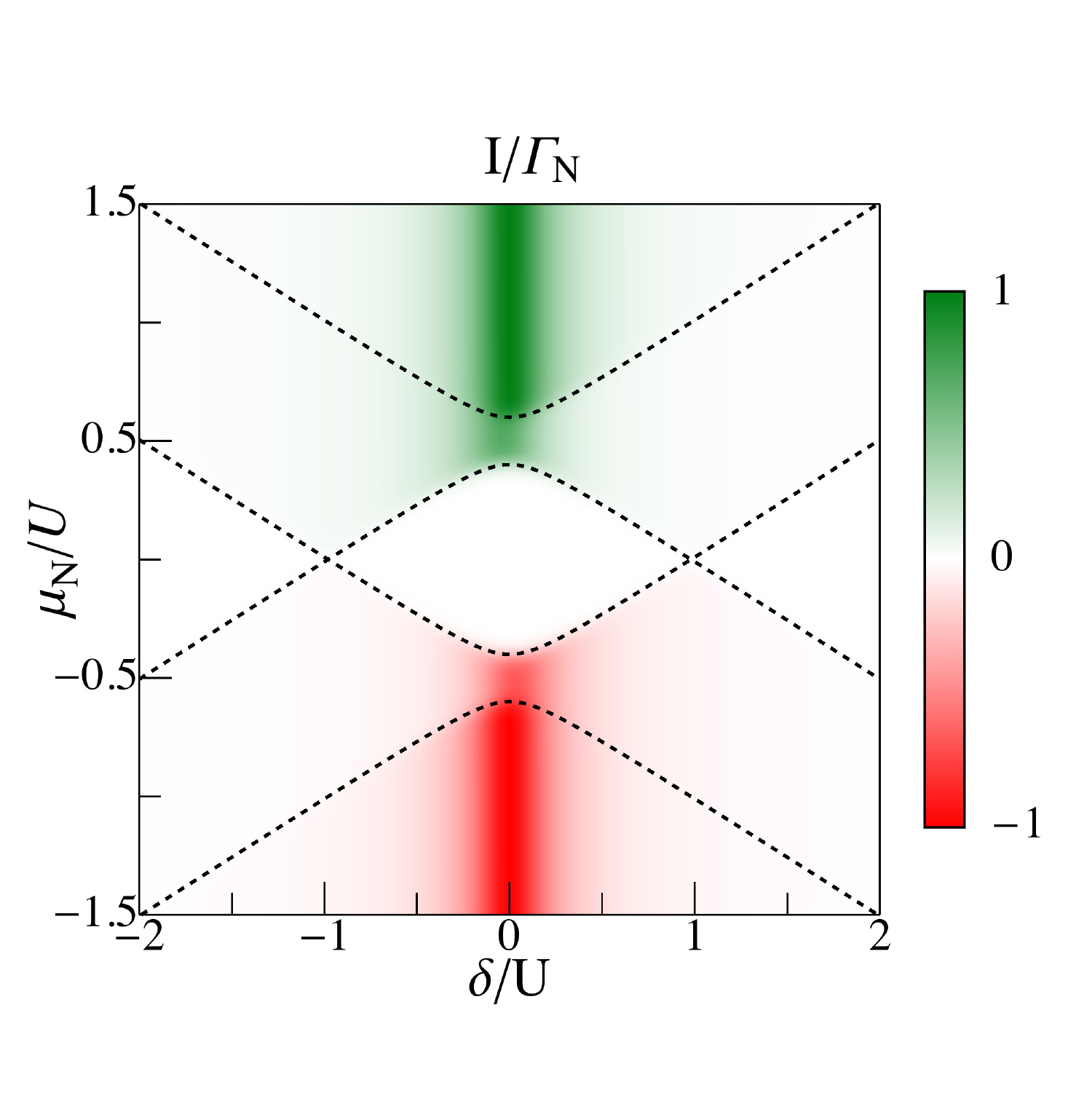}
   \caption{(color online). Density plot of the Andreev current as a function of the detuning $\delta$ and the chemical potential of the normal lead $\mu_\text{N}$ both in units of the Coulomb interaction strength $U$. The other parameters are $\Gamma_\text{S}=0.2U$, $\Gamma_\text{N}=0.002U$ and $k_\text{B}T=0.02U$.}
	\label{fig:current}
\end{figure}
Before discussing the finite-frequency current noise, we give a brief overview over the properties of the Andreev current. When a finite bias voltage is applied across the quantum dot with one superconducting and one normal-conducting lead, a so called Andreev current flows across the structure, which is due to Cooper-pair tunnelling between quantum dot and superconductor caused by Andreev reflection processes.~\cite{Pala07,governale08} 

We determine the current through the single-level quantum dot by using Eq.~\eqref{eq:current_average}, and show the result  in Fig.~\ref{fig:current} as a function of the chemical potential of the normal lead, $\mu_\text{N}$, and the detuning, $\delta$. 
The current is largest when superconducting correlations on the dot are strong and it furthermore shows features at the Andreev addition energies.

The Andreev addition energies (dashed lines in Fig.~\ref{fig:current}) are symmetric around zero bias voltage, $\mu_\text{N}=0$, and with respect to zero detuning, $\delta=0$.
In the region around zero bias voltage the system is mainly in one of the singly-occupied states, $|\sigma\rangle$, since the charging energy
 suppresses transitions from $|\sigma\rangle$ to the $|\pm\rangle$ states. The  Andreev current is thus zero. Only when the bias voltage is large enough, such that one of the conditions $\mu_{\text{N}} \gtrsim E_{+-}$ or $\mu_{\text{N}} \lesssim E_{-+}$ is fulfilled, 
 the quantum dot has a finite probability to be either empty or doubly occupied and the Andreev current sets in. A further increase of the Andreev current is observed, when also the other two addition energies, $E_{++}$ and $E_{--}$, enter the bias window.

Outside the region where the current is suppressed due to the charging energy, the current is largest for $\delta\sim 0$, the regime of strongest superconducting correlation.  We obtain a simple analytic result for the Andreev current in the unidirectional transport regime, namely when $\mu_\text{N}\gg E_{+,+}$, where the applied bias voltage to the normal conducting lead is much larger than all other energy scales in the system apart from the superconducting gap $\Delta$,
\begin{equation}
I_\text{uni}=\Gamma_\text{N}\frac{\Gamma^{2}_\text{S}}{4\epsilon^{2}_\text{A}}. 
\label{eq:current_high_bias}
\end{equation}
As required, this result matches the current displayed in Fig.~\ref{fig:current}, for $\mu_\text{N}\gg E_{++}$.
Indeed, the current in the unidirectional regime, Eq.~\eqref{eq:current_high_bias}, is maximal for $|\delta|\ll\Gamma_\text{S}$, namely when $2\epsilon_\text{A}$, the splitting between the ABSs, is minimal and just given by the coupling strength to the superconducting lead $\Gamma_\text{S}$. 
In this situation, the empty $|0\rangle$ and doubly-occupied state $|d\rangle$ are nearly degenerate and hence the mixing between them is maximal (the proximity effect is on resonance). 
If the detuning $\delta$ becomes large, i.e. $|\delta|\gg\Gamma_{\text{S}}$, the superconducting correlations on the dot are almost zero (the proximity effect  is off-resonance) and the Andreev current goes to zero as shown in Fig.~\ref{fig:current}.

The value of the Andreev addition energies can however only roughly be extracted from the current. We will show in Secs. \ref{sec:unidirectional}, \ref{sec:quantum_noise} and \ref{sec:high_bias} that they lead to sharp features in the noise spectrum.

\section{Results for the finite-frequency noise} \label{results}

In this section we come to the actual focus of the present paper, the finite-frequency noise associated to the current flow through the hybrid quantum-dot system, which we calculate based on the diagrammatic real-time approach introduced before.

 The discussion of the finite-frequency noise is divided into three parts: the unidirectional transport regime, where the applied bias voltage $\mu_\text{N}$ is chosen such that no back tunnelling to the normal lead is allowed,  the finite bias regime, where the applied bias voltage can be of the same order of the Andreev addition energies and the noise frequency, and the low-bias regime, where the current through the dot is suppressed. 
All regimes are shown to provide direct access to the internal dynamics of the system. 

Depending on the applied bias voltage different frequency regimes of the noise are accessible. Table~\ref{table:regimes} gives an overview over the different types of noise depending on the characteristic energy scales of the system, the thermal energy $k_\text{B}T$, the energy related to the noise frequency $\omega$, and the applied bias voltage $\mu_\text{N}$, which we are going to discuss within the different parts of this section. 
\begin{table}[t]
 \begin{tabular}{  | p{2.5cm} | p{1.85cm} | p{1.85cm} | p{1.75cm}  | }
    \hline
    \diaghead{\theadfont Frequency Frequency}{Bias\\ regime}{Frequency} & \thead{Low} & \thead{Intermediate} & \thead{High, $\omega>$} \newline $\mu_\text{N}$,$k_\text{B}T,\Gamma_\text{N}$\\ \hline
    Unidirectional, $\mu_\text{N}$ largest energy scale  &  $\omega\leq\Gamma_\text{N}$ \newline (shot noise) \newline Sec.~\ref{sec:low_freq} &  $\omega>\Gamma_\text{N}$ \newline(shot noise) \newline Sec.~\ref{sec:inter_freq}&\ \phantom{N/A}\newline \ \phantom{N/A}\newline N/A \\ \hline
       Finite bias & $\omega<\mu_\text{N}$ $\&$\newline $\omega\leq k_\text{B}T,\Gamma_\text{N}$\newline (thermal $\&$ shot noise)  \newline  Sec.~\ref{sec:high_bias} & $\mu_\text{N}>\omega$ $\&$\newline $\omega>k_\text{B}T,\Gamma_\text{N}$\newline (shot noise) \newline  \ \newline Sec.~\ref{sec:high_bias} &   \ \newline\ \newline (quantum noise)\newline  Sec.~\ref{sec:high_bias}  \\  \hline
     Zero and low bias & $\omega$$\leq$$ k_\text{B}T$ \newline(thermal noise) \newline Sec.~\ref{sec:zero_bias} &  \ \newline\ \newline \ \newline N/A &  \ \newline(quantum noise) \newline Sec.~\ref{sec:zero_bias} \\ \hline
 \end{tabular}
           \caption{Table sumarising different noise regimes depending on noise frequency (increasing from left to right) and applied bias voltage (decreasing from up to down). }
       \label{table:regimes}
           \end{table}

At low and intermediate frequencies the noise is dominated by time-dependent fluctuations in the conductance. In the limit of zero frequency ($\omega\rightarrow0$) the noise spectrum exhibits the information of a long-time measurement. 
In this frequency range, the equilibrium thermal noise, due to thermal fluctuations in the occupation number of the leads, is dominant in the spectrum if $k_\text{B}T\gg \mu_\text{N},\omega$. 
In contrast, the so-called non-equilibrium shot noise, which is due to charge quantization, is dominant for $\mu_\text{N} \gg k_\text{B}T,\omega$. 

The quantum noise, which arises from zero-point fluctuations in the device, is dominant for high frequencies $\omega\gg k_\text{B}T, \mu_\text{N}$. It is a measure of the ability of the system to absorb or to emit a certain energy $\omega$,~\cite{Clerk10} and will therefore allow to visualize transport processes which are enabled or blocked by energy absorption or emission.

Although we here consider the symmetrized noise, we refer to the regime of high frequencies $\omega\gg k_\text{B}T, \mu_\text{N}$ as quantum noise, as discussed e.g. in Ref.~\onlinecite{Emary11}.
\subsection{Noise in the unidirectional transport regime}\label{sec:unidirectional}

We start our analysis with the unidirectional transport regime, where we set the chemical potential of the normal lead $\mu_\text{N}$ to be much larger than all relevant energy scales of the system (apart from the superconducting gap $\Delta$). 
In particular, since $\mu_\text{N}\gg E_{+,+}$, all Andreev levels are in the transport window and sufficiently far away from the chemical potential $\mu_\text{N}$, thus allowing no back tunnelling from the dot into the normal conducting lead. 
 A sketch of this situation is shown in Fig.~\ref{fig:model}.

\begin{figure}[htbp]
\centering
  \includegraphics[width=0.4\textwidth]{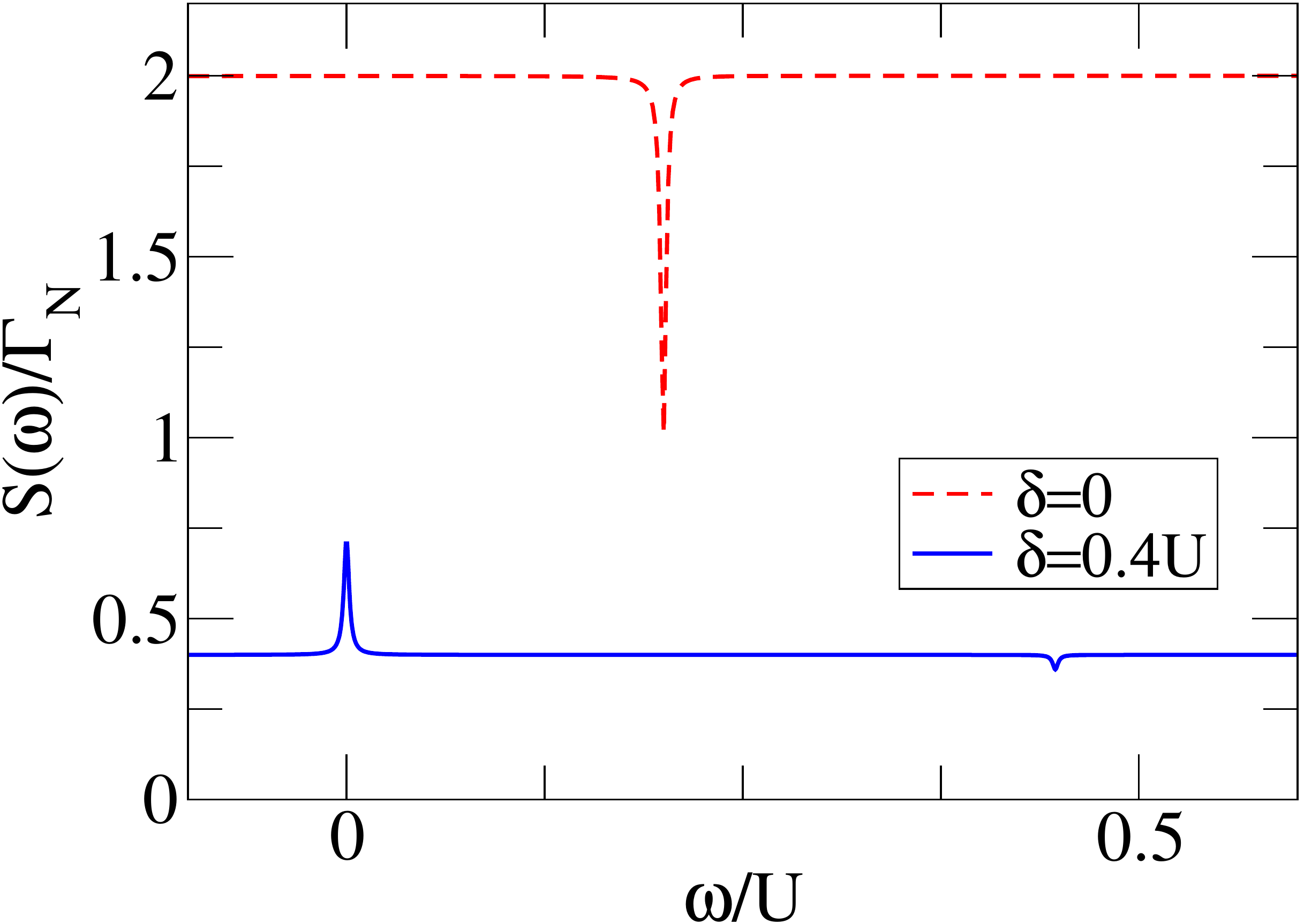}
  \caption{(color online). Finite-frequency noise $S(\omega)$ in the unidirectional transport regime for $\Gamma_\text{S}=0.2U$, $\Gamma_\text{N}=0.002U$ both for the case where the proximity effect is on resonance, $\delta=0$, and off resonance, $\delta=0.4U$. The peaks and dips are located at $\omega=0$ and at $\omega=\pm2\epsilon_\text{A}$, the oscillation frequency of the Cooper pairs. Here and in the following figures we concentrate on the positive-frequency part of the spectrum.}
	\label{fig:Fano_high_coherences}
\end{figure}

The finite-frequency noise in the unidirectional transport regime is given by
\begin{widetext}
\begin{equation}\label{eq:Fano_high_bias}
\begin{split}
\frac{S(\omega)}{2 I_\mathrm{uni}} = 
 1+\frac{\Gamma^{2}_\text{N}\delta^{2}}{4\epsilon^{2}_\text{A}(\Gamma^{2}_\text{N}+\omega^{2})}-\frac{1}{2}\frac{\Gamma_\text{S}^2}{4\epsilon^{2}_\text{A}}\left[\frac{\Gamma^{2}_\text{N}}{\Gamma_\text{N}^{2}+(\omega-2\epsilon_\text{A})^{2}}\left(1-\frac{\omega-2\epsilon_\text{A}}{\epsilon_\text{A}}\right)+\frac{\Gamma^{2}_\text{N}}{\Gamma_\text{N}^{2}+(2\epsilon_\text{A}+\omega)^{2}}\left(1+\frac{\omega+2\epsilon_\text{A}}{\epsilon_\text{A}}\right)\right]  .
\end{split}
\end{equation}
\end{widetext}
Fig.~\ref{fig:Fano_high_coherences} shows the finite-frequency noise in units of $\Gamma_\text{N}$ as a function of the noise frequency $\omega$. Two limiting cases are displayed, the one where the proximity effect on the dot is on resonance, $|\delta|\ll\Gamma_\text{S}$ (red dashed line), and the one where the proximity effect on the dot is off resonance, $|\delta|\gg\Gamma_\text{S}$ (blue solid line). In different frequency regimes, the spectrum shows sharp features, which depend on the strength of the detuning, $\delta$. In the following subsections we discuss first the low-frequency noise, followed by a discussion of the intermediate-frequency regime. 

\subsubsection{Low-frequency noise, $\omega\lesssim\Gamma_\text{N}$}\label{sec:low_freq}
The first part of Eq.~\eqref{eq:Fano_high_bias}  is the low-frequency contribution to the current noise, 
\begin{equation}
S(\omega)=2\Gamma_\text{N}\frac{\Gamma^{2}_\text{S}}{4\epsilon^{2}_\text{A}}\left(1+\frac{\delta^{2}\Gamma^{2}_\text{N}}{4\epsilon^{2}_\text{A}(\Gamma^{2}_\text{N}+\omega^{2})} \right)\ .
\label{eq:Fano_low_highbias}
\end{equation}
It is indeed the only contribution to the noise, when the noise frequency is of the order of the coupling strength to the normal-conducting lead, $\omega\lesssim\Gamma_\text{N}$. Note that in the low-frequency noise for unidirectional transport only shot noise is present. 

On resonance ($\delta\approx0$), when the Andreev current is maximal in the high-bias regime, see Fig.~\ref{fig:current}, the noise is frequency independent and is given by two times the Andreev current $2 I_\text{uni}$, which in the limit of zero-detuning discussed here equals $2\Gamma_\text{N}$. This means that the noise equals the long-time measurement result ($\omega\rightarrow 0$) over the whole low-frequency range. This effect has previously been discussed in Ref.~\onlinecite{Braggio11}: if the proximity effect is on resonance, the superconducting correlations on the dot are maximal and Cooper pairs oscillate rapidly between dot and superconductor.  This oscillation of Cooper pairs is only interrupted by single-electron tunnel events from the normal conducting lead to the dot. It is these independent charge injections which give rise to a Poissonian transfer of single electrons. 

When the proximity effect is off resonance ($|\delta|\gg\Gamma_\text{S}$), the low-frequency noise can be approximated by
\begin{equation}
S(\omega)\approx 2\Gamma_\text{N}\frac{\Gamma^{2}_\text{S}}{\delta^2}\left(1+\frac{\Gamma^{2}_\text{N}}{\Gamma^{2}_\text{N}+\omega^{2}}\right) .
\end{equation}
The noise spectrum shows a Lorentzian dependence on the frequency $\omega$, as shown by the low-frequency contribution of Fig.~\ref{fig:Fano_high_coherences} (solid blue line). This maximum has a  width given by the coupling strength $\Gamma_\text{N}$ and a height scaling with the magnitude of the Andreev current. Except for this maximum, the noise is overall suppressed with respect to the case on resonance. The reason for this is that in the unidirectional transport regime if $\delta\gg\Gamma_\text{S}$, the Andreev current becomes negligibly small with increasing detuning, as depicted in Fig.~\ref{fig:current}.

This behaviour is similar to the case of a quantum dot with normal conducting leads only,~\cite{Blanter00,Braun06} as presented in the Appendix~\ref{subsec:unidirectional_Anderson_dot}. In this purely normal-conducting case the low-frequency noise shows a Lorentzian behaviour if the coupling to the two leads is asymmetric (similar to what we observe in the case of finite detuning in the hybrid system). The noise is frequency-independent, when the coupling to the leads is symmetric. However the constant is only half as big as in the hybrid case discussed here, due to the absence of Cooper pairs in the system.

\subsubsection{Intermediate-frequency regime, $\omega\gg\Gamma_\text{N}$}\label{sec:inter_freq}
The remaining part of Eq.~\eqref{eq:Fano_high_bias} carries the information of the noise in the intermediate-frequency regime, $\omega\gg\Gamma_\text{N}$, where the noise frequency becomes larger than the coupling strength to the normal-conducting lead. In this regime,  the noise starts to reveal the internal dynamics of the quantum dot. Indeed, the finite-frequency noise shows resonance dips, whose position and size depend on the splitting of the Andreev bound states (equal to the frequency of the coherent oscillation between the empty and doubly-occupied dot states~\cite{Rajabi13}), $2\epsilon_\text{A}$,  and hence on the strength of the proximity effect, see Fig.~\ref{fig:Fano_high_coherences}. Such dips are characteristic for the noise spectrum of multilevel quantum dots.~\cite{Wohlman07,Gabdank11,Marcos11} In the case of a hybrid quantum-dot system studied here, the dips arise from a coherent destructive interference between the ABS, leading to features at  $\omega=\pm|\epsilon_{+}-\epsilon_{-}|=\pm2\epsilon_\text{A}$.   In the following we describe the properties of these dips.

Similar to the noise-enhancing peak in the low-frequency regime, the shape of these noise-suppressing resonance dips is Lorentzian, see Eq.~\eqref{eq:Fano_high_bias}, with a width given by $\Gamma_\text{N}$.  
The depth of the resonance dip depends on the strength of the proximity effect and is equal to $I_\mathrm{uni}\Gamma_\text{S}^2/4\epsilon^{2}_\text{A}$, the prefactor of the intermediate-frequency contribution in Eq.~\eqref{eq:Fano_high_bias}.
The resonance dip becomes most prominent if the proximity effect is on resonance and its depth is maximally equal to $\Gamma_\text{N}$ for $\delta\approx0$. In contrast, the dip vanishes if the detuning $\delta$ becomes much larger than the coupling strength to the superconducting lead $\Gamma_\text{S}$, namely when the superconducting correlations on the dot are almost zero, see Fig.~\ref{fig:Fano_high_coherences}.  
The resonance dip in the spectrum hence indicates the strength of the proximity effect. It is a signature of the coherent oscillation of Cooper pairs between dot and superconductor. 

For even higher frequencies, $\omega\gg2\epsilon_\text{A}$, the noise in the unidirectional transport regime is given by two times the Andreev current, which depends strongly on the detuning $\delta$, see Figs.~\ref{fig:current} and \ref{fig:Fano_high_coherences}.

In the unidirectional transport regime the noise spectrum can be used to extract the splitting of the ABS, but not the individual Andreev addition energies. In order to get the information of the excitation energies and the effective coupling strengths of the ABSs, also back tunnelling to the normal-conducting lead must be allowed. Hence, in the next two subsections we will  consider the regime where the electrochemical potential of the normal lead is not the largest energy scale any more and the frequency can become larger than the distance between the Andreev addition energies and the transport voltage, $\omega> |E_{\pm,\pm}-\mu_\text{N}|$. In this regime quantum noise can become the dominant contribution to noise. 

\subsection{Zero and low bias regime, $\mu_\text{N}<E_{+,-}$}\label{sec:quantum_noise}\label{sec:zero_bias}

\subsubsection{Zero bias regime}

\begin{figure}[t]
\subfigure{ \includegraphics[height=0.25\textwidth]{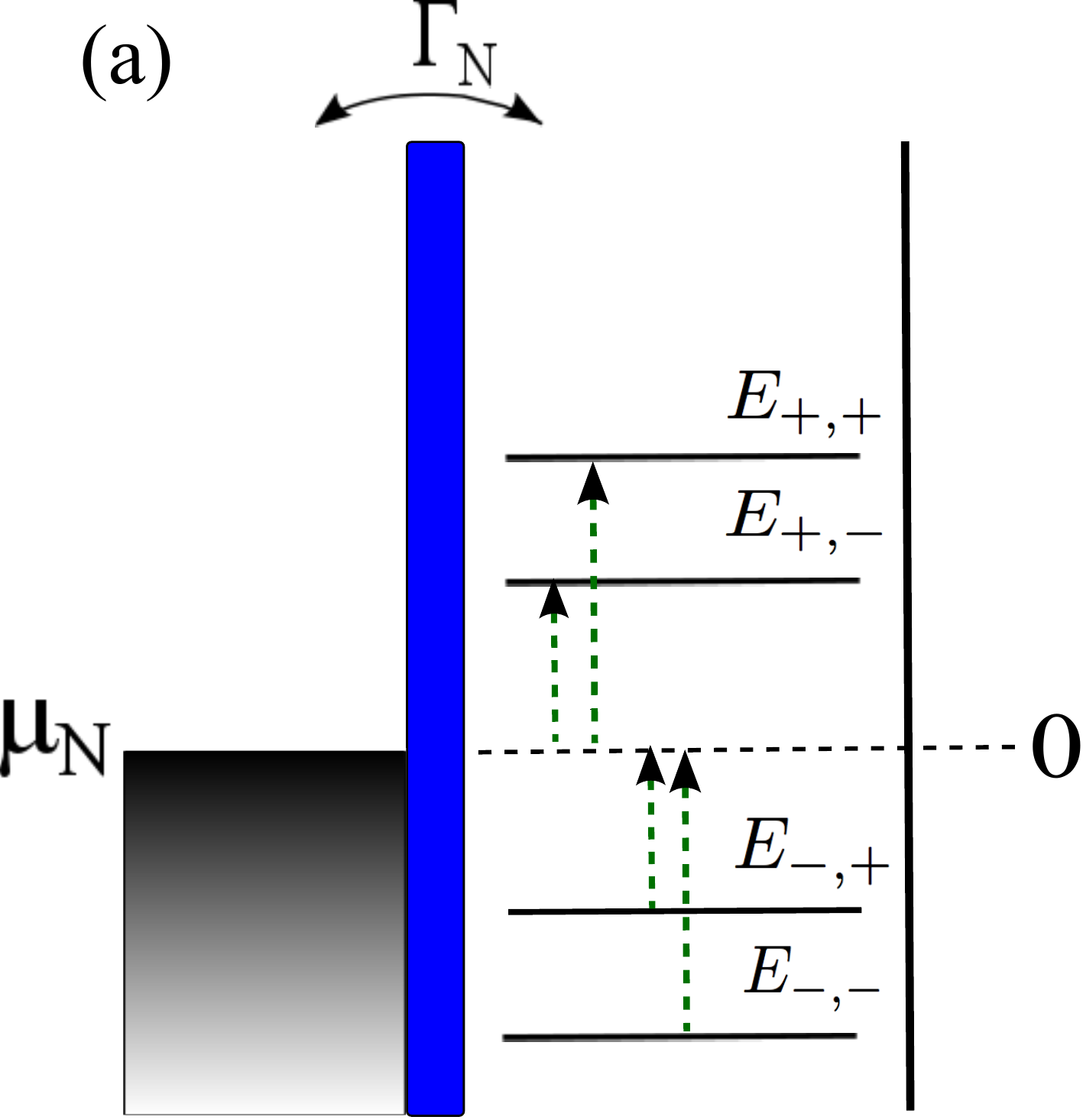}}
\subfigure{ \includegraphics[height=0.3\textwidth]{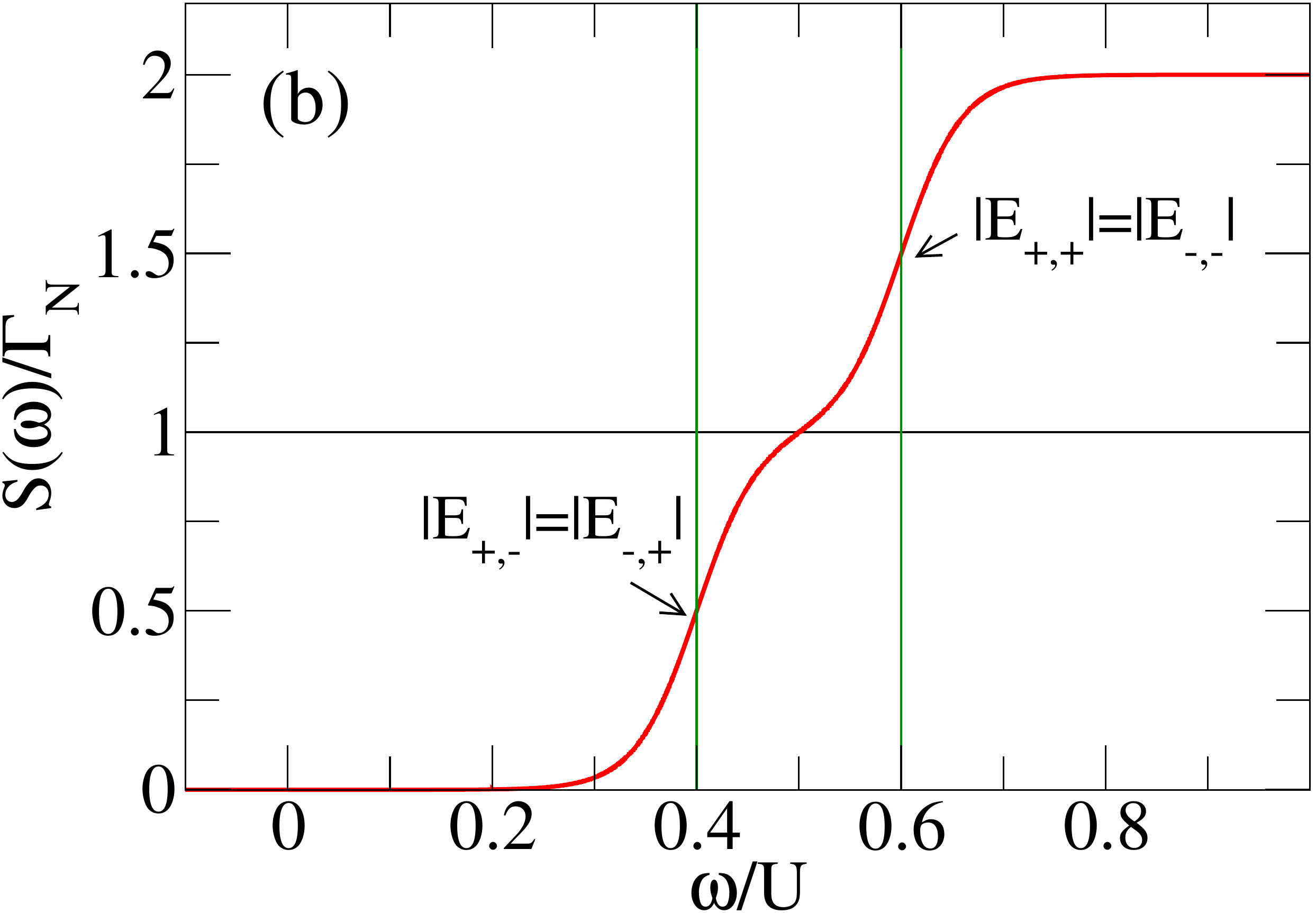}}
\caption{(color online). (a) Sketch of the energy landscape of the  proximized single-level quantum dot for $\mu_\text{N}=0$. (b) Finite-frequency noise $S(\omega)$ with $\Gamma_\text{N}=0.0002U$, $\Gamma_\text{S}=0.2U$, $\delta=0$, $k_\text{B}T=0.03U$; the spectrum has quantum noise steps at $\omega=E_{+,-}$ and $\omega=E_{+,+}$ indicated by the green vertical lines. }
\label{fig:V0}
\end{figure}

If the transport voltage of the normal lead goes to zero, $\mu_\text{N} \rightarrow 0$, the dot is in the singly-occupied state $|\sigma\rangle$, see Fig.~\ref{fig:V0}~(a) for a sketch of the energy landscape of the system. 
Hence the transport excitation energies $E_{\pm,\pm}$ are outside the bias window and transport is blocked, meaning that the Andreev current is zero as shown in Fig.~\ref{fig:current}.
In this limit shot noise is negligible and quantum noise is dominant in the spectrum, since $\omega\gg k_\text{B}T$, $\mu_\text{N}$. 
Also the thermal noise, which is expected to be dominant at zero bias up to a noise frequency of $\omega=k_\text{B} T$, is here suppressed, since no dot excitation energy is close enough to the Fermi energy to allow for thermal excitations of the system. 

We show the finite-frequency noise in this regime in Fig.~\ref{fig:V0} and we observe that the noise spectrum has steps at frequencies $\omega=|E_{\pm,\pm}|$ equal to the Andreev addition energies of the system. This behaviour is typical for the high-frequency noise of a system in which transport is blocked. At certain noise frequencies the effect of new "noisy" channels becomes visible, leading to steps increasing the noise. The steps thus reflect the internal structure of the energy levels on the quantum dot. Here they occur at frequencies equal to the Andreev addition energies of the dot-superconductor subsystem as described in detail in the following. An analogy to the normal-conducting case with symmetrically and asymmetrically coupled leads can be found in Appendix~\ref{app_lowbias}.

In the limit $\mu_\text{N} \rightarrow 0$, the noise is suppressed until the noise frequency is equal to the energy which is necessary to excite the dot from the singly-occupied state $|\sigma\rangle$ to the $|-\rangle$ state, $\omega=|E_{+,-}|$. Equally, also the inverse process, namely the excitation from the ABS $|-\rangle$ into the singly-occupied state $|\sigma\rangle$ yields a contribution to the noise. It takes place at $\omega=|E_{-,+}|$. These excitation energies are however degenerate in the zero-bias limit, see Fig.~\ref{fig:V0}~(a), and consequently only one step occurs at the noise frequency $\omega=|E_{-,+}|=|E_{+,-}|$. The step height is given by the sum of the respective effective coupling strengths, see Eqs.~\eqref{eq:Gamma_effective1}  and \eqref{eq:Gamma_effective2}, $\Gamma_{\sigma\rightarrow-}+\Gamma_{-\rightarrow\sigma}=\Gamma_\text{N}$.~\footnote{Note that we here consider a symmetrized noise spectrum. Therefore both contributions, independently on whether the excitation energy of a process is positive or negative, enter the noise spectrum with equal weight at positive frequencies.} 

A second step takes place at $\omega=|E_{+,+}|=|E_{-,-}|$, the energy necessary for the excitation between a singly-occupied state and the ABS $|+\rangle$. The height of the second step is again given by  $\Gamma_\text{N}$, the sum of the effective coupling strengths, $\Gamma_{\sigma\rightarrow+}+\Gamma_{+\rightarrow\sigma}=\Gamma_\text{N}$.

Consequently, when the noise frequency is larger than the energy which needs to be provided to excite between any of  the singly-occupied states and the ABSs, the noise is constant and given by the sum of all four effective coupling strengths, $2\Gamma_\text{N}$.~\cite{Engel04,Jin12,Dong13} 

\subsubsection{Low bias regime, $0<\mu_\text{N}<E_{+,-}$}

\begin{figure}[t]
\subfigure{ \includegraphics[height=0.25\textwidth]{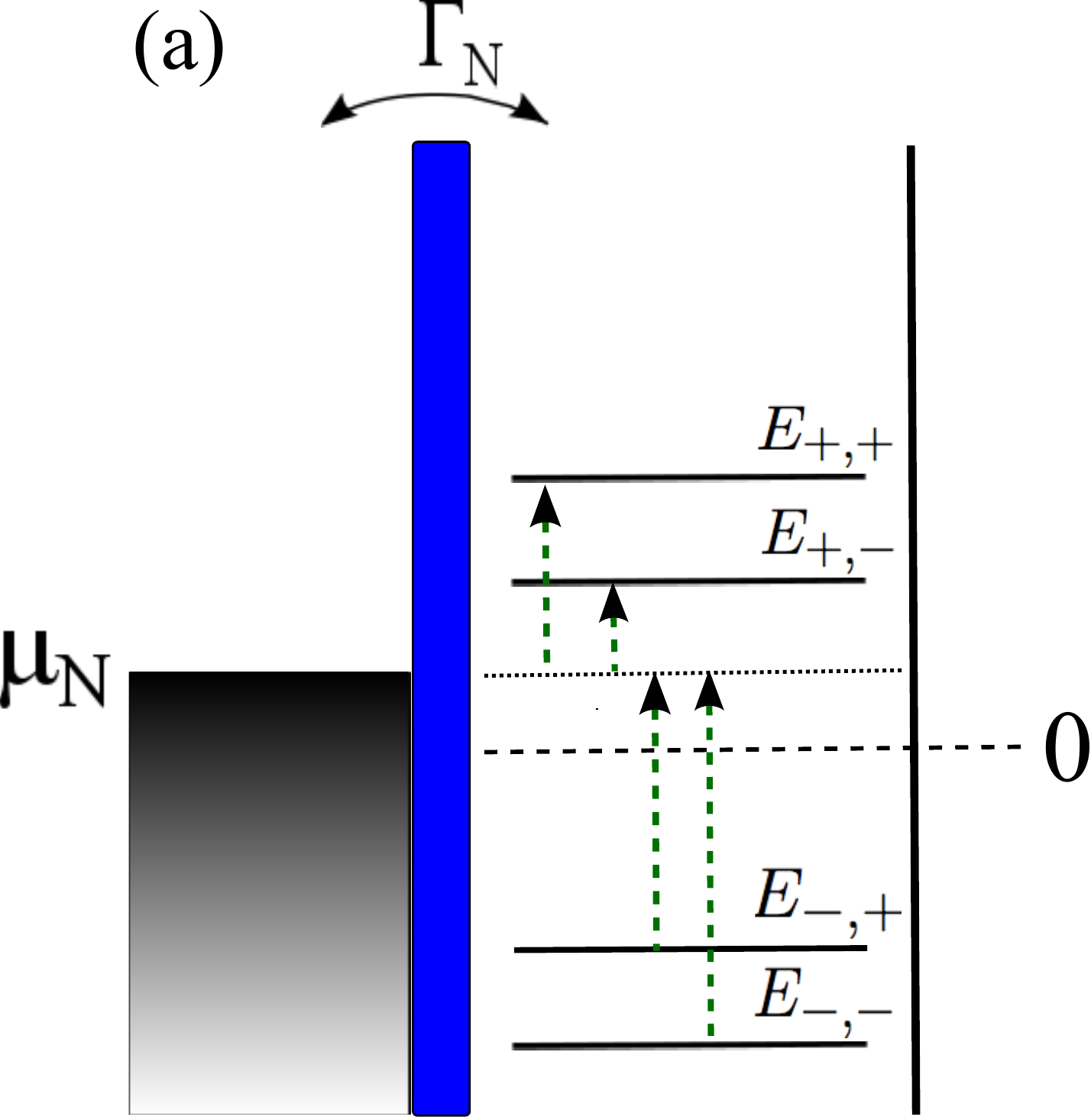}}
\subfigure{ \includegraphics[height=0.3\textwidth]{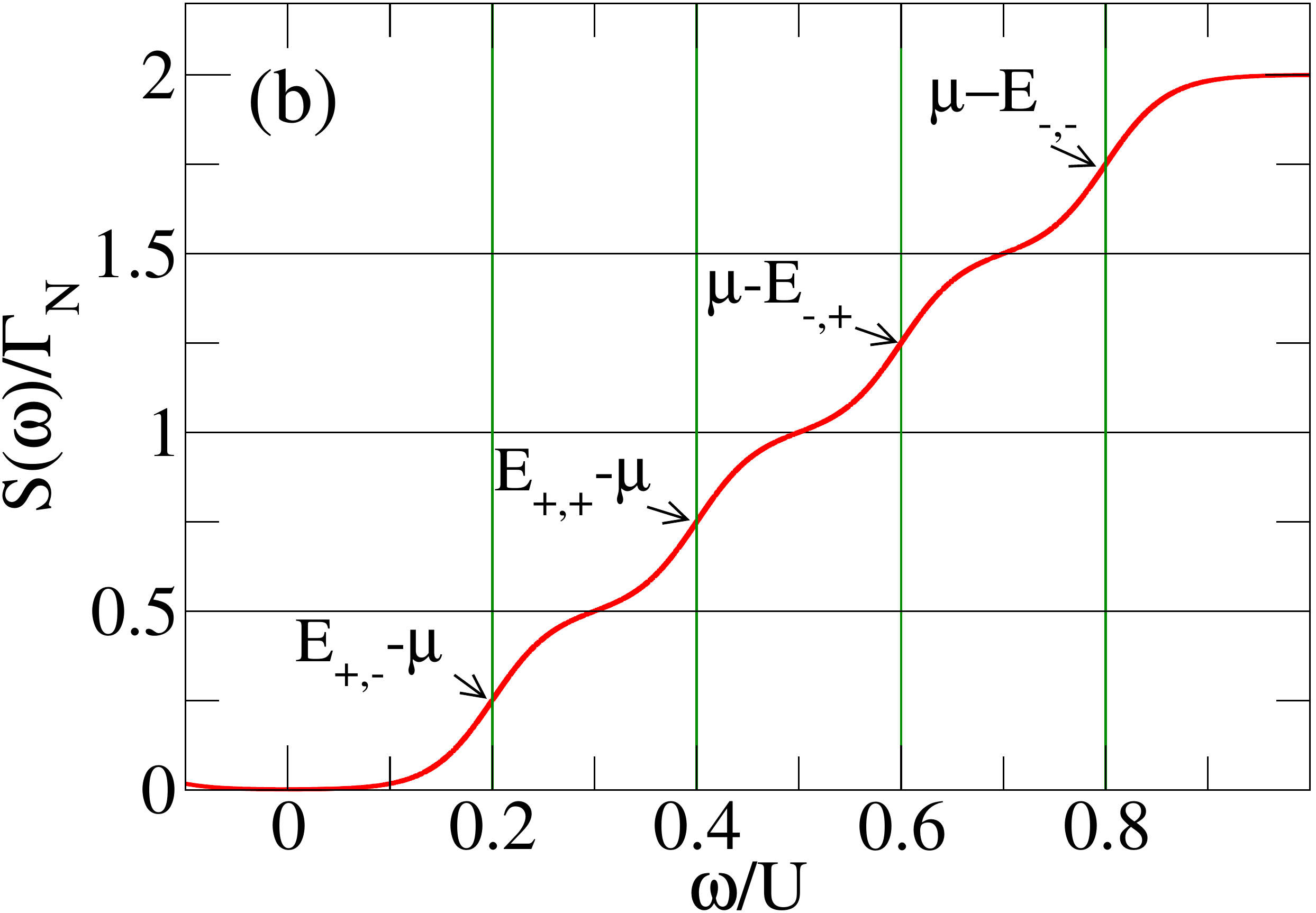}}
\caption{(color online). (a) Sketch of the energy landscape of the proximized single-level quantum dot for $\mu_\text{N}=0.2U$. (b)  Finite-frequency noise $S(\omega)$ with $\Gamma_\text{N}=0.0002U$, $\Gamma_\text{S}=0.2U$, $\delta=0$, $k_\text{B}T=0.03U$. }
\label{fig:V02}
\end{figure}
\begin{figure}[htbp]
\subfigure{ \includegraphics[height=0.25\textwidth]{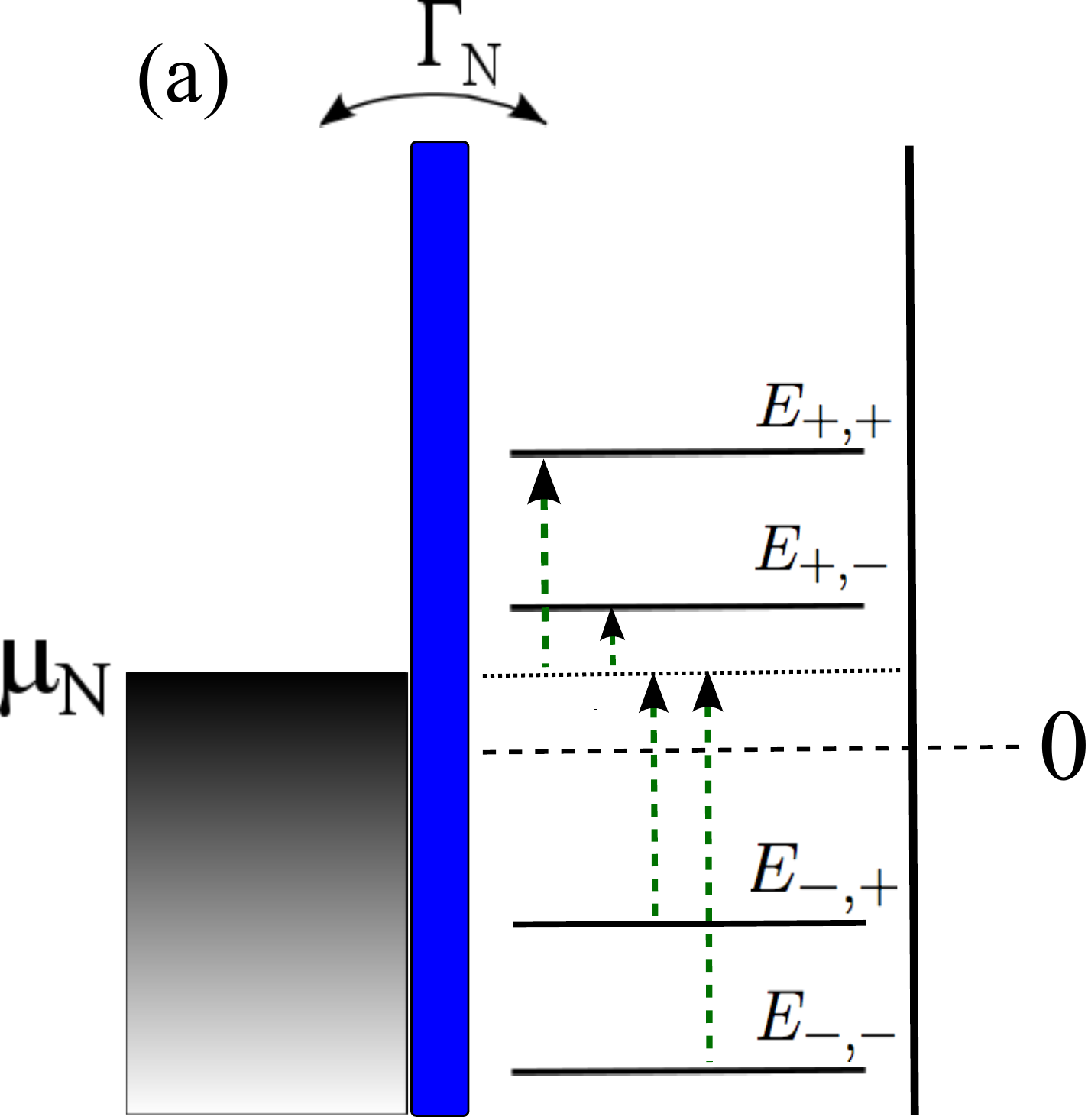}}
\subfigure{ \includegraphics[height=0.3\textwidth]{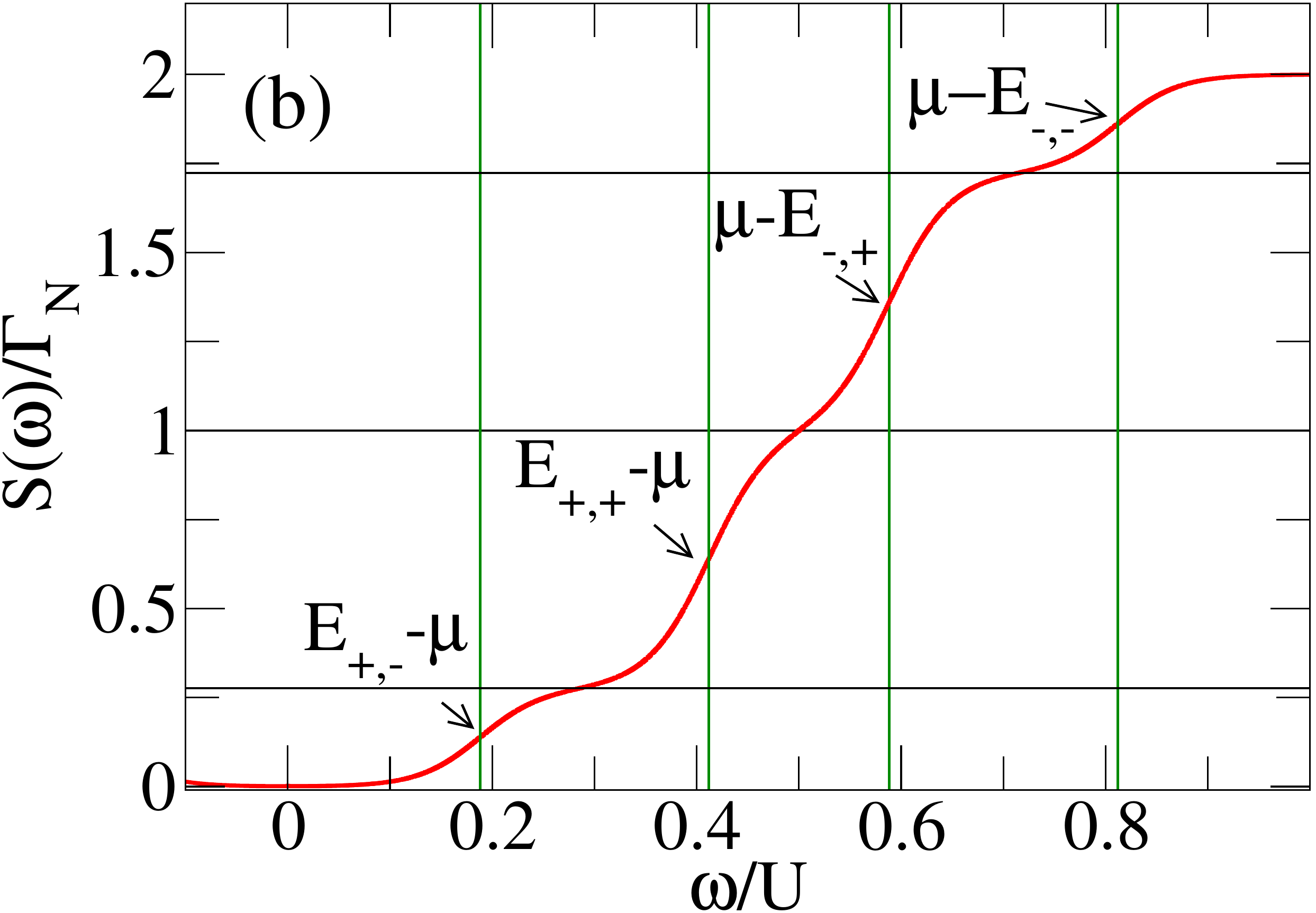}}
\caption{(color online). (a) Sketch of the energy landscape of the proximized single-level quantum dot for $\mu_\text{N}=0.2U$ and $\delta=0.1U$. (b)  Finite-frequency noise $S(\omega)$ with $\Gamma_\text{N}=0.0002U$, $\Gamma_\text{S}=0.2U$, $k_\text{B}T=0.03U$.}
\label{fig:V02d}
\end{figure}

If a small (positive) bias voltage is applied to the normal-conducting lead, but the bias is still smaller than the energy  $E_{+,-}$, necessary to excite from the singly-occupied state $|\sigma\rangle$ into the ABS $|-\rangle$, the dot is singly occupied and the system is hence still in the region where the current is suppressed (see  Figs.~\ref{fig:V02}~(a) and \ref{fig:V02d}~(a) for the energy landscape of the system and Fig.~\ref{fig:current} for the respective behaviour of the current).

In this bias regime the noise spectrum exhibits four steps at noise frequencies $\omega=|E_{\pm,\pm}-\mu_\text{N}|$ as shown in Figs.~\ref{fig:V02}~(b) and \ref{fig:V02d}~(b). The reason for this is that a finite transport voltage breaks the degeneracy between the excitation energies that  is present for $\mu_\text{N}= 0$. This is indicated by the different lengths of the green dashed arrows in the energy-landscape sketches of the system.
The height of each of the four steps is given by the respective effective coupling strength, Eq.~\eqref{eq:Gamma_effective1}  and \eqref{eq:Gamma_effective2}. 

In the limit of zero detuning, $\delta=0$, the effective coupling to all four levels is equal, $\Gamma_{\sigma\rightarrow+}=\Gamma_{+\rightarrow\sigma}=\Gamma_{\sigma\rightarrow-}=\Gamma_{-\rightarrow\sigma}=\Gamma_\text{N}/2$. The step heights shown in the noise spectrum in Fig.~\ref{fig:V02}~(b) are therefore all equal to $\Gamma_\text{N}/2$. 

In contrast, if the detuning $\delta$ is finite, the effective coupling strengths differ.
For $\delta>0$, the coupling for the excitation to go from the singly-occupied state $|\sigma\rangle$ to the $|+\rangle$ state and to excite from the $|-\rangle$ state into the singly-occupied state $|\sigma\rangle$ is stronger than for the other two excitations, $\Gamma_{\sigma\rightarrow +}=\Gamma_{-\rightarrow\sigma}>\Gamma_{\sigma\rightarrow -}=\Gamma_{+\rightarrow\sigma}$. Consequently these first excitations give a larger contribution to the noise than the latter ones and the noise spectrum exhibits steps with different heights, see Fig.~\ref{fig:V02d}~(b). 

This behavior holds only as long as the detuning does not become much larger than the coupling strength to the superconducting lead. As soon as $\delta\gg\Gamma_\text{S}$, the noise spectrum exhibits again only two steps because the superconducting correlations on the quantum dot vanish. The effective coupling strengths of the Andreev levels corresponding to the excitation from $|\sigma\rangle$ to $|-\rangle$ and from $|+\rangle$ to $|\sigma\rangle$ go to zero, $\Gamma_{\sigma\rightarrow-},\Gamma_{+\rightarrow\sigma}\rightarrow0$, while the other two  excitations are coupled with the effective tunnel-coupling strength $\Gamma_\text{N}$ for $\delta\gg\Gamma_\text{S}$, see Eq.~\eqref{eq:Gamma_effective1}  and \eqref{eq:Gamma_effective2}.
When setting the bias voltage to 0, we find the previous result, as shown in Fig.~\ref{fig:V0}~(b). This is due to the fact that the coupling strength is here twice as large as in the case of zero detuning, however only half of the excitations contribute to the current and to the noise when $\delta\gg\Gamma_\text{S}$.

The finite-frequency noise spectrum in this low-bias regime provides a spectroscopy of the Andreev levels as well as the effective coupling strengths.  

Note, that in this case, namely when the Andreev levels are outside the bias window, the noise steps always lead to an increase of the noise, regardless of whether the noise step is related to a tunnel process between the reservoir and a strongly or a weakly coupled dot resonance. The reason for this is that we here observe features at noise frequencies, which always correspond to energies necessary to \textit{excite otherwise blocked transport channels} between dot resonance and reservoir.
This is different if some of the Andreev levels are in the bias window, as we will discuss in the next section.

\subsection{Finite-bias regime}\label{sec:high_bias}

We consider in this section two different situations for a finite transport voltage applied to the normal-conducting lead: first, the voltage is applied such, that all Andreev energies are in the bias window (high-bias regime) and second, such that part of the excitation energies lie outside the transport window (intermediate-bias regime). This allows us to study the full noise spectrum with its different contributions, similar to what we observed separately in the previous sections, Sec.~\ref{sec:unidirectional} and Sec.~\ref{sec:quantum_noise}.

The finite-frequency noise spectrum will be shown to provide a full spectroscopy of the system.

\subsubsection{High-bias regime, $\mu_\text{N}>E_{+,+}$ }
\begin{figure*}
\centering
\includegraphics[height=0.5\textwidth]{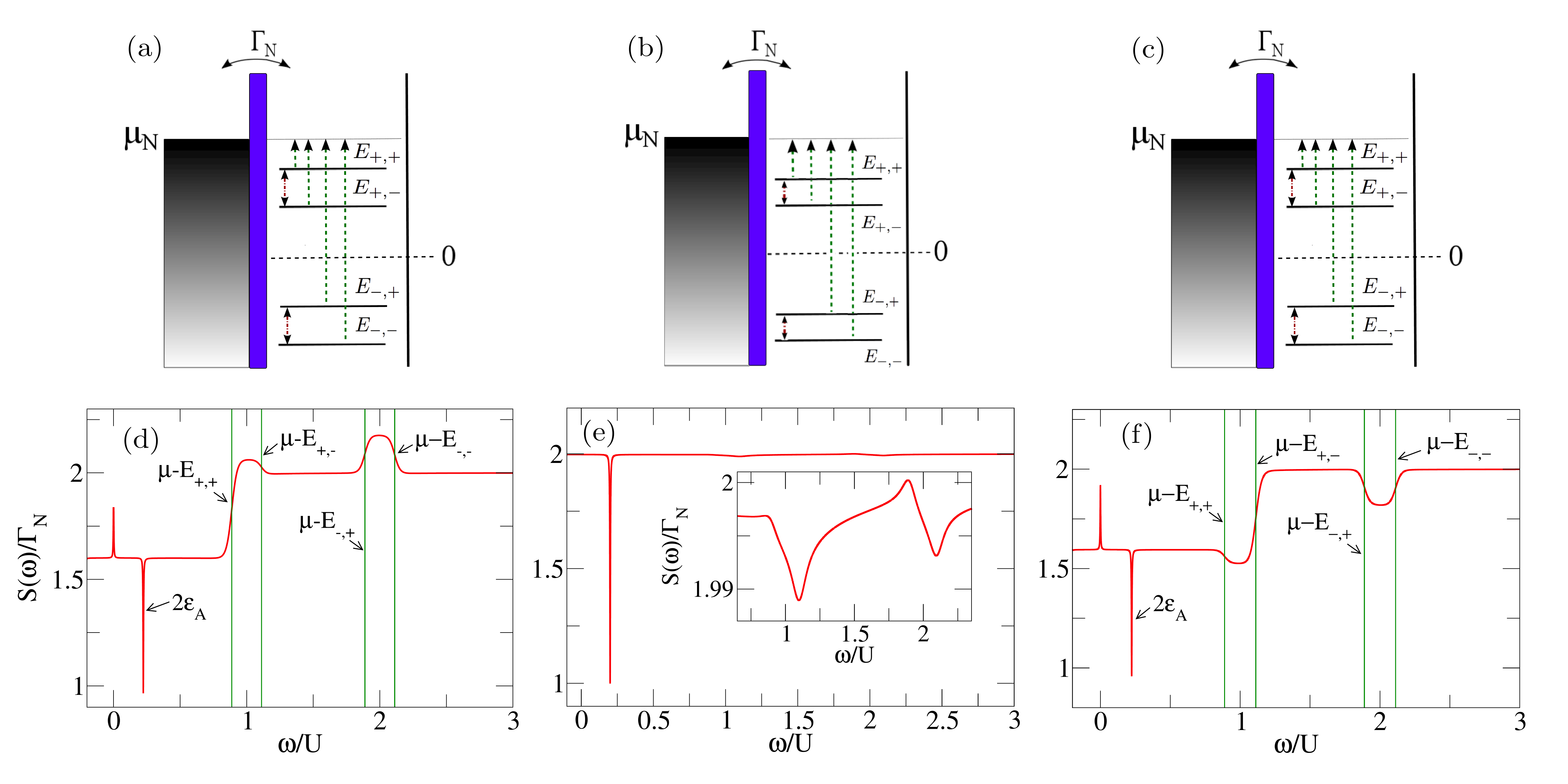}
\caption{(color online). Sketch of the energy landscape of the proximized single-level quantum dot for $\mu_\text{N}=1.5U$ and (a) $\delta=0.1U$, (b) $\delta=0$ and (c) $\delta=-0.1U$. Finite-frequency noise $S(\omega)$ with $\mu_\text{N}=1.5U$, $\Gamma_\text{N}=0.002U$, $\Gamma_\text{S}=0.2U$, $k_\text{B}T=0.02U$ and (d) $\delta=0.1U$, (e) $\delta=0$ and (f) $\delta=-0.1U$.}
\label{fig:finite_bias}
\end{figure*}

We now assume the dot to be in a regime where the bias, $\mu_\text{N}=1.5U$, is chosen such that all Andreev energies are in the bias window, see the sketches in Fig.~\ref{fig:finite_bias}~(a), (b) and (c). We consider the case where the superconducting correlations on the dot are strong, which is realized for the detuning being smaller than the coupling to the superconductor, $\delta<\Gamma_\mathrm{S}$. The results for the noise spectrum in this regime are shown in Fig.~\ref{fig:finite_bias}, for $\delta=0.1U$ (d), $\delta=0$ (e), or $\delta=-0.1U$ (f). In this regime the Andreev current is close to maximal, corresponding to the upper edge of the density plot in Fig.~\ref{fig:current}.

In the low- and intermediate-frequency regimes, $\omega<|E_{+,+}-\mu_\text{N}|$, the noise spectrum shows Lorentzian-shaped features, as discussed in Sec.~\ref{sec:unidirectional}. These are a  peak for a non-zero detuning $\delta$ in the low-frequency regime, $\omega<\Gamma_\text{N}$, and dips at a noise frequency equal to the splitting of the ABSs,  $\omega=|\epsilon_+ - \epsilon_-|= 2\epsilon_\text{A}$, due to a coherent destructive interference of the ABSs.

The high-frequency part of the noise spectrum of Fig.~\ref{fig:finite_bias}~(d) and (f) exhibits quantum noise steps at frequencies $\omega=|E_{\pm,\pm}-\mu_\text{N}|$, similar to what was discussed in Sec.~\ref{sec:quantum_noise}. 
However, in contrast to the previous section, where all steps lead to an increase of the noise, the quantum noise steps found here show different signs \textit{depending on the effective coupling strength}. A noise process related to a strongly coupled Andreev level leads to an increase of the noise, while a process between the electronic reservoir and a weaker coupled Andreev level decreases the noise.
The steps occurring in the finite-bias regime at high frequencies can be understood from an analogy to the ones obtained for a quantum dot coupled to normal-conducting leads. In this case an asymmetric coupling to the two normal-conducting leads takes the role of the differently coupled Andreev levels in the hybridised dot. See Appendix~\ref{sec:Anderson_noise} for a detailed discussion of this simpler case.

From this we deduce that the varying directions of the noise steps namely originate from the competition between different noise contributions of opposite sign. 
When a certain noise frequency $\omega$ is reached, an ensemble of transport processes becomes visible which can in principle involve different types of transitions, $|\sigma\rangle\leftrightarrow|\pm\rangle$, due to tunneling with the normal lead.
The contribution to the noise which stems from correlations of tunnelling processes  involving only one type of transitions tends to increase the noise, while the noise contribution stemming from correlations of tunnelling processes with different transitions tends to decrease the noise. Depending on which of these contributions has the larger magnitude, the step is positive or negative. The magnitude of the correlations in turn depends on the coupling strength of the involved processes. The addition of these two noise contributions hence yields the noise spectra shown in Fig.~\ref{fig:finite_bias}~(d),~(e) and (f). In the following we discuss the implications of this effect for different magnitudes of the detuning $\delta$.

In Fig.~\ref{fig:finite_bias}~(d), we observe a large increase of the noise at the noise frequency $\omega=|\mu_\text{N}-E_{+,+}|$, because the related Andreev level is coupled strongly to the normal-conducting lead. This noise frequency corresponds to the energy which an electron on the dot needs to absorb in order to tunnel out of the dot. The second step at $\omega=|\mu_\text{N}-E_{+,-}|$ occurs when the the noise frequency provides the energy for an electron to tunnel out of the weaker coupled Andreev level $E_{+,-}$. The noise step stemming from the process with the weaker coupled Andreev level has a negative sign. In Fig.~~\ref{fig:finite_bias}~(d) the noise increases again at $\omega=|\mu_\text{N}-E_{-,+}|$ and decreases at $\omega=|\mu_\text{N}-E_{-,-}|$. 
For even higher frequencies, $\omega\gg\mu_\text{N}$, the  noise is always given by the sum of the effective coupling strengths, $2\Gamma_\text{N}$, since the noise frequency provides enough energy to excite from the singly-occupied state into either one of the ABSs and vice versa. 

If we invert the order of the excitation energies or the strength of their effective couplings, the broad maxima are transformed into troughs. This can for example be achieved by inverting the bias $\mu_\text{N}\rightarrow -\mu_\text{N}$ or the detuning $\delta\rightarrow -\delta$.  Figure \ref{fig:finite_bias}~(c) shows the energy landscape and (f) the finite-frequency noise spectrum for $\delta=-0.1U$. The detuning is reversed 
compared to the previously discussed case in Fig.~\ref{fig:finite_bias}~(d). Consequently, the noise is first suppressed, when a process with a weakly coupled Andreev level takes place, and then enhanced to the value $2\Gamma_\text{N}$ at a noise frequency related to a strongly coupled Andreev level, see Fig.~\ref{fig:finite_bias}~(f).

The quantum steps in the noise spectrum occur only in a regime of intermediate detuning $\delta$.
If the detuning $\delta$ becomes larger than the coupling to the superconducting lead $\Gamma_\text{S}$, the step structure as shown in Fig.~\ref{fig:finite_bias} gets suppressed, because the superconducting correlations become weaker.

If the proximity effect is on resonance and the detuning is exactly equal to zero ($\delta=0$) the effective coupling strengths to the different Andreev levels are equal. The finite-frequency noise spectrum for this case is displayed in Fig.~\ref{fig:finite_bias}~(e) with the corresponding energy landscape of the proximized dot, Fig.~\ref{fig:finite_bias}~(b). In this regime, when the probabilities of the dot to be in any of the ABSs are equal, no steps but only shallow dips appear in the spectrum due to an almost complete compensation of the different noise contributions. The high-frequency noise spectrum is given by the sum of the effective coupling strengths,  $2\Gamma_\text{N}$. Only small features at $\omega=|\mu-E_{\pm,\pm}|$ remain as shown in the inset of Fig.~\ref{fig:finite_bias}~(e).

\subsubsection{Intermediate-bias regime, $E_{+,-}<\mu_\text{N}<E_{+,+}$}

\begin{figure}[t]
\subfigure{ \includegraphics[height=0.25\textwidth]{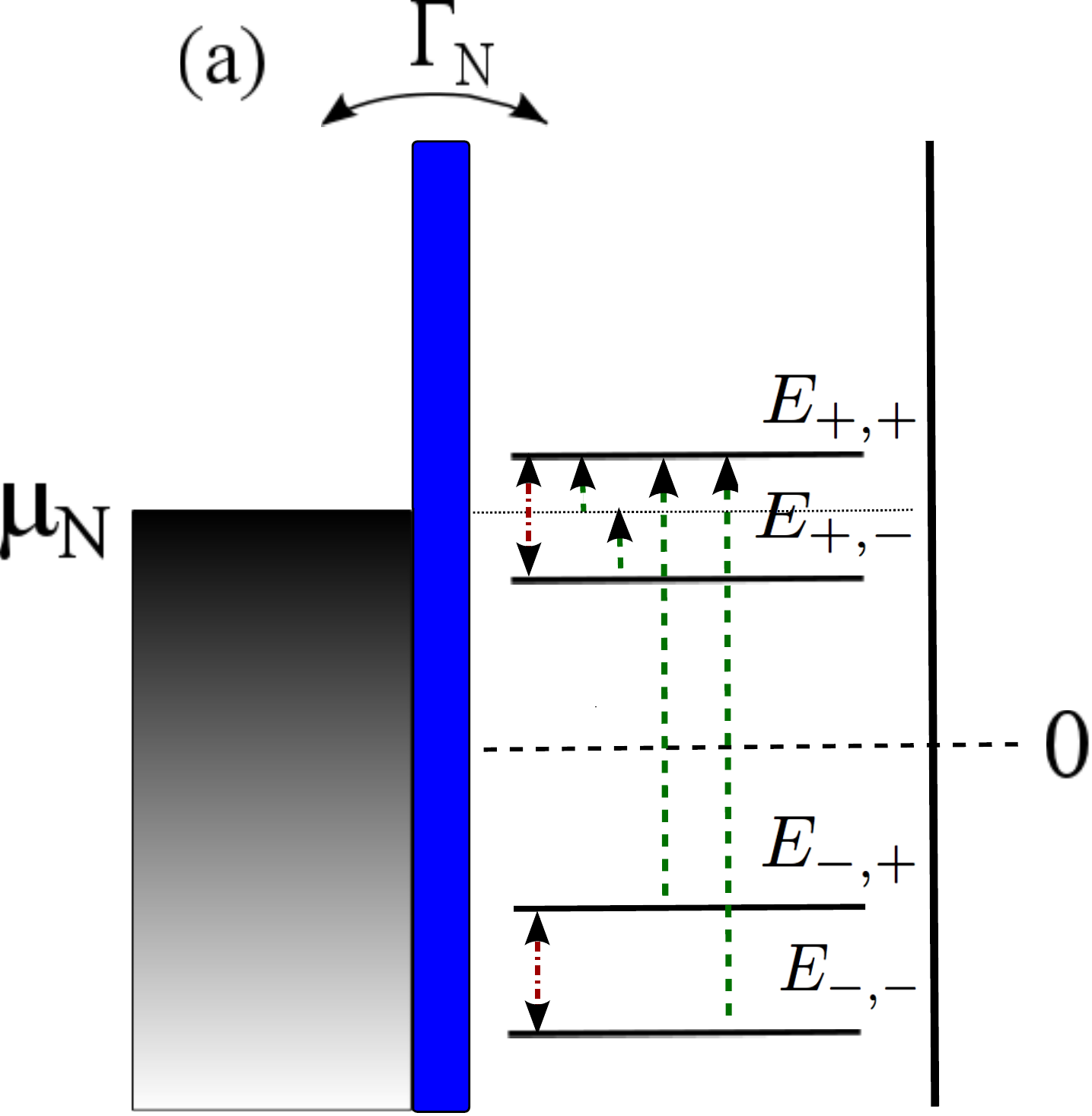}}
\subfigure{ \includegraphics[height=0.3\textwidth]{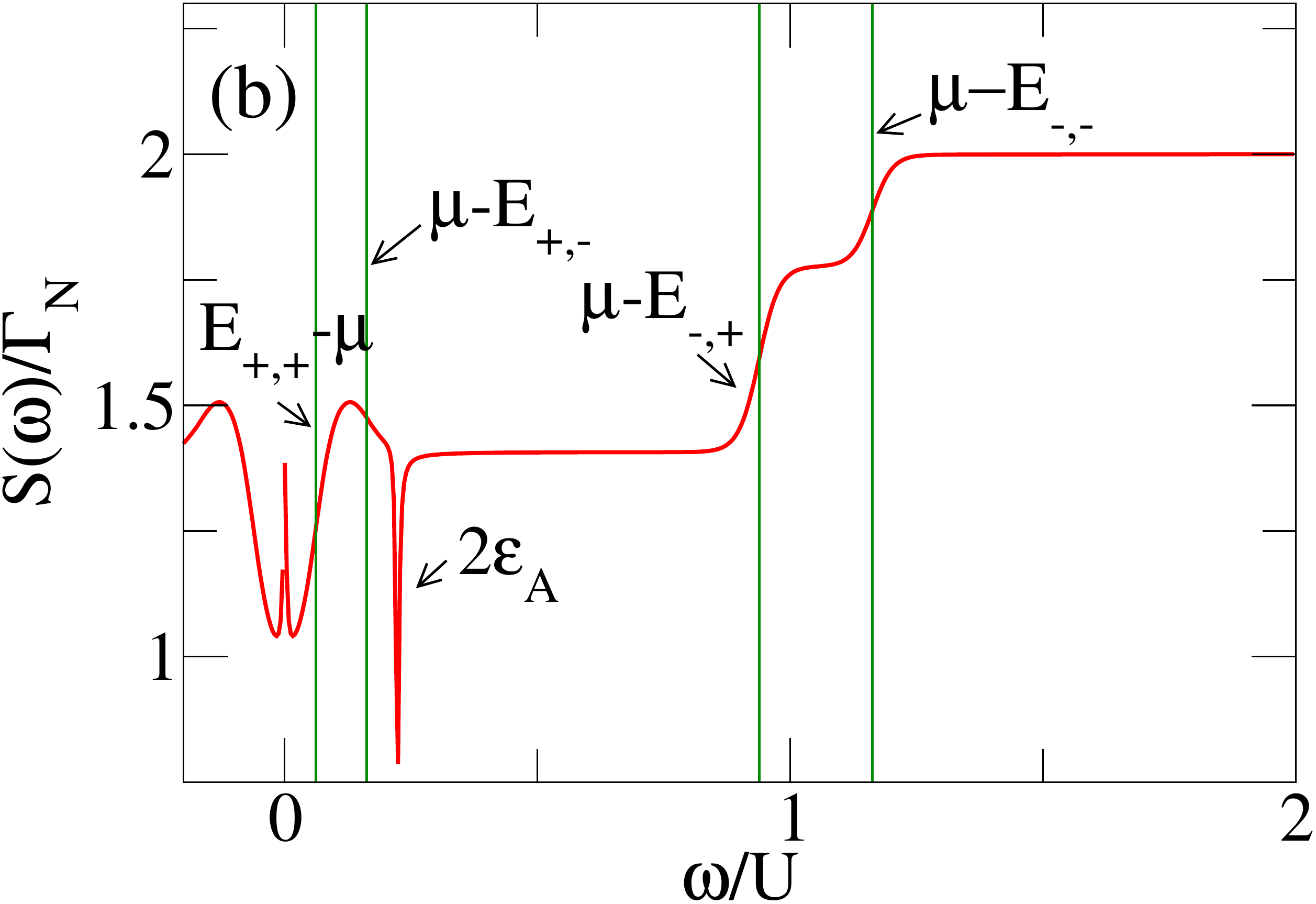}}
\caption{(color online). (a) Sketch of the proximized single-level quantum dot for $\mu_\text{N}=0.55U$ and $\delta=0.1U$. (b)  Finite-frequency noise $S(\omega)$ with $\Gamma_\text{N}=0.002U$, $\Gamma_\text{S}=0.2U$, $k_\text{B}T=0.02U$.}
\label{fig:V05}
\end{figure}

We finally also address the case of the intermediate bias regime, where only a part of the levels is in the bias window. In Fig.~\ref{fig:V05}~(a), we show the  energy landscape of the dot considered here, with $\mu_\text{N}=0.55U$ and $\delta=0.1U$, where the excitation energy $E_{+,+}$ is outside the bias window. The noise spectrum of the intermediate-bias regime, see  Fig.~\ref{fig:V05}~(b), shows a mixture of the previously observed effects in the unidirectional, low-bias and high-bias regime. We can identify in the spectrum the features discussed in Sec.~\ref{sec:unidirectional}, namely the Lorentzian dependence in the low-frequency regime and the resonance dips at $\omega=2\epsilon_\text{A}$, which are a signature of the coherent transfer of Cooper pairs between dot and superconducting lead. 

Furthermore, the quantum noise steps show an overlap of the features discussed in the previous subsections: 
the steps at $\omega=|E_{+,+}-\mu_\text{N}|$ and $\omega=|\mu_\text{N}-E_{-,-}|$  both lead to an increase of the noise but with a different step height due to the fact that the corresponding Andreev levels couple with different effective coupling strengths to the reservoir. The first of these steps overlaps with the intermediate-frequency regime, namely where the resonance dips due to the internal dynamics occur. 
The spectrum furthermore shows two steps at frequencies $\omega=|\mu_\text{N}-E_{-,+}|$ and $\omega=|\mu_\text{N}-E_{+,-}|$, where the direction of the steps tells us if it is a noise process between the normal conducting lead and a strongly or weakly coupled Andreev level. Note that also the resulting trough is here partly found in the intermediate-frequency regime. 

At even higher frequencies, $\omega\gg\mu_\text{N}$, the noise is again given by the sum of the effective coupling strengths.

\section{Conclusions}\label{conclusions}

We have presented a calculation of the noise spectrum of a system composed by a
a single-level quantum dot tunnel-coupled to a superconductor and a normal-conducting lead. We found that the noise spectrum reflects the internal spectrum of the proximized dot.  Resonance dips occur at a frequency equal to the splitting of the ABSs, $\omega=2\epsilon_\text{A}$. 
This feature is a signature of the coherent oscillation of Cooper pairs between quantum dot and superconductor. The effect is strongest if the superconducting correlations on the dot are maximal, which happens when the proximity effect is on resonance, $\delta=0$. In oder to observe this effect experimentally, the frequency of the resonance dip, approximately $\Gamma_\text{S}$ close to resonance, needs to be within the GHz frequency range. For example, in recent experiments~\cite{Deacon10,Deacon10b} the coupling to the superconductor is $\Gamma_\text{S}\approx 50-250$ GHz. 

The high-frequency regime of the noise spectrum shows quantum-noise steps at frequencies $\omega=|E_{\pm,\pm}-\mu_\text{N}|$. The quantum-noise steps provide not only information on the  Andreev addition energies of the system, but also on the effective coupling strength of the Andreev levels to the normal conducting lead. The height (in the low bias regime) and sign (in the finite-bias regime) of the steps tell the strength of the effective coupling of each Andreev level to the reservoir. 
Therefore, we conclude that the finite-frequency noise spectrum provides a full spectroscopy of the proximized quantum dot. 

\acknowledgments 
We thank Christina P\"oltl for helpful discussions. Part of this work has been done at Institute for Theorie of Statistical Physics, RWTH Aachen University, with a financial support by the Ministry of Innovation, NRW. S.D. appreciates the hospitality at the RWTH Aachen University during a long-term visit. 

\appendix

\section{Diagrammatic rules to calculate irreducible blocks $W_{\chi,\chi'}$} \label{App:rules}
In this section we summarize the rules to determine diagrammatically the different contributions to the kernel and current kernel as given in Refs.~\onlinecite{Koenig96,Koenig96B} and \onlinecite{Braun06}. We adapt the rules for the system studied here, which has only one normal lead. The rules for the kernel $\boldsymbol{W}(\omega)$ are:

\begin{enumerate}
\item Draw all topologically different diagrams with $n$ directed tunnelling lines connecting pairs of vertices containing lead electron operators. Assign spin index $\sigma$ and energy $z$ to every tunnelling line. Additionally, assign state index $\chi$ and the corresponding energy $E_{\chi}$ to each element of the Keldysh contour connecting two vertices. Also, add an external horizontal bosonic energy line transporting the energy $\omega$ to each diagram, which results from the Fourier transform.

\item For each time segment between two adjacent vertices write a resolvent $1/(\Delta E(t)+i 0^+)$ with 
$\Delta E$ being the difference between all backward-going minus forward-going energies, including tunnelling lines transporting the energy $z$ as well as the external line transporting the energy $\omega$.

\item Each vertex containing a dot operator $d^{(\dagger)}_{\sigma}$ gives rise to a matrix element $\langle \chi' |d^{(\dagger)}_{\sigma}|\chi\rangle$ where $\chi$ ($\chi'$) is the dot state entering (leaving) the vertex with respect to the Keldysh contour. Consequently, for each vertex connecting a doubly-occupied state $d$ to the up state $\uparrow$, the diagram acquires a factor (-1).

\item Each tunnelling line contributes with a factor $\frac{1}{2\pi}\Gamma_{\text{N}}f_{\text{N}}(z)$ for a backward-going line with respect to the closed time path and a factor $\frac{1}{2\pi}\Gamma_{\text{N}}\left[1-f_{\text{N}}(z)\right]$ for a forward-going contribution.

\item Each diagram has an overall prefactor $\quad\quad(-i)(-1)^{b+c}$, where
$b$ is the total number of vertices on the backward propagator and 
$c$ is the number of crossings of tunnelling lines.

\item Finally, sum over the spin $\sigma$ and integrate over the energies $z$ of tunnelling lines and sum over all diagrams that contribute to the same kernel element. 
\\
\\
As a next step we provide the additional rules to determine the blocks containing one or two current operators $\boldsymbol{W}_{\RNum{1}}(\omega)$ and $\boldsymbol{W}_{\RNum{2}}(\omega)$.

\item Replace one (two) tunnel vertex by a current vertex to calculate diagrams contributing to the kernels $\boldsymbol{W}_{\RNum{1}}(\omega)$ ($\boldsymbol{W}_{\RNum{2}}(\omega)$). Note, that the current vertex (open circle) might also be placed on the start or end point of the diagram.

\item Multiply each diagram by a prefactor, determining the position of the current vertex inside the diagram: we have to multiply each diagram by a factor of $(-1)$ for a current vertex on the upper (lower) Keldysh time branch and a particle tunnelling into (out of) the normal lead. In the two other cases multiply the diagram with a factor of  $(+1)$. 

\item The diagrams contributing to $\boldsymbol{W}_{\RNum{1}>}(\omega)$, $\boldsymbol{W}_{\RNum{1}<}(\omega)$ have open external frequency lines to the right or left side attached to the current vertex. 
Diagrams with frequency lines leaving the diagram to the right contribute to the kernel $\boldsymbol{W}_{\RNum{1}>}(\omega)$, while diagrams with frequency lines coming from the left contribute to $\boldsymbol{W}_{\RNum{1}<}(\omega)$.

\end{enumerate}

\section{Finite-frequency noise of a single-level quantum dot coupled to normal-conducting leads}\label{sec:Anderson_noise}

In this section of the appendix we present results for the current and the noise in a noninteracting single-level quantum dot coupled to two normal-conducting leads. The presentation of these known results, see Refs.~\onlinecite{Averin93,Rothstein09,Engel04,Gabdank11,Marcos11,Jin12,Dong13}, is helpful as a comparison for the understanding of the more complex results for the interacting proximized dot studied in this paper. In several cases, the effect of the differently coupled Andreev levels studied in the main part of this paper can be mimicked by considering asymmetric coupling of the dot to the two normal-conducting leads. The system studied in this appendix is shown in Fig.~\ref{fig:app}~(a). Note, that in this section, we consider the total current and its noise  rather than that in a single contact.

\subsection{Unidirectional transport regime}\label{subsec:unidirectional_Anderson_dot}
In the unidirectional transport regime, when the applied bias voltage is such that $\epsilon<V/2$, the current is given by
\begin{equation}
I_\text{uni}=\frac{2\Gamma_\text{L}\Gamma_\text{R}}{\Gamma}, 
\label{eq:current_b}
\end{equation}
with $\Gamma=\Gamma_\text{L}+\Gamma_\text{R}$.
The finite-frequency noise in this unidirectional transport regime, where also $V>\omega$ is fulfilled, is given by
\begin{equation}
S_\text{uni}(\omega)=I_\text{uni}\left[1+\frac{(\Gamma_\text{L}-\Gamma_\text{R})^{2}}{\Gamma^{2}+\omega^{2}}\right]\ .
\label{eq:Fano_low_b}
\end{equation}

The noise shows a Lorentzian dependence on the noise frequency $\omega$.~\cite{Blanter00,Braun06}
For a symmetric coupling of the dot to the normal conducting leads $\Gamma_\text{L}=\Gamma_\text{R}$, the noise equals $\Gamma/2$ and is hence independent of the noise frequency. 

\subsection{Low- and finite-bias regime}\label{app_lowbias}

In order to get an insight into the parameters controlling the height of the steps occurring in the quantum noise regime, we here analyse the high-frequency noise spectrum in the regime of low and finite bias, where quantum noise is dominant. 
For $\omega\gg\Gamma$  the noise in the high-frequency regime is found to be given by
\begin{eqnarray}
S_\text{fin}(\omega) & = & \frac{1}{2}\frac{\Gamma^{2}_\text{L}}{\Gamma}\left[f^{+}_\text{L}(\epsilon)f^{-}_\text{L}(\epsilon+\omega)+f^{+}_\text{L}(\epsilon-\omega)f^{-}_\text{L}(\epsilon)\right]\nonumber\\
& + & \frac{1}{2}\frac{\Gamma^{2}_\text{R}}{\Gamma}\left[f^{+}_\text{R}(\epsilon)f^{-}_\text{R}(\epsilon+\omega)+f^{+}_\text{R}(\epsilon-\omega)f^{-}_\text{R}(\epsilon)\right]\nonumber\\
& + & \frac{1}{2}\frac{\Gamma_\text{L}\Gamma_\text{R}}{\Gamma}\left[f^{+}_\text{L}(\epsilon+\omega)f^{-}_\text{R}(\epsilon)+f^{+}_\text{L}(\epsilon)f^{-}_\text{R}(\epsilon-\omega)\right]\nonumber\\
& + &  \frac{1}{2}\frac{\Gamma_\text{L}\Gamma_\text{R}}{\Gamma}\left[f^{+}_\text{R}(\epsilon)f^{-}_\text{L}(\epsilon+\omega)+f^{+}_\text{R}(\epsilon-\omega)f^{-}_\text{L}(\epsilon)\right]\nonumber\\
& + & \omega\rightarrow -\omega,
\label{eq:high_freq_n}
\end{eqnarray}
with the Fermi function $f^+_\alpha(\omega)=1/(1+e^{(\omega-\mu_\alpha)/k_\text{B}T})$ for the two leads $\alpha=\text{L,R}$ and $f^-_\alpha(\omega)=1-f_\alpha^+(\omega)$. While the first two contributions result from correlations in the same lead, the latter two are related to correlations between different leads.
In the following we will analyse the noise systematically for the different bias regimes and investigate the effect of an asymmetric coupling to the reservoirs ($\Gamma_\text{L}\neq \Gamma_\text{R}$) on the noise spectrum. 

\subsubsection{Zero bias}

\begin{figure*}
\centering
\includegraphics[height=0.6\textwidth]{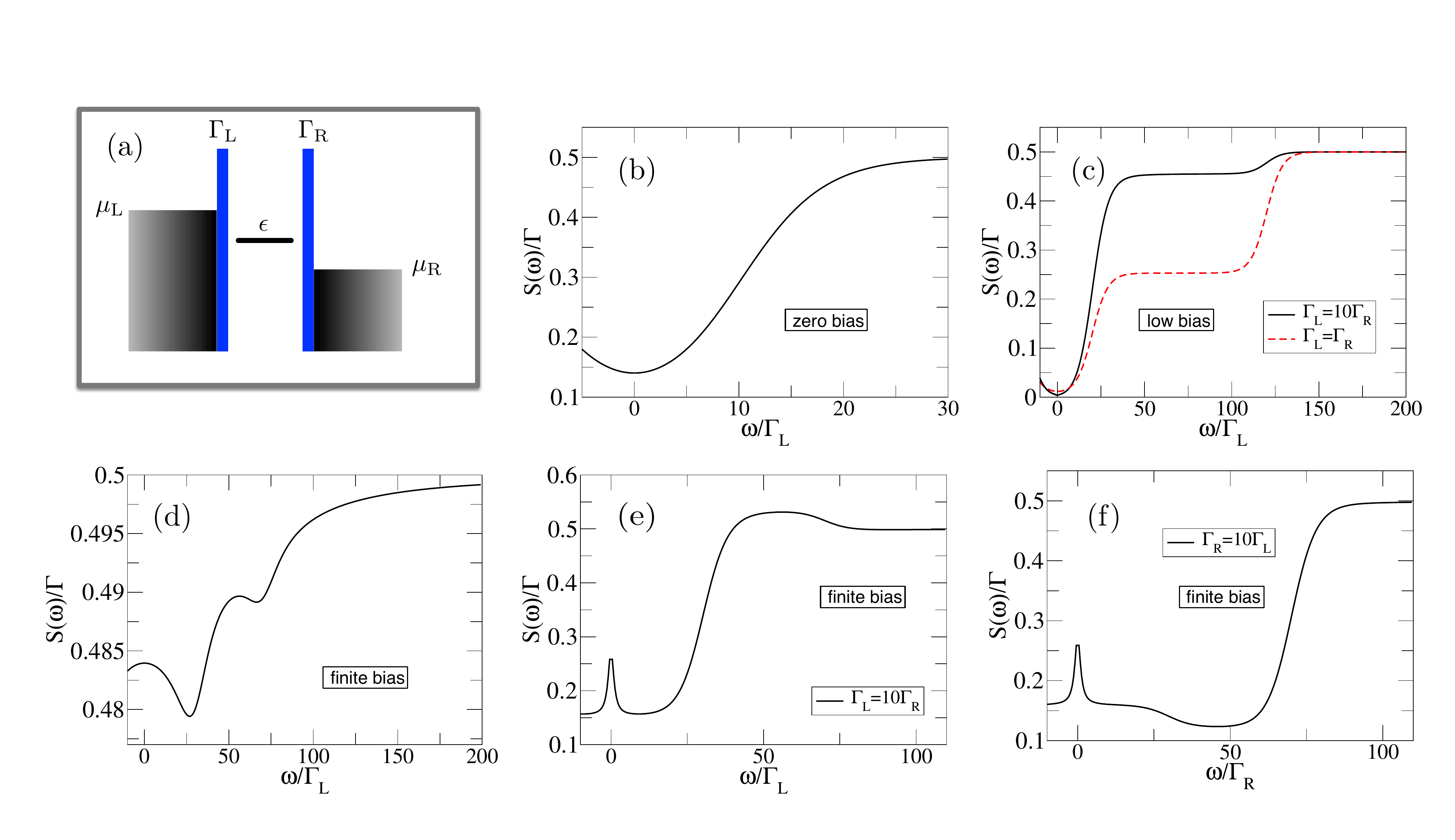}
\caption{(color online). (a) Sketch of the energy landscape of a noninteracting single-level quantum dot (with level energy $\epsilon$) coupled to normal-conducting leads (with electrochemical potentials $\mu_\text{L}$, $\mu_\text{R}$ and $\mu_\text{L}-\mu_\text{R}=V$). The coupling strengths to the leads are given by $\Gamma_\text{L}$ and $\Gamma_\text{R}$. The other panels show the finite-frequency noise $S(\omega)$ for a quantum dot coupled to normal leads in units of $\Gamma$. (b) Zero-bias regime with $\Gamma_\text{L}=\Gamma_\text{R}$, $\epsilon=10\Gamma_\text{L}$, $k_\text{B}T=4\Gamma_\text{L}$. (c) Low-bias regime for different coupling strengths with $\epsilon=70\Gamma_\text{L}$, $k_\text{B}T=4.5\Gamma_\text{L}$ and $V=100\Gamma_\text{L}$. (d) Finite bias regime with symmetric coupling $\Gamma_\text{L}=\Gamma_\text{R}$, and $\epsilon=20\Gamma_\text{L}$, $k_\text{B}T=4\Gamma_\text{L}$ and $V=100\Gamma_\text{L}$. (e) Finite bias regime with asymmetric coupling  $\Gamma_\text{L}=10\Gamma_\text{R}$, $\epsilon=20\Gamma_\text{L}$, $k_\text{B}T=4\Gamma_\text{L}$ and $V=100\Gamma_\text{L}$ and (f) with inverted asymmetry $\Gamma_\text{R}=10\Gamma_\text{L}$, $\epsilon=20\Gamma_\text{R}$, $k_\text{B}T=4\Gamma_\text{R}$ and $V=100\Gamma_\text{R}$.}
\label{fig:app}
\end{figure*}

In the limit of $V\rightarrow0$, shot noise is negligible. Thermal noise, which  is generally cut off at $\omega=k_\text{B}T$, is here suppressed due to $\epsilon\neq0$. Hence quantum noise is dominant in this regime. 

The noise spectrum, Fig.~\ref{fig:app}~(b), has one step at $\omega=|\epsilon|$. Since at zero bias all factors in Eq.~(\ref{eq:high_freq_n}) containing Fermi functions are equal, the step height is given by $\Gamma/2$. An asymmetric coupling to the leads does hence not influence the shape of the noise spectrum. 

\subsubsection{Low bias, $\epsilon>V/2$}

When a finite transport voltage is applied, but the transport level is outside the bias window, quantum noise is still the dominant noise contribution and  the noise spectrum exhibits two steps. 

For the situation shown in Fig.~\ref{fig:app}~(c), when $\epsilon>\mu_\text{L}>\mu_\text{R}$, the quantum dot is unoccupied in the stationary regime and all factors in Eq.~(\ref{eq:high_freq_n}) containing $f_\alpha^+(\epsilon)$ are zero. Then the first step stems from the contributions of the first and the third term of Eq.~(\ref{eq:high_freq_n}) for which the factor containing Fermi functions is equal. It occurs at $\omega=|\epsilon-\mu_\text{L}|$ when the excitation of the dot from the left lead becomes visible and it has the height $\Gamma_\text{L}/2$. Analogously, the second step at $\omega=|\epsilon-\mu_\text{R}|$ has height $\Gamma_\text{R}/2$. 
In both places an increase of the noise is observed  as long as the dot level is outside the bias window, because in both cases the effect of an otherwise blocked transport channel becomes visible.  
Fig.~\ref{fig:app}~(c) shows two noise spectra, for a symmetrically coupled quantum dot $\Gamma_\text{L}=\Gamma_\text{R}$ (red dashed line) and an asymmetrically coupled dot $\Gamma_\text{L}>\Gamma_\text{R}$ (black solid line).  

 \subsubsection{Finite bias, $\epsilon<V/2$}\label{app_finitebias}

We finally consider the case, where the energy level lies inside the bias window and shot noise as well as quantum noise is present. 
When choosing asymmetric coupling to the leads, we find a situation which can be compared to the proximized quantum dot with finite detuning as discussed in the main text. 

In Fig.~\ref{fig:app}~(e) the noise is displayed for a situation where the left lead is coupled much stronger to the quantum dot $\Gamma_\text{L}=10\Gamma_\text{R}$. 
The Lorentzian behaviour of the low-frequency noise, $\omega\ll(\mu_\text{L}-\epsilon)$, is described with the expression given in Eq.~(\ref{eq:Fano_low_b}). 
Furthermore, steps occur at  $\omega=|\mu_\text{L}-\epsilon|$ and $\omega=|\mu_\text{R}-\epsilon|$. The first step at $\omega=|\mu_\text{L}-\epsilon|$ increases the noise. At this frequency back-tunnelling of an electron to the strongly coupled left lead, emptying the quantum dot, becomes visible.  
Its height is given by $\Gamma_\text{L}(\Gamma_\text{L}-\Gamma_\text{R})/2(\Gamma_\text{R}+\Gamma_\text{L})$. 
The second step at $\omega=|\epsilon-\mu_\text{R}|$, which occurs when an electron can tunnel back onto the dot from the right lead, results in  a decrease of the noise. Its depth is given by $-\Gamma_\text{R}(\Gamma_\text{R}-\Gamma_\text{L})/2(\Gamma_\text{R}+\Gamma_\text{L})$. 
The high frequency noise is again given by $\Gamma/2$. 

Whether a step in the noise leads to an increase or a decrease of the total noise depends on the coupling strength of the different excitations. The reason for this is that, in the regime where the level is in the bias window, the part of the noise stemming from correlations in the same lead increases at the frequencies equal to the excitation energies with an amount given by the square of the respective coupling strength, while the noise due to correlations between different leads decreases by an amount which is always proportional to $\Gamma_\text{L}\Gamma_\text{R}$. This can be directly read off from Eq.~(\ref{eq:high_freq_n}). Depending on whether the positive contribution multiplied by $\Gamma_\alpha^2$ or the negative contribution multiplied by $\Gamma_\text{L}\Gamma_\text{R}$ is larger the step hence changes direction.

Fig.~\ref{fig:app}~(f) shows the noise for the same applied bias voltage but with the right reservoir coupled stronger than the left reservoir, $\Gamma_\text{R}=10\Gamma_\text{L}$. The noise spectrum therefore shows a reversed order of the steps with respect to the result shown in Fig.~\ref{fig:app} (e), leading to an occurrence of troughs rather than plateaus in the noise. 

The situation for a symmetrically coupled quantum dot ($\Gamma_\mathrm{L}=\Gamma_\mathrm{R}$) where the probabilities of the quantum dot to be empty or singly occupied are equal, is shown in Fig.~\ref{fig:app}~(d). The noise spectrum shows no quantum noise steps at $\omega=|\epsilon-\mu_\text{L/R}|$ due to an almost complete compensation of the noise stemming from correlations in the same lead and correlations between the two leads. 

The sum of these contributions to the noise leads to the shallow dip structure in the symmetrized noise spectrum as displayed in Fig.~\ref{fig:app}~(d).
For frequencies larger than the bias voltage, the noise takes again the value $\Gamma/2$. 
The case of a quantum dot symmetrically coupled to normal-conducting leads is equivalent to the situation of zero detuning in the high bias regime in the main text.

\section{Displacement current for the single-level quantum dot coupled to a normal and a superconducting lead }\label{App:diss_current}

We consider a simple capacitive model and denote with $C_\text{N}$ and $C_\text{S}$ the capacitances of the tunnel barriers with the normal and superconducting lead, respectively. 
The number of electrons in the dot is $\hat{n}=\sum_\sigma d_{\sigma}^\dagger d_{\sigma}$.
Within this model the displacement current in the normal lead is\cite{Blanter00}
\begin{equation}
\hat{ I}_{\text{N},\text{displ}}=-\frac{C_\text{N}}{C_\text{N}+C_\text{S}} \dot{\hat{n}}. 
\end{equation}
The total current is simply the sum of the tunnelling current and the displacement current and reads
\begin{equation}
\hat{I}_{\text{N},\text{tot}}=\hat{I}_{\text{N},\text{tunn}}- \frac{C_\text{N}}{C_\text{N}+C_\text{S}} \dot{\hat{n}}.
\end{equation}
with the tunneling current given by
\begin{equation}
\begin{split}
\hat{I}_{\text{N},\text{tunn}} & =-\dot{\hat{N}}=-\frac{1}{i\hbar}[\hat{N},H] \\
& =\frac{i}{\hbar}\sum_{k,\sigma}\left(t_\mathrm{N}c^{\dagger}_{k\sigma}d_{\sigma}-t^{*}_\mathrm{N}d^{\dagger}_{\sigma}c_{k\sigma}\right) \ .
\end{split}
\end{equation}
 Note, that the tunneling current $\hat{I}_{\text{N},\text{tunn}}$ is equal the current $\hat{I}$ introduced in Sec.~\ref{sec:current}. The current $I_\text{N}$ is positive when flowing out of the normal lead. 
At this stage, it is worth mentioning  that if $C_\text{S}\gg C_\text{N}$, the displacement current in the normal lead can be neglected. This assumption is consistent with $\Gamma_\text{S}\gg \Gamma_\text{N}$. 

Now, we proceed to evaluate $\dot{\hat{n}}$:
\begin{equation}
\begin{split}
\dot{\hat{n}} & =\frac{1}{i\hbar}[\hat{n} ,H]=\hat{I}_{\text{N},\text{tunn}}+\frac{1}{i\hbar}[\hat{n},H_{\text{eff}}]\\
                   &=\hat{I}_{\text{N},\text{tunn}}+\hat{I}_{\text{S},\text{tunn}},
\end{split}
\end{equation}
where the tunnelling current with the superconductor reads
\begin{equation}
\hat{I}_{\text{S},\text{tunn}}=\frac{i}{\hbar}\Gamma_\text{S}\left(d^{\dagger}_{\uparrow}d^{\dagger}_{\downarrow}-d^{}_{\downarrow}d^{}_{\uparrow}\right).
\end{equation}
Putting everything together we obtain the Ramo-Shockley theorem for the N-dot-S system: 
\begin{equation}
\label{Itot}
\hat{I}_{\text{N},\text{tot}}=\frac{C_\text{S}}{C_\text{N}+C_\text{S}}\hat{I}_{\text{N},\text{tunn}}-\frac{C_\text{N}}{C_\text{N}+C_\text{S}}\hat{I}_{\text{S},\text{tunn}}.
\end{equation}

\bibliographystyle{apsrev4-1}

\begin{thebibliography}{74}%
\makeatletter
\providecommand \@ifxundefined [1]{%
 \@ifx{#1\undefined}
}%
\providecommand \@ifnum [1]{%
 \ifnum #1\expandafter \@firstoftwo
 \else \expandafter \@secondoftwo
 \fi
}%
\providecommand \@ifx [1]{%
 \ifx #1\expandafter \@firstoftwo
 \else \expandafter \@secondoftwo
 \fi
}%
\providecommand \natexlab [1]{#1}%
\providecommand \enquote  [1]{``#1''}%
\providecommand \bibnamefont  [1]{#1}%
\providecommand \bibfnamefont [1]{#1}%
\providecommand \citenamefont [1]{#1}%
\providecommand \href@noop [0]{\@secondoftwo}%
\providecommand \href [0]{\begingroup \@sanitize@url \@href}%
\providecommand \@href[1]{\@@startlink{#1}\@@href}%
\providecommand \@@href[1]{\endgroup#1\@@endlink}%
\providecommand \@sanitize@url [0]{\catcode `\\12\catcode `\$12\catcode
  `\&12\catcode `\#12\catcode `\^12\catcode `\_12\catcode `\%12\relax}%
\providecommand \@@startlink[1]{}%
\providecommand \@@endlink[0]{}%
\providecommand \url  [0]{\begingroup\@sanitize@url \@url }%
\providecommand \@url [1]{\endgroup\@href {#1}{\urlprefix }}%
\providecommand \urlprefix  [0]{URL }%
\providecommand \Eprint [0]{\href }%
\providecommand \doibase [0]{http://dx.doi.org/}%
\providecommand \selectlanguage [0]{\@gobble}%
\providecommand \bibinfo  [0]{\@secondoftwo}%
\providecommand \bibfield  [0]{\@secondoftwo}%
\providecommand \translation [1]{[#1]}%
\providecommand \BibitemOpen [0]{}%
\providecommand \bibitemStop [0]{}%
\providecommand \bibitemNoStop [0]{.\EOS\space}%
\providecommand \EOS [0]{\spacefactor3000\relax}%
\providecommand \BibitemShut  [1]{\csname bibitem#1\endcsname}%
\let\auto@bib@innerbib\@empty
\bibitem [{\citenamefont {Franceschi}\ \emph {et~al.}(2010)\citenamefont
  {Franceschi}, \citenamefont {Kouwenhoven}, \citenamefont {Schoenenberger},\
  and\ \citenamefont {Wernsdorfer}}]{DeFranceschi10}%
  \BibitemOpen
  \bibfield  {author} {\bibinfo {author} {\bibfnamefont {S.~D.}\ \bibnamefont
  {Franceschi}}, \bibinfo {author} {\bibfnamefont {L.}~\bibnamefont
  {Kouwenhoven}}, \bibinfo {author} {\bibfnamefont {C.}~\bibnamefont
  {Schoenenberger}}, \ and\ \bibinfo {author} {\bibfnamefont {W.}~\bibnamefont
  {Wernsdorfer}},\ }\href@noop {} {\bibfield  {journal} {\bibinfo  {journal}
  {Nature Nanotechnol.}\ }\textbf {\bibinfo {volume} {5}},\ \bibinfo {pages}
  {703} (\bibinfo {year} {2010})}\BibitemShut {NoStop}%
\bibitem [{\citenamefont {Martin-Rodero}\ and\ \citenamefont
  {Yeyati}(2011)}]{Martin-Rodero11}%
  \BibitemOpen
  \bibfield  {author} {\bibinfo {author} {\bibfnamefont {A.}~\bibnamefont
  {Martin-Rodero}}\ and\ \bibinfo {author} {\bibfnamefont {A.~L.}\ \bibnamefont
  {Yeyati}},\ }\href@noop {} {\bibfield  {journal} {\bibinfo  {journal} {Adv.
  Phys.}\ }\textbf {\bibinfo {volume} {60}},\ \bibinfo {pages} {899} (\bibinfo
  {year} {2011})}\BibitemShut {NoStop}%
\bibitem [{\citenamefont {Leijnse}\ and\ \citenamefont
  {Flensberg}(2012)}]{Leijnse12}%
  \BibitemOpen
  \bibfield  {author} {\bibinfo {author} {\bibfnamefont {M.}~\bibnamefont
  {Leijnse}}\ and\ \bibinfo {author} {\bibfnamefont {K.}~\bibnamefont
  {Flensberg}},\ }\href {\doibase 10.1103/PhysRevB.86.134528} {\bibfield
  {journal} {\bibinfo  {journal} {Phys. Rev. B}\ }\textbf {\bibinfo {volume}
  {86}},\ \bibinfo {pages} {134528} (\bibinfo {year} {2012})}\BibitemShut
  {NoStop}%
\bibitem [{\citenamefont {Sothmann}\ \emph {et~al.}(2013)\citenamefont
  {Sothmann}, \citenamefont {Li},\ and\ \citenamefont
  {B\"uttiker}}]{sothmann_2013}%
  \BibitemOpen
  \bibfield  {author} {\bibinfo {author} {\bibfnamefont {B.}~\bibnamefont
  {Sothmann}}, \bibinfo {author} {\bibfnamefont {J.}~\bibnamefont {Li}}, \ and\
  \bibinfo {author} {\bibfnamefont {M.}~\bibnamefont {B\"uttiker}},\
  }\href@noop {} {\bibfield  {journal} {\bibinfo  {journal} {New. J. Phys.}\
  }\textbf {\bibinfo {volume} {15}},\ \bibinfo {pages} {085018} (\bibinfo
  {year} {2013})}\BibitemShut {NoStop}%
\bibitem [{\citenamefont {Wright}\ and\ \citenamefont
  {Veldhorst}(2013)}]{Wright13}%
  \BibitemOpen
  \bibfield  {author} {\bibinfo {author} {\bibfnamefont {A.~R.}\ \bibnamefont
  {Wright}}\ and\ \bibinfo {author} {\bibfnamefont {M.}~\bibnamefont
  {Veldhorst}},\ }\href@noop {} {\bibfield  {journal} {\bibinfo  {journal}
  {Phys. Rev. Lett.}\ }\textbf {\bibinfo {volume} {111}},\ \bibinfo {pages}
  {096801} (\bibinfo {year} {2013})}\BibitemShut {NoStop}%
\bibitem [{\citenamefont {Brunetti}\ \emph {et~al.}(2013)\citenamefont
  {Brunetti}, \citenamefont {Zazunov}, \citenamefont {Kundu},\ and\
  \citenamefont {Egger}}]{Brunetti13}%
  \BibitemOpen
  \bibfield  {author} {\bibinfo {author} {\bibfnamefont {A.}~\bibnamefont
  {Brunetti}}, \bibinfo {author} {\bibfnamefont {A.}~\bibnamefont {Zazunov}},
  \bibinfo {author} {\bibfnamefont {A.}~\bibnamefont {Kundu}}, \ and\ \bibinfo
  {author} {\bibfnamefont {R.}~\bibnamefont {Egger}},\ }\href@noop {}
  {\bibfield  {journal} {\bibinfo  {journal} {Phys. Rev. B}\ }\textbf {\bibinfo
  {volume} {88}},\ \bibinfo {pages} {144515} (\bibinfo {year}
  {2013})}\BibitemShut {NoStop}%
\bibitem [{\citenamefont {Sothmann}\ \emph {et~al.}(2014)\citenamefont
  {Sothmann}, \citenamefont {Weiss}, \citenamefont {Governale},\ and\
  \citenamefont {K\"onig}}]{sothmann_2014}%
  \BibitemOpen
  \bibfield  {author} {\bibinfo {author} {\bibfnamefont {B.}~\bibnamefont
  {Sothmann}}, \bibinfo {author} {\bibfnamefont {S.}~\bibnamefont {Weiss}},
  \bibinfo {author} {\bibfnamefont {M.}~\bibnamefont {Governale}}, \ and\
  \bibinfo {author} {\bibfnamefont {J.}~\bibnamefont {K\"onig}},\ }\href@noop
  {} {\bibfield  {journal} {\bibinfo  {journal} {Phys. Rev. B}\ }\textbf
  {\bibinfo {volume} {90}},\ \bibinfo {pages} {220501} (\bibinfo {year}
  {2014})}\BibitemShut {NoStop}%
\bibitem [{\citenamefont {Hofstetter}\ \emph {et~al.}(2009)\citenamefont
  {Hofstetter}, \citenamefont {Csonka}, \citenamefont {Nyg{\aa}rd},\ and\
  \citenamefont {Sch\"onenberger}}]{hofstetter09}%
  \BibitemOpen
  \bibfield  {author} {\bibinfo {author} {\bibfnamefont {L.}~\bibnamefont
  {Hofstetter}}, \bibinfo {author} {\bibfnamefont {S.}~\bibnamefont {Csonka}},
  \bibinfo {author} {\bibfnamefont {J.}~\bibnamefont {Nyg{\aa}rd}}, \ and\
  \bibinfo {author} {\bibfnamefont {C.}~\bibnamefont {Sch\"onenberger}},\
  }\href@noop {} {\bibfield  {journal} {\bibinfo  {journal} {Nature}\ }\textbf
  {\bibinfo {volume} {461}},\ \bibinfo {pages} {960} (\bibinfo {year}
  {2009})}\BibitemShut {NoStop}%
\bibitem [{\citenamefont {Herrmann}\ \emph {et~al.}(2010)\citenamefont
  {Herrmann}, \citenamefont {Portier}, \citenamefont {Roche}, \citenamefont
  {Yeyati}, \citenamefont {Kontos},\ and\ \citenamefont {Strunk}}]{hermann10}%
  \BibitemOpen
  \bibfield  {author} {\bibinfo {author} {\bibfnamefont {L.~G.}\ \bibnamefont
  {Herrmann}}, \bibinfo {author} {\bibfnamefont {F.}~\bibnamefont {Portier}},
  \bibinfo {author} {\bibfnamefont {P.}~\bibnamefont {Roche}}, \bibinfo
  {author} {\bibfnamefont {A.~L.}\ \bibnamefont {Yeyati}}, \bibinfo {author}
  {\bibfnamefont {T.}~\bibnamefont {Kontos}}, \ and\ \bibinfo {author}
  {\bibfnamefont {C.}~\bibnamefont {Strunk}},\ }\href {\doibase
  10.1103/PhysRevLett.104.026801} {\bibfield  {journal} {\bibinfo  {journal}
  {Phys. Rev. Lett.}\ }\textbf {\bibinfo {volume} {104}},\ \bibinfo {pages}
  {026801} (\bibinfo {year} {2010})}\BibitemShut {NoStop}%
\bibitem [{\citenamefont {Governale}\ \emph {et~al.}(2008)\citenamefont
  {Governale}, \citenamefont {Pala},\ and\ \citenamefont
  {K\"onig}}]{governale08}%
  \BibitemOpen
  \bibfield  {author} {\bibinfo {author} {\bibfnamefont {M.}~\bibnamefont
  {Governale}}, \bibinfo {author} {\bibfnamefont {M.~G.}\ \bibnamefont {Pala}},
  \ and\ \bibinfo {author} {\bibfnamefont {J.}~\bibnamefont {K\"onig}},\
  }\href@noop {} {\bibfield  {journal} {\bibinfo  {journal} {Phys. Rev. B}\
  }\textbf {\bibinfo {volume} {77}},\ \bibinfo {pages} {134513} (\bibinfo
  {year} {2008})}\BibitemShut {NoStop}%
\bibitem [{\citenamefont {Braggio}\ \emph {et~al.}(2011)\citenamefont
  {Braggio}, \citenamefont {Governale}, \citenamefont {Pala},\ and\
  \citenamefont {K\"onig}}]{Braggio11}%
  \BibitemOpen
  \bibfield  {author} {\bibinfo {author} {\bibfnamefont {A.}~\bibnamefont
  {Braggio}}, \bibinfo {author} {\bibfnamefont {M.}~\bibnamefont {Governale}},
  \bibinfo {author} {\bibfnamefont {M.~G.}\ \bibnamefont {Pala}}, \ and\
  \bibinfo {author} {\bibfnamefont {J.}~\bibnamefont {K\"onig}},\ }\href@noop
  {} {\bibfield  {journal} {\bibinfo  {journal} {Solid State Comm.}\ }\textbf
  {\bibinfo {volume} {151}},\ \bibinfo {pages} {155} (\bibinfo {year}
  {2011})}\BibitemShut {NoStop}%
\bibitem [{\citenamefont {Sand-Jespersen}\ \emph {et~al.}(2007)\citenamefont
  {Sand-Jespersen}, \citenamefont {Paaske}, \citenamefont {Andersen},
  \citenamefont {Grove-Rasmussen}, \citenamefont {J\o{}rgensen}, \citenamefont
  {Aagesen}, \citenamefont {S\o{}rensen}, \citenamefont {Lindelof},
  \citenamefont {Flensberg},\ and\ \citenamefont {Nyg\aa{}rd}}]{Jespersen07}%
  \BibitemOpen
  \bibfield  {author} {\bibinfo {author} {\bibfnamefont {T.}~\bibnamefont
  {Sand-Jespersen}}, \bibinfo {author} {\bibfnamefont {J.}~\bibnamefont
  {Paaske}}, \bibinfo {author} {\bibfnamefont {B.~M.}\ \bibnamefont
  {Andersen}}, \bibinfo {author} {\bibfnamefont {K.}~\bibnamefont
  {Grove-Rasmussen}}, \bibinfo {author} {\bibfnamefont {H.~I.}\ \bibnamefont
  {J\o{}rgensen}}, \bibinfo {author} {\bibfnamefont {M.}~\bibnamefont
  {Aagesen}}, \bibinfo {author} {\bibfnamefont {C.~B.}\ \bibnamefont
  {S\o{}rensen}}, \bibinfo {author} {\bibfnamefont {P.~E.}\ \bibnamefont
  {Lindelof}}, \bibinfo {author} {\bibfnamefont {K.}~\bibnamefont {Flensberg}},
  \ and\ \bibinfo {author} {\bibfnamefont {J.}~\bibnamefont {Nyg\aa{}rd}},\
  }\href@noop {} {\bibfield  {journal} {\bibinfo  {journal} {Phys. Rev. Lett}\
  }\textbf {\bibinfo {volume} {99}},\ \bibinfo {pages} {126603} (\bibinfo
  {year} {2007})}\BibitemShut {NoStop}%
\bibitem [{\citenamefont {Eichler}\ \emph {et~al.}(2007)\citenamefont
  {Eichler}, \citenamefont {Weiss}, \citenamefont {Oberholzer}, \citenamefont
  {Sch\"onenberger}, \citenamefont {Levy~Yeyati}, \citenamefont {Cuevas},\ and\
  \citenamefont {Martin-Rodero}}]{Eichler07}%
  \BibitemOpen
  \bibfield  {author} {\bibinfo {author} {\bibfnamefont {A.}~\bibnamefont
  {Eichler}}, \bibinfo {author} {\bibfnamefont {M.}~\bibnamefont {Weiss}},
  \bibinfo {author} {\bibfnamefont {S.}~\bibnamefont {Oberholzer}}, \bibinfo
  {author} {\bibfnamefont {C.}~\bibnamefont {Sch\"onenberger}}, \bibinfo
  {author} {\bibfnamefont {A.}~\bibnamefont {Levy~Yeyati}}, \bibinfo {author}
  {\bibfnamefont {J.}~\bibnamefont {Cuevas}}, \ and\ \bibinfo {author}
  {\bibfnamefont {A.}~\bibnamefont {Martin-Rodero}},\ }\href@noop {} {\bibfield
   {journal} {\bibinfo  {journal} {Phys. Rev. Lett}\ }\textbf {\bibinfo
  {volume} {99}},\ \bibinfo {pages} {126602} (\bibinfo {year}
  {2007})}\BibitemShut {NoStop}%
\bibitem [{\citenamefont {Deacon}\ \emph
  {et~al.}(2010{\natexlab{a}})\citenamefont {Deacon}, \citenamefont {Tanaka},
  \citenamefont {Oiwa}, \citenamefont {Sakano}, \citenamefont {Yoshida},
  \citenamefont {Shibata}, \citenamefont {Hirakawa},\ and\ \citenamefont
  {Tarucha}}]{Deacon10}%
  \BibitemOpen
  \bibfield  {author} {\bibinfo {author} {\bibfnamefont {R.~S.}\ \bibnamefont
  {Deacon}}, \bibinfo {author} {\bibfnamefont {Y.}~\bibnamefont {Tanaka}},
  \bibinfo {author} {\bibfnamefont {A.}~\bibnamefont {Oiwa}}, \bibinfo {author}
  {\bibfnamefont {R.}~\bibnamefont {Sakano}}, \bibinfo {author} {\bibfnamefont
  {K.}~\bibnamefont {Yoshida}}, \bibinfo {author} {\bibfnamefont
  {K.}~\bibnamefont {Shibata}}, \bibinfo {author} {\bibfnamefont
  {K.}~\bibnamefont {Hirakawa}}, \ and\ \bibinfo {author} {\bibfnamefont
  {S.}~\bibnamefont {Tarucha}},\ }\href@noop {} {\bibfield  {journal} {\bibinfo
   {journal} {Phys. Rev. Lett}\ }\textbf {\bibinfo {volume} {104}},\ \bibinfo
  {pages} {076805} (\bibinfo {year} {2010}{\natexlab{a}})}\BibitemShut
  {NoStop}%
\bibitem [{\citenamefont {Pillet}\ \emph {et~al.}(2010)\citenamefont {Pillet},
  \citenamefont {Quay}, \citenamefont {Morfin}, \citenamefont {Bena},
  \citenamefont {Yeyati},\ and\ \citenamefont {Joyez}}]{Pillet10}%
  \BibitemOpen
  \bibfield  {author} {\bibinfo {author} {\bibfnamefont {J.-D.}\ \bibnamefont
  {Pillet}}, \bibinfo {author} {\bibfnamefont {C.~H.~L.}\ \bibnamefont {Quay}},
  \bibinfo {author} {\bibfnamefont {P.}~\bibnamefont {Morfin}}, \bibinfo
  {author} {\bibfnamefont {C.}~\bibnamefont {Bena}}, \bibinfo {author}
  {\bibfnamefont {A.~L.}\ \bibnamefont {Yeyati}}, \ and\ \bibinfo {author}
  {\bibfnamefont {P.}~\bibnamefont {Joyez}},\ }\href
  {http://dx.doi.org/10.1038/nphys1811} {\bibfield  {journal} {\bibinfo
  {journal} {Nat. Phys.}\ }\textbf {\bibinfo {volume} {6}},\ \bibinfo {pages}
  {965} (\bibinfo {year} {2010})}\BibitemShut {NoStop}%
\bibitem [{\citenamefont {Dirks}\ \emph {et~al.}(2011)\citenamefont {Dirks},
  \citenamefont {Hughes}, \citenamefont {Lal}, \citenamefont {Uchoa},
  \citenamefont {Chen}, \citenamefont {Chialvo}, \citenamefont {Goldbart},\
  and\ \citenamefont {Mason}}]{Dirks11}%
  \BibitemOpen
  \bibfield  {author} {\bibinfo {author} {\bibfnamefont {T.}~\bibnamefont
  {Dirks}}, \bibinfo {author} {\bibfnamefont {T.~L.}\ \bibnamefont {Hughes}},
  \bibinfo {author} {\bibfnamefont {S.}~\bibnamefont {Lal}}, \bibinfo {author}
  {\bibfnamefont {B.}~\bibnamefont {Uchoa}}, \bibinfo {author} {\bibfnamefont
  {Y.-F.}\ \bibnamefont {Chen}}, \bibinfo {author} {\bibfnamefont
  {C.}~\bibnamefont {Chialvo}}, \bibinfo {author} {\bibfnamefont {P.~M.}\
  \bibnamefont {Goldbart}}, \ and\ \bibinfo {author} {\bibfnamefont
  {N.}~\bibnamefont {Mason}},\ }\href {http://dx.doi.org/10.1038/nphys1911}
  {\bibfield  {journal} {\bibinfo  {journal} {Nat. Phys.}\ }\textbf {\bibinfo
  {volume} {7}},\ \bibinfo {pages} {386} (\bibinfo {year} {2011})}\BibitemShut
  {NoStop}%
\bibitem [{\citenamefont {Lee}\ \emph {et~al.}(2012)\citenamefont {Lee},
  \citenamefont {Jiang}, \citenamefont {Aguado}, \citenamefont {Katsaros},
  \citenamefont {Lieber},\ and\ \citenamefont {De~Franceschi}}]{Lee12}%
  \BibitemOpen
  \bibfield  {author} {\bibinfo {author} {\bibfnamefont {E.}~\bibnamefont
  {Lee}}, \bibinfo {author} {\bibfnamefont {X.}~\bibnamefont {Jiang}}, \bibinfo
  {author} {\bibfnamefont {R.}~\bibnamefont {Aguado}}, \bibinfo {author}
  {\bibfnamefont {G.}~\bibnamefont {Katsaros}}, \bibinfo {author}
  {\bibfnamefont {C.}~\bibnamefont {Lieber}}, \ and\ \bibinfo {author}
  {\bibfnamefont {S.}~\bibnamefont {De~Franceschi}},\ }\href@noop {} {\bibfield
   {journal} {\bibinfo  {journal} {Phys. Rev. Lett}\ }\textbf {\bibinfo
  {volume} {109}},\ \bibinfo {pages} {186802} (\bibinfo {year}
  {2012})}\BibitemShut {NoStop}%
\bibitem [{\citenamefont {Pillet}\ \emph {et~al.}(2013)\citenamefont {Pillet},
  \citenamefont {Joyez}, \citenamefont {\ifmmode~\check{Z}\else
  \v{Z}\fi{}itko},\ and\ \citenamefont {Goffman}}]{Pillet13}%
  \BibitemOpen
  \bibfield  {author} {\bibinfo {author} {\bibfnamefont {J.-D.}\ \bibnamefont
  {Pillet}}, \bibinfo {author} {\bibfnamefont {P.}~\bibnamefont {Joyez}},
  \bibinfo {author} {\bibfnamefont {R.}~\bibnamefont {\ifmmode~\check{Z}\else
  \v{Z}\fi{}itko}}, \ and\ \bibinfo {author} {\bibfnamefont {M.}~\bibnamefont
  {Goffman}},\ }\href@noop {} {\bibfield  {journal} {\bibinfo  {journal} {Phys.
  Rev. B}\ }\textbf {\bibinfo {volume} {88}},\ \bibinfo {pages} {045101}
  (\bibinfo {year} {2013})}\BibitemShut {NoStop}%
\bibitem [{\citenamefont {Chang}\ \emph {et~al.}(2013)\citenamefont {Chang},
  \citenamefont {Manucharyan}, \citenamefont {Jespersen}, \citenamefont
  {Nyg\aa{}rd},\ and\ \citenamefont {Marcus}}]{Chang13}%
  \BibitemOpen
  \bibfield  {author} {\bibinfo {author} {\bibfnamefont {W.}~\bibnamefont
  {Chang}}, \bibinfo {author} {\bibfnamefont {V.}~\bibnamefont {Manucharyan}},
  \bibinfo {author} {\bibfnamefont {T.}~\bibnamefont {Jespersen}}, \bibinfo
  {author} {\bibfnamefont {J.}~\bibnamefont {Nyg\aa{}rd}}, \ and\ \bibinfo
  {author} {\bibfnamefont {C.}~\bibnamefont {Marcus}},\ }\href@noop {}
  {\bibfield  {journal} {\bibinfo  {journal} {Phys. Rev. Lett}\ }\textbf
  {\bibinfo {volume} {110}},\ \bibinfo {pages} {217005} (\bibinfo {year}
  {2013})}\BibitemShut {NoStop}%
\bibitem [{\citenamefont {Lee}\ \emph {et~al.}(2014)\citenamefont {Lee},
  \citenamefont {Jiang}, \citenamefont {Houzet}, \citenamefont {Aguado},
  \citenamefont {Lieber},\ and\ \citenamefont {De~Franceschi}}]{Lee14}%
  \BibitemOpen
  \bibfield  {author} {\bibinfo {author} {\bibfnamefont {E.~J.~H.}\
  \bibnamefont {Lee}}, \bibinfo {author} {\bibfnamefont {X.}~\bibnamefont
  {Jiang}}, \bibinfo {author} {\bibfnamefont {M.}~\bibnamefont {Houzet}},
  \bibinfo {author} {\bibfnamefont {R.}~\bibnamefont {Aguado}}, \bibinfo
  {author} {\bibfnamefont {C.~M.}\ \bibnamefont {Lieber}}, \ and\ \bibinfo
  {author} {\bibfnamefont {S.}~\bibnamefont {De~Franceschi}},\ }\href
  {http://dx.doi.org/10.1038/nnano.2013.267} {\bibfield  {journal} {\bibinfo
  {journal} {Nat. Nanotechnol.}\ }\textbf {\bibinfo {volume} {9}},\ \bibinfo {pages}
  {79} (\bibinfo {year} {2014})}\BibitemShut {NoStop}%
\bibitem [{\citenamefont {Kumar}\ \emph {et~al.}(2014)\citenamefont {Kumar},
  \citenamefont {Gaim}, \citenamefont {Steininger}, \citenamefont {Yeyati},
  \citenamefont {Martin-Rodero}, \citenamefont {H\"uttel},\ and\ \citenamefont
  {Strunk}}]{Kumar14}%
  \BibitemOpen
  \bibfield  {author} {\bibinfo {author} {\bibfnamefont {A.}~\bibnamefont
  {Kumar}}, \bibinfo {author} {\bibfnamefont {M.}~\bibnamefont {Gaim}},
  \bibinfo {author} {\bibfnamefont {D.}~\bibnamefont {Steininger}}, \bibinfo
  {author} {\bibfnamefont {A.~L.}\ \bibnamefont {Yeyati}}, \bibinfo {author}
  {\bibfnamefont {A.}~\bibnamefont {Martin-Rodero}}, \bibinfo {author}
  {\bibfnamefont {A.~K.}\ \bibnamefont {H\"uttel}}, \ and\ \bibinfo {author}
  {\bibfnamefont {C.}~\bibnamefont {Strunk}},\ }\href@noop {} {\bibfield
  {journal} {\bibinfo  {journal} {Phys. Rev. B}\ }\textbf {\bibinfo {volume}
  {89}},\ \bibinfo {pages} {075428} (\bibinfo {year} {2014})}\BibitemShut
  {NoStop}%
\bibitem [{\citenamefont {Schindele}\ \emph {et~al.}(2014)\citenamefont
  {Schindele}, \citenamefont {Baumgartner}, \citenamefont {Maurand},
  \citenamefont {Weiss},\ and\ \citenamefont {Sch\"onenberger}}]{Schindele14}%
  \BibitemOpen
  \bibfield  {author} {\bibinfo {author} {\bibfnamefont {J.}~\bibnamefont
  {Schindele}}, \bibinfo {author} {\bibfnamefont {A.}~\bibnamefont
  {Baumgartner}}, \bibinfo {author} {\bibfnamefont {R.}~\bibnamefont
  {Maurand}}, \bibinfo {author} {\bibfnamefont {M.}~\bibnamefont {Weiss}}, \
  and\ \bibinfo {author} {\bibfnamefont {C.}~\bibnamefont {Sch\"onenberger}},\
  }\href@noop {} {\bibfield  {journal} {\bibinfo  {journal} {Phys. Rev. B}\
  }\textbf {\bibinfo {volume} {89}},\ \bibinfo {pages} {045422} (\bibinfo
  {year} {2014})}\BibitemShut {NoStop}%
\bibitem [{\citenamefont {Brandes}(2008)}]{Brandes08}%
  \BibitemOpen
  \bibfield  {author} {\bibinfo {author} {\bibfnamefont {T.}~\bibnamefont
  {Brandes}},\ }\href@noop {} {\bibfield  {journal} {\bibinfo  {journal} {Ann.
  Phys.}\ }\textbf {\bibinfo {volume} {17}},\ \bibinfo {pages} {477} (\bibinfo
  {year} {2008})}\BibitemShut {NoStop}%
\bibitem [{\citenamefont {Albert}\ \emph {et~al.}(2012)\citenamefont {Albert},
  \citenamefont {Haack}, \citenamefont {Flindt},\ and\ \citenamefont
  {B\"uttiker}}]{Albert12}%
  \BibitemOpen
  \bibfield  {author} {\bibinfo {author} {\bibfnamefont {M.}~\bibnamefont
  {Albert}}, \bibinfo {author} {\bibfnamefont {G.}~\bibnamefont {Haack}},
  \bibinfo {author} {\bibfnamefont {C.}~\bibnamefont {Flindt}}, \ and\ \bibinfo
  {author} {\bibfnamefont {M.}~\bibnamefont {B\"uttiker}},\ }\href@noop {}
  {\bibfield  {journal} {\bibinfo  {journal} {Phys. Rev. Lett}\ }\textbf
  {\bibinfo {volume} {108}},\ \bibinfo {pages} {186806} (\bibinfo {year}
  {2012})}\BibitemShut {NoStop}%
\bibitem [{\citenamefont {Thomas}\ and\ \citenamefont
  {Flindt}(2013)}]{Thomas13}%
  \BibitemOpen
  \bibfield  {author} {\bibinfo {author} {\bibfnamefont {K.~H.}\ \bibnamefont
  {Thomas}}\ and\ \bibinfo {author} {\bibfnamefont {C.}~\bibnamefont
  {Flindt}},\ }\href@noop {} {\bibfield  {journal} {\bibinfo  {journal} {Phys.
  Rev. B}\ }\textbf {\bibinfo {volume} {87}},\ \bibinfo {pages} {121405}
  (\bibinfo {year} {2013})}\BibitemShut {NoStop}%
\bibitem [{\citenamefont {Rajabi}\ \emph {et~al.}(2013)\citenamefont {Rajabi},
  \citenamefont {P\"oltl},\ and\ \citenamefont {Governale}}]{Rajabi13}%
  \BibitemOpen
  \bibfield  {author} {\bibinfo {author} {\bibfnamefont {L.}~\bibnamefont
  {Rajabi}}, \bibinfo {author} {\bibfnamefont {C.}~\bibnamefont {P\"oltl}}, \
  and\ \bibinfo {author} {\bibfnamefont {M.}~\bibnamefont {Governale}},\
  }\href@noop {} {\bibfield  {journal} {\bibinfo  {journal} {Phys. Rev. Lett}\
  }\textbf {\bibinfo {volume} {111}},\ \bibinfo {pages} {067002} (\bibinfo
  {year} {2013})}\BibitemShut {NoStop}%
\bibitem [{\citenamefont {Koch}\ \emph {et~al.}(1982)\citenamefont {Koch},
  \citenamefont {van Harlingen},\ and\ \citenamefont {Clarke}}]{Koch82}%
  \BibitemOpen
  \bibfield  {author} {\bibinfo {author} {\bibfnamefont {R.~H.}\ \bibnamefont
  {Koch}}, \bibinfo {author} {\bibfnamefont {D.}~\bibnamefont {van Harlingen}},
  \ and\ \bibinfo {author} {\bibfnamefont {J.}~\bibnamefont {Clarke}},\
  }\href@noop {} {\bibfield  {journal} {\bibinfo  {journal} {Phys. Rev. B}\
  }\textbf {\bibinfo {volume} {26}},\ \bibinfo {pages} {74} (\bibinfo {year}
  {1982})}\BibitemShut {NoStop}%
\bibitem [{\citenamefont {Schoelkopf}\ \emph {et~al.}(1997)\citenamefont
  {Schoelkopf}, \citenamefont {Burke}, \citenamefont {Kozhevnikov},
  \citenamefont {Prober},\ and\ \citenamefont {Rooks}}]{Schoelkopf97}%
  \BibitemOpen
  \bibfield  {author} {\bibinfo {author} {\bibfnamefont {R.~J.}\ \bibnamefont
  {Schoelkopf}}, \bibinfo {author} {\bibfnamefont {P.~J.}\ \bibnamefont
  {Burke}}, \bibinfo {author} {\bibfnamefont {A.~A.}\ \bibnamefont
  {Kozhevnikov}}, \bibinfo {author} {\bibfnamefont {D.~E.}\ \bibnamefont
  {Prober}}, \ and\ \bibinfo {author} {\bibfnamefont {M.~J.}\ \bibnamefont
  {Rooks}},\ }\href@noop {} {\bibfield  {journal} {\bibinfo  {journal} {Phys.
  Rev. Lett}\ }\textbf {\bibinfo {volume} {78}},\ \bibinfo {pages} {3370}
  (\bibinfo {year} {1997})}\BibitemShut {NoStop}%
\bibitem [{\citenamefont {Deblock}\ \emph {et~al.}(2003)\citenamefont
  {Deblock}, \citenamefont {Onac}, \citenamefont {Gurevich},\ and\
  \citenamefont {Kouwenhoven}}]{Deblock03}%
  \BibitemOpen
  \bibfield  {author} {\bibinfo {author} {\bibfnamefont {R.}~\bibnamefont
  {Deblock}}, \bibinfo {author} {\bibfnamefont {E.}~\bibnamefont {Onac}},
  \bibinfo {author} {\bibfnamefont {L.}~\bibnamefont {Gurevich}}, \ and\
  \bibinfo {author} {\bibfnamefont {L.~P.}\ \bibnamefont {Kouwenhoven}},\
  }\href@noop {} {\bibfield  {journal} {\bibinfo  {journal} {Science}\ }\textbf
  {\bibinfo {volume} {301}},\ \bibinfo {pages} {203} (\bibinfo {year}
  {2003})}\BibitemShut {NoStop}%
\bibitem [{\citenamefont {Onac}\ \emph {et~al.}(2006)\citenamefont {Onac},
  \citenamefont {Balestro}, \citenamefont {van Beveren}, \citenamefont
  {Hartmann}, \citenamefont {Nazarov},\ and\ \citenamefont
  {Kouwenhoven}}]{Onac06}%
  \BibitemOpen
  \bibfield  {author} {\bibinfo {author} {\bibfnamefont {E.}~\bibnamefont
  {Onac}}, \bibinfo {author} {\bibfnamefont {F.}~\bibnamefont {Balestro}},
  \bibinfo {author} {\bibfnamefont {L.~H.~W.}\ \bibnamefont {van Beveren}},
  \bibinfo {author} {\bibfnamefont {U.}~\bibnamefont {Hartmann}}, \bibinfo
  {author} {\bibfnamefont {Y.~V.}\ \bibnamefont {Nazarov}}, \ and\ \bibinfo
  {author} {\bibfnamefont {L.~P.}\ \bibnamefont {Kouwenhoven}},\ }\href@noop {}
  {\bibfield  {journal} {\bibinfo  {journal} {Phys. Rev. Lett}\ }\textbf
  {\bibinfo {volume} {96}},\ \bibinfo {pages} {176601} (\bibinfo {year}
  {2006})}\BibitemShut {NoStop}%
\bibitem [{\citenamefont {Billangeon}\ \emph {et~al.}(2006)\citenamefont
  {Billangeon}, \citenamefont {Pierre}, \citenamefont {Bouchiat},\ and\
  \citenamefont {Deblock}}]{Billangeon06}%
  \BibitemOpen
  \bibfield  {author} {\bibinfo {author} {\bibfnamefont {P.-M.}\ \bibnamefont
  {Billangeon}}, \bibinfo {author} {\bibfnamefont {F.}~\bibnamefont {Pierre}},
  \bibinfo {author} {\bibfnamefont {H.}~\bibnamefont {Bouchiat}}, \ and\
  \bibinfo {author} {\bibfnamefont {R.}~\bibnamefont {Deblock}},\ }\href@noop
  {} {\bibfield  {journal} {\bibinfo  {journal} {Phys. Rev. Lett}\ }\textbf
  {\bibinfo {volume} {96}},\ \bibinfo {pages} {136804} (\bibinfo {year}
  {2006})}\BibitemShut {NoStop}%
\bibitem [{\citenamefont {Billangeon}\ \emph {et~al.}(2007)\citenamefont
  {Billangeon}, \citenamefont {Pierre}, \citenamefont {Bouchiat},\ and\
  \citenamefont {Deblock}}]{Billangeon07}%
  \BibitemOpen
  \bibfield  {author} {\bibinfo {author} {\bibfnamefont {P.-M.}\ \bibnamefont
  {Billangeon}}, \bibinfo {author} {\bibfnamefont {F.}~\bibnamefont {Pierre}},
  \bibinfo {author} {\bibfnamefont {H.}~\bibnamefont {Bouchiat}}, \ and\
  \bibinfo {author} {\bibfnamefont {R.}~\bibnamefont {Deblock}},\ }\href@noop
  {} {\bibfield  {journal} {\bibinfo  {journal} {Phys. Rev. Lett}\ }\textbf
  {\bibinfo {volume} {98}},\ \bibinfo {pages} {126802} (\bibinfo {year}
  {2007})}\BibitemShut {NoStop}%
\bibitem [{\citenamefont {Zakka-Bajjani}\ \emph {et~al.}(2007)\citenamefont
  {Zakka-Bajjani}, \citenamefont {S{\'e}gala}, \citenamefont {Portier},
  \citenamefont {Roche}, \citenamefont {Glattli}, \citenamefont {Cavanna},\
  and\ \citenamefont {Jin}}]{Zakka07}%
  \BibitemOpen
  \bibfield  {author} {\bibinfo {author} {\bibfnamefont {E.}~\bibnamefont
  {Zakka-Bajjani}}, \bibinfo {author} {\bibfnamefont {J.}~\bibnamefont
  {S{\'e}gala}}, \bibinfo {author} {\bibfnamefont {F.}~\bibnamefont {Portier}},
  \bibinfo {author} {\bibfnamefont {P.}~\bibnamefont {Roche}}, \bibinfo
  {author} {\bibfnamefont {D.~C.}\ \bibnamefont {Glattli}}, \bibinfo {author}
  {\bibfnamefont {A.}~\bibnamefont {Cavanna}}, \ and\ \bibinfo {author}
  {\bibfnamefont {Y.}~\bibnamefont {Jin}},\ }\href@noop {} {\bibfield
  {journal} {\bibinfo  {journal} {Phys. Rev. Lett}\ }\textbf {\bibinfo {volume}
  {99}},\ \bibinfo {pages} {236803} (\bibinfo {year} {2007})}\BibitemShut
  {NoStop}%
\bibitem [{\citenamefont {Gabelli}\ and\ \citenamefont
  {Reulet}(2008)}]{Gabelli08}%
  \BibitemOpen
  \bibfield  {author} {\bibinfo {author} {\bibfnamefont {J.}~\bibnamefont
  {Gabelli}}\ and\ \bibinfo {author} {\bibfnamefont {B.}~\bibnamefont
  {Reulet}},\ }\href@noop {} {\bibfield  {journal} {\bibinfo  {journal} {Phys.
  Rev. Lett}\ }\textbf {\bibinfo {volume} {100}},\ \bibinfo {pages} {026601}
  (\bibinfo {year} {2008})}\BibitemShut {NoStop}%
\bibitem [{\citenamefont {Ubbelohde}\ \emph {et~al.}(2012)\citenamefont
  {Ubbelohde}, \citenamefont {Fricke}, \citenamefont {Flindt}, \citenamefont
  {Hohls},\ and\ \citenamefont {Haug}}]{Ubbelohde12}%
  \BibitemOpen
  \bibfield  {author} {\bibinfo {author} {\bibfnamefont {N.}~\bibnamefont
  {Ubbelohde}}, \bibinfo {author} {\bibfnamefont {C.}~\bibnamefont {Fricke}},
  \bibinfo {author} {\bibfnamefont {C.}~\bibnamefont {Flindt}}, \bibinfo
  {author} {\bibfnamefont {F.}~\bibnamefont {Hohls}}, \ and\ \bibinfo {author}
  {\bibfnamefont {R.~J.}\ \bibnamefont {Haug}},\ }\href@noop {} {\bibfield
  {journal} {\bibinfo  {journal} {Nat. Commun.}\ }\textbf {\bibinfo {volume}
  {3}},\ \bibinfo {pages} {612} (\bibinfo {year} {2012})}\BibitemShut {NoStop}%
\bibitem [{\citenamefont {Maisi}\ \emph {et~al.}(2014)\citenamefont {Maisi},
  \citenamefont {Kambly}, \citenamefont {Flindt},\ and\ \citenamefont
  {Pekola}}]{Maisi14}%
  \BibitemOpen
  \bibfield  {author} {\bibinfo {author} {\bibfnamefont {V.~F.}\ \bibnamefont
  {Maisi}}, \bibinfo {author} {\bibfnamefont {D.}~\bibnamefont {Kambly}},
  \bibinfo {author} {\bibfnamefont {C.}~\bibnamefont {Flindt}}, \ and\ \bibinfo
  {author} {\bibfnamefont {J.~P.}\ \bibnamefont {Pekola}},\ }\href@noop {}
  {\bibfield  {journal} {\bibinfo  {journal} {Phys. Rev. Lett}\ }\textbf
  {\bibinfo {volume} {112}},\ \bibinfo {pages} {036801} (\bibinfo {year}
  {2014})}\BibitemShut {NoStop}%
\bibitem [{\citenamefont {Averin}(1993)}]{Averin93}%
  \BibitemOpen
  \bibfield  {author} {\bibinfo {author} {\bibfnamefont {D.~V.}\ \bibnamefont
  {Averin}},\ }\href@noop {} {\bibfield  {journal} {\bibinfo  {journal} {J.
  Appl. Phys.}\ }\textbf {\bibinfo {volume} {73}},\ \bibinfo {pages} {2593}
  (\bibinfo {year} {1993})}\BibitemShut {NoStop}%
\bibitem [{\citenamefont {Ding}\ and\ \citenamefont {Ng}(1997)}]{Ding97}%
  \BibitemOpen
  \bibfield  {author} {\bibinfo {author} {\bibfnamefont {G.-H.}\ \bibnamefont
  {Ding}}\ and\ \bibinfo {author} {\bibfnamefont {T.-K.}\ \bibnamefont {Ng}},\
  }\href@noop {} {\bibfield  {journal} {\bibinfo  {journal} {Phys. Rev. B}\
  }\textbf {\bibinfo {volume} {56}},\ \bibinfo {pages} {15521 (R)} (\bibinfo
  {year} {1997})}\BibitemShut {NoStop}%
\bibitem [{\citenamefont {Aguado}\ and\ \citenamefont
  {Kouwenhoven}(2000)}]{Aguado00}%
  \BibitemOpen
  \bibfield  {author} {\bibinfo {author} {\bibfnamefont {R.}~\bibnamefont
  {Aguado}}\ and\ \bibinfo {author} {\bibfnamefont {L.~P.}\ \bibnamefont
  {Kouwenhoven}},\ }\href@noop {} {\bibfield  {journal} {\bibinfo  {journal}
  {Phys. Rev. Lett}\ }\textbf {\bibinfo {volume} {84}},\ \bibinfo {pages}
  {1986} (\bibinfo {year} {2000})}\BibitemShut {NoStop}%
\bibitem [{\citenamefont {Blanter}\ and\ \citenamefont
  {B\"uttiker}(2000)}]{Blanter00}%
  \BibitemOpen
  \bibfield  {author} {\bibinfo {author} {\bibfnamefont {Y.}~\bibnamefont
  {Blanter}}\ and\ \bibinfo {author} {\bibfnamefont {M.}~\bibnamefont
  {B\"uttiker}},\ }\href@noop {} {\bibfield  {journal} {\bibinfo  {journal}
  {Phys. Rep.}\ }\textbf {\bibinfo {volume} {336}},\ \bibinfo {pages} {1}
  (\bibinfo {year} {2000})}\BibitemShut {NoStop}%
\bibitem [{\citenamefont {Engel}\ and\ \citenamefont {Loss}(2004)}]{Engel04}%
  \BibitemOpen
  \bibfield  {author} {\bibinfo {author} {\bibfnamefont {H.~A.}\ \bibnamefont
  {Engel}}\ and\ \bibinfo {author} {\bibfnamefont {D.}~\bibnamefont {Loss}},\
  }\href@noop {} {\bibfield  {journal} {\bibinfo  {journal} {Phys. Rev. Lett}\
  }\textbf {\bibinfo {volume} {93}},\ \bibinfo {pages} {136602} (\bibinfo
  {year} {2004})}\BibitemShut {NoStop}%
\bibitem [{\citenamefont {Braun}\ \emph {et~al.}(2006)\citenamefont {Braun},
  \citenamefont {K\"onig},\ and\ \citenamefont {Martinek}}]{Braun06}%
  \BibitemOpen
  \bibfield  {author} {\bibinfo {author} {\bibfnamefont {M.}~\bibnamefont
  {Braun}}, \bibinfo {author} {\bibfnamefont {J.}~\bibnamefont {K\"onig}}, \
  and\ \bibinfo {author} {\bibfnamefont {J.}~\bibnamefont {Martinek}},\
  }\href@noop {} {\bibfield  {journal} {\bibinfo  {journal} {Phys. Rev. B}\
  }\textbf {\bibinfo {volume} {74}},\ \bibinfo {pages} {075328} (\bibinfo
  {year} {2006})}\BibitemShut {NoStop}%
\bibitem [{\citenamefont {Entin-Wohlman}\ \emph {et~al.}(2007)\citenamefont
  {Entin-Wohlman}, \citenamefont {Imry}, \citenamefont {Gurvitz},\ and\
  \citenamefont {Aharony}}]{Wohlman07}%
  \BibitemOpen
  \bibfield  {author} {\bibinfo {author} {\bibfnamefont {O.}~\bibnamefont
  {Entin-Wohlman}}, \bibinfo {author} {\bibfnamefont {Y.}~\bibnamefont {Imry}},
  \bibinfo {author} {\bibfnamefont {S.~A.}\ \bibnamefont {Gurvitz}}, \ and\
  \bibinfo {author} {\bibfnamefont {A.}~\bibnamefont {Aharony}},\ }\href@noop
  {} {\bibfield  {journal} {\bibinfo  {journal} {Phys. Rev. B}\ }\textbf
  {\bibinfo {volume} {75}},\ \bibinfo {pages} {193308} (\bibinfo {year}
  {2007})}\BibitemShut {NoStop}%
\bibitem [{\citenamefont {Rothstein}\ \emph {et~al.}(2009)\citenamefont
  {Rothstein}, \citenamefont {Entin-Wohlman},\ and\ \citenamefont
  {Aharony}}]{Rothstein09}%
  \BibitemOpen
  \bibfield  {author} {\bibinfo {author} {\bibfnamefont {E.~A.}\ \bibnamefont
  {Rothstein}}, \bibinfo {author} {\bibfnamefont {O.}~\bibnamefont
  {Entin-Wohlman}}, \ and\ \bibinfo {author} {\bibfnamefont {A.}~\bibnamefont
  {Aharony}},\ }\href@noop {} {\bibfield  {journal} {\bibinfo  {journal} {Phys.
  Rev. B}\ }\textbf {\bibinfo {volume} {79}},\ \bibinfo {pages} {075307}
  (\bibinfo {year} {2009})}\BibitemShut {NoStop}%
\bibitem [{\citenamefont {Gabdank}\ \emph {et~al.}(2011)\citenamefont
  {Gabdank}, \citenamefont {Rothstein}, \citenamefont {Entin-Wohlman},\ and\
  \citenamefont {Aharony}}]{Gabdank11}%
  \BibitemOpen
  \bibfield  {author} {\bibinfo {author} {\bibfnamefont {N.}~\bibnamefont
  {Gabdank}}, \bibinfo {author} {\bibfnamefont {E.~A.}\ \bibnamefont
  {Rothstein}}, \bibinfo {author} {\bibfnamefont {O.}~\bibnamefont
  {Entin-Wohlman}}, \ and\ \bibinfo {author} {\bibfnamefont {A.}~\bibnamefont
  {Aharony}},\ }\href@noop {} {\bibfield  {journal} {\bibinfo  {journal} {Phys.
  Rev. B}\ }\textbf {\bibinfo {volume} {84}},\ \bibinfo {pages} {235435}
  (\bibinfo {year} {2011})}\BibitemShut {NoStop}%
\bibitem [{\citenamefont {Orth}\ \emph {et~al.}(2012)\citenamefont {Orth},
  \citenamefont {Urban},\ and\ \citenamefont {Komnik}}]{Orth12}%
  \BibitemOpen
  \bibfield  {author} {\bibinfo {author} {\bibfnamefont {C.~P.}\ \bibnamefont
  {Orth}}, \bibinfo {author} {\bibfnamefont {D.~F.}\ \bibnamefont {Urban}}, \
  and\ \bibinfo {author} {\bibfnamefont {A.}~\bibnamefont {Komnik}},\
  }\href@noop {} {\bibfield  {journal} {\bibinfo  {journal} {Phys. Rev. B}\
  }\textbf {\bibinfo {volume} {86}},\ \bibinfo {pages} {125324} (\bibinfo
  {year} {2012})}\BibitemShut {NoStop}%
\bibitem [{\citenamefont {Bransch\"adel}\ \emph {et~al.}(2010)\citenamefont
  {Bransch\"adel}, \citenamefont {Boulat}, \citenamefont {Saleur},\ and\
  \citenamefont {Schmitteckert}}]{Brandschaedel11}%
  \BibitemOpen
  \bibfield  {author} {\bibinfo {author} {\bibfnamefont {A.}~\bibnamefont
  {Bransch\"adel}}, \bibinfo {author} {\bibfnamefont {E.}~\bibnamefont
  {Boulat}}, \bibinfo {author} {\bibfnamefont {H.}~\bibnamefont {Saleur}}, \
  and\ \bibinfo {author} {\bibfnamefont {P.}~\bibnamefont {Schmitteckert}},\
  }\href@noop {} {\bibfield  {journal} {\bibinfo  {journal} {Phys. Rev. Lett}\
  }\textbf {\bibinfo {volume} {105}},\ \bibinfo {pages} {146805} (\bibinfo
  {year} {2010})}\BibitemShut {NoStop}%
\bibitem [{\citenamefont {Marcos}\ \emph {et~al.}(2011)\citenamefont {Marcos},
  \citenamefont {Emary}, \citenamefont {Brandes},\ and\ \citenamefont
  {Aguado}}]{Marcos11}%
  \BibitemOpen
  \bibfield  {author} {\bibinfo {author} {\bibfnamefont {D.}~\bibnamefont
  {Marcos}}, \bibinfo {author} {\bibfnamefont {C.}~\bibnamefont {Emary}},
  \bibinfo {author} {\bibfnamefont {T.}~\bibnamefont {Brandes}}, \ and\
  \bibinfo {author} {\bibfnamefont {R.}~\bibnamefont {Aguado}},\ }\href@noop {}
  {\bibfield  {journal} {\bibinfo  {journal} {Phys. Rev. B}\ }\textbf {\bibinfo
  {volume} {83}},\ \bibinfo {pages} {125426} (\bibinfo {year}
  {2011})}\BibitemShut {NoStop}%
\bibitem [{\citenamefont {Joho}\ \emph {et~al.}(2012)\citenamefont {Joho},
  \citenamefont {Maier},\ and\ \citenamefont {Komnik}}]{Joho12}%
  \BibitemOpen
  \bibfield  {author} {\bibinfo {author} {\bibfnamefont {K.}~\bibnamefont
  {Joho}}, \bibinfo {author} {\bibfnamefont {S.}~\bibnamefont {Maier}}, \ and\
  \bibinfo {author} {\bibfnamefont {A.}~\bibnamefont {Komnik}},\ }\href@noop {}
  {\bibfield  {journal} {\bibinfo  {journal} {Phys. Rev. B}\ }\textbf {\bibinfo
  {volume} {86}},\ \bibinfo {pages} {155304} (\bibinfo {year}
  {2012})}\BibitemShut {NoStop}%
\bibitem [{\citenamefont {Sothmann}\ \emph {et~al.}(2010)\citenamefont
  {Sothmann}, \citenamefont {K\"onig},\ and\ \citenamefont
  {Kadigrobov}}]{Sothmann10}%
  \BibitemOpen
  \bibfield  {author} {\bibinfo {author} {\bibfnamefont {B.}~\bibnamefont
  {Sothmann}}, \bibinfo {author} {\bibfnamefont {J.}~\bibnamefont {K\"onig}}, \
  and\ \bibinfo {author} {\bibfnamefont {A.}~\bibnamefont {Kadigrobov}},\
  }\href@noop {} {\bibfield  {journal} {\bibinfo  {journal} {Phys. Rev. B}\
  }\textbf {\bibinfo {volume} {82}},\ \bibinfo {pages} {205314} (\bibinfo
  {year} {2010})}\BibitemShut {NoStop}%
\bibitem [{\citenamefont {Jin}\ \emph {et~al.}(2012)\citenamefont {Jin},
  \citenamefont {Zhang}, \citenamefont {Li},\ and\ \citenamefont
  {Yan}}]{Jin12}%
  \BibitemOpen
  \bibfield  {author} {\bibinfo {author} {\bibfnamefont {J.}~\bibnamefont
  {Jin}}, \bibinfo {author} {\bibfnamefont {W.-M.}\ \bibnamefont {Zhang}},
  \bibinfo {author} {\bibfnamefont {X.-Q.}\ \bibnamefont {Li}}, \ and\ \bibinfo
  {author} {\bibfnamefont {Y.~J.}\ \bibnamefont {Yan}},\ }\href@noop {}
  {\bibfield  {journal} {\bibinfo  {journal} {arXiv:cond-mat/1105.0136v2}\ }
  (\bibinfo {year} {2012})}\BibitemShut {NoStop}%
\bibitem [{\citenamefont {M\"uller}\ \emph {et~al.}(2013)\citenamefont
  {M\"uller}, \citenamefont {Pletyukhov}, \citenamefont {Schuricht},\ and\
  \citenamefont {Andergassen}}]{Mueller13}%
  \BibitemOpen
  \bibfield  {author} {\bibinfo {author} {\bibfnamefont {S.~Y.}\ \bibnamefont
  {M\"uller}}, \bibinfo {author} {\bibfnamefont {M.}~\bibnamefont
  {Pletyukhov}}, \bibinfo {author} {\bibfnamefont {D.}~\bibnamefont
  {Schuricht}}, \ and\ \bibinfo {author} {\bibfnamefont {S.}~\bibnamefont
  {Andergassen}},\ }\href@noop {} {\bibfield  {journal} {\bibinfo  {journal}
  {Phys. Rev. B}\ }\textbf {\bibinfo {volume} {87}},\ \bibinfo {pages} {245115}
  (\bibinfo {year} {2013})}\BibitemShut {NoStop}%
\bibitem [{\citenamefont {Ding}\ and\ \citenamefont {Dong}(2013)}]{Dong13}%
  \BibitemOpen
  \bibfield  {author} {\bibinfo {author} {\bibfnamefont {G.-H.}\ \bibnamefont
  {Ding}}\ and\ \bibinfo {author} {\bibfnamefont {B.}~\bibnamefont {Dong}},\
  }\href@noop {} {\bibfield  {journal} {\bibinfo  {journal} {Phys. Rev. B}\
  }\textbf {\bibinfo {volume} {87}},\ \bibinfo {pages} {235303} (\bibinfo
  {year} {2013})}\BibitemShut {NoStop}%
\bibitem [{\citenamefont {Kirton}\ \emph {et~al.}(2012)\citenamefont {Kirton},
  \citenamefont {Armour}, \citenamefont {Houzet},\ and\ \citenamefont
  {Pistolesi}}]{Kirton12}%
  \BibitemOpen
  \bibfield  {author} {\bibinfo {author} {\bibfnamefont {P.~G.}\ \bibnamefont
  {Kirton}}, \bibinfo {author} {\bibfnamefont {A.~D.}\ \bibnamefont {Armour}},
  \bibinfo {author} {\bibfnamefont {M.}~\bibnamefont {Houzet}}, \ and\ \bibinfo
  {author} {\bibfnamefont {F.}~\bibnamefont {Pistolesi}},\ }\href@noop {}
  {\bibfield  {journal} {\bibinfo  {journal} {Phys. Rev. B}\ }\textbf {\bibinfo
  {volume} {86}},\ \bibinfo {pages} {081305(R)} (\bibinfo {year}
  {2012})}\BibitemShut {NoStop}%
\bibitem [{\citenamefont {Soller}\ and\ \citenamefont
  {Komnik}(2014)}]{Soller14}%
  \BibitemOpen
  \bibfield  {author} {\bibinfo {author} {\bibfnamefont {H.}~\bibnamefont
  {Soller}}\ and\ \bibinfo {author} {\bibfnamefont {A.}~\bibnamefont
  {Komnik}},\ }\href@noop {} {\bibfield  {journal} {\bibinfo  {journal}
  {EPL}\ }\textbf {\bibinfo {volume} {106}},\ \bibinfo {pages} {37009} (\bibinfo {year} {2014})}\BibitemShut
  {NoStop}%
\bibitem [{\citenamefont {Moca}\ \emph {et~al.}(2014)\citenamefont {Moca},
  \citenamefont {Simon}, \citenamefont {Chung},\ and\ \citenamefont
  {Zarand}}]{Moca14}%
  \BibitemOpen
  \bibfield  {author} {\bibinfo {author} {\bibfnamefont {C.~P.}\ \bibnamefont
  {Moca}}, \bibinfo {author} {\bibfnamefont {P.}~\bibnamefont {Simon}},
  \bibinfo {author} {\bibfnamefont {C.-H.}\ \bibnamefont {Chung}}, \ and\
  \bibinfo {author} {\bibfnamefont {G.}~\bibnamefont {Zarand}},\ }\href@noop {}
  {\bibfield  {journal} {\bibinfo  {journal} {Phys. Rev. B}\ }\textbf {\bibinfo
  {volume} {89}},\ \bibinfo {pages} {155138} (\bibinfo {year}
  {2014})}\BibitemShut {NoStop}%
\bibitem [{\citenamefont {K\"onig}\ \emph
  {et~al.}(1996{\natexlab{a}})\citenamefont {K\"onig}, \citenamefont
  {Schoeller},\ and\ \citenamefont {Sch\"on}}]{Koenig96}%
  \BibitemOpen
  \bibfield  {author} {\bibinfo {author} {\bibfnamefont {J.}~\bibnamefont
  {K\"onig}}, \bibinfo {author} {\bibfnamefont {H.}~\bibnamefont {Schoeller}},
  \ and\ \bibinfo {author} {\bibfnamefont {G.}~\bibnamefont {Sch\"on}},\
  }\href@noop {} {\bibfield  {journal} {\bibinfo  {journal} {Phys. Rev. Lett}\
  }\textbf {\bibinfo {volume} {76}},\ \bibinfo {pages} {1715} (\bibinfo {year}
  {1996}{\natexlab{a}})}\BibitemShut {NoStop}%
\bibitem [{\citenamefont {K\"onig}\ \emph
  {et~al.}(1996{\natexlab{b}})\citenamefont {K\"onig}, \citenamefont {Schmid},
  \citenamefont {Schoeller},\ and\ \citenamefont {Sch\"on}}]{Koenig96B}%
  \BibitemOpen
  \bibfield  {author} {\bibinfo {author} {\bibfnamefont {J.}~\bibnamefont
  {K\"onig}}, \bibinfo {author} {\bibfnamefont {J.}~\bibnamefont {Schmid}},
  \bibinfo {author} {\bibfnamefont {H.}~\bibnamefont {Schoeller}}, \ and\
  \bibinfo {author} {\bibfnamefont {G.}~\bibnamefont {Sch\"on}},\ }\href@noop
  {} {\bibfield  {journal} {\bibinfo  {journal} {Phys. Rev. B}\ }\textbf
  {\bibinfo {volume} {54}},\ \bibinfo {pages} {16820} (\bibinfo {year}
  {1996}{\natexlab{b}})}\BibitemShut {NoStop}%
\bibitem [{\citenamefont {Thielmann}\ \emph {et~al.}(2003)\citenamefont
  {Thielmann}, \citenamefont {Hettler}, \citenamefont {K\"onig},\ and\
  \citenamefont {Sch\"on}}]{Thielmann03}%
  \BibitemOpen
  \bibfield  {author} {\bibinfo {author} {\bibfnamefont {A.}~\bibnamefont
  {Thielmann}}, \bibinfo {author} {\bibfnamefont {M.~H.}\ \bibnamefont
  {Hettler}}, \bibinfo {author} {\bibfnamefont {J.}~\bibnamefont {K\"onig}}, \
  and\ \bibinfo {author} {\bibfnamefont {G.}~\bibnamefont {Sch\"on}},\
  }\href@noop {} {\bibfield  {journal} {\bibinfo  {journal} {Phys. Rev. B}\
  }\textbf {\bibinfo {volume} {68}},\ \bibinfo {pages} {115105} (\bibinfo
  {year} {2003})}\BibitemShut {NoStop}%
\bibitem [{\citenamefont {Cuevas}\ \emph {et~al.}(2001)\citenamefont {Cuevas},
  \citenamefont {Levy~Yeyati},\ and\ \citenamefont {Martin-Rodero}}]{Cuevas01}%
  \BibitemOpen
  \bibfield  {author} {\bibinfo {author} {\bibfnamefont {J.}~\bibnamefont
  {Cuevas}}, \bibinfo {author} {\bibfnamefont {A.}~\bibnamefont {Levy~Yeyati}},
  \ and\ \bibinfo {author} {\bibfnamefont {A.}~\bibnamefont {Martin-Rodero}},\
  }\href@noop {} {\bibfield  {journal} {\bibinfo  {journal} {Phys. Rev. B}\
  }\textbf {\bibinfo {volume} {63}},\ \bibinfo {pages} {094515} (\bibinfo
  {year} {2001})}\BibitemShut {NoStop}%
\bibitem [{\citenamefont {Tanaka1}\ \emph {et~al.}(2007)\citenamefont
  {Tanaka1}, \citenamefont {Kawakami},\ and\ \citenamefont {Oguri}}]{Tanaka07}%
  \BibitemOpen
  \bibfield  {author} {\bibinfo {author} {\bibfnamefont {Y.}~\bibnamefont
  {Tanaka1}}, \bibinfo {author} {\bibfnamefont {N.}~\bibnamefont {Kawakami}}, \
  and\ \bibinfo {author} {\bibfnamefont {A.}~\bibnamefont {Oguri}},\
  }\href@noop {} {\bibfield  {journal} {\bibinfo  {journal} {J. Phys. Soc.
  Jpn.}\ }\textbf {\bibinfo {volume} {76}},\ \bibinfo {pages} {074701}
  (\bibinfo {year} {2007})}\BibitemShut {NoStop}%
\bibitem [{\citenamefont {Yamada}\ \emph {et~al.}(2010)\citenamefont {Yamada},
  \citenamefont {Tanaka},\ and\ \citenamefont {Kawakami}}]{Yamada10}%
  \BibitemOpen
  \bibfield  {author} {\bibinfo {author} {\bibfnamefont {Y.}~\bibnamefont
  {Yamada}}, \bibinfo {author} {\bibfnamefont {Y.}~\bibnamefont {Tanaka}}, \
  and\ \bibinfo {author} {\bibfnamefont {N.}~\bibnamefont {Kawakami}},\
  }\href@noop {} {\bibfield  {journal} {\bibinfo  {journal} {J. Phys. Soc.
  Jpn.}\ }\textbf {\bibinfo {volume} {79}},\ \bibinfo {pages} {043705}
  (\bibinfo {year} {2010})}\BibitemShut {NoStop}%
\bibitem [{\citenamefont {Yamada}\ \emph {et~al.}(2011)\citenamefont {Yamada},
  \citenamefont {Tanaka},\ and\ \citenamefont {Kawakami}}]{Yamada11}%
  \BibitemOpen
  \bibfield  {author} {\bibinfo {author} {\bibfnamefont {Y.}~\bibnamefont
  {Yamada}}, \bibinfo {author} {\bibfnamefont {Y.}~\bibnamefont {Tanaka}}, \
  and\ \bibinfo {author} {\bibfnamefont {N.}~\bibnamefont {Kawakami}},\
  }\href@noop {} {\bibfield  {journal} {\bibinfo  {journal} {Phys. Rev. B}\
  }\textbf {\bibinfo {volume} {84}},\ \bibinfo {pages} {075484} (\bibinfo
  {year} {2011})}\BibitemShut {NoStop}%
\bibitem [{\citenamefont {Zapalska}\ and\ \citenamefont
  {Domanski}(2014)}]{Zapalska14}%
  \BibitemOpen
  \bibfield  {author} {\bibinfo {author} {\bibfnamefont {M.}~\bibnamefont
  {Zapalska}}\ and\ \bibinfo {author} {\bibfnamefont {T.}~\bibnamefont
  {Domanski}},\ }\href@noop {} {\bibfield  {journal} {\bibinfo  {journal}
  {arXiv:1402.1291}\ } (\bibinfo {year} {2014})}\BibitemShut {NoStop}%
\bibitem [{\citenamefont {\ifmmode~\check{Z}\else \v{Z}\fi{}itko}\ \emph
  {et~al.}(2015)\citenamefont {\ifmmode~\check{Z}\else \v{Z}\fi{}itko},
  \citenamefont {Lim}, \citenamefont {L\'opez},\ and\ \citenamefont
  {Aguado}}]{zitko}%
  \BibitemOpen
  \bibfield  {author} {\bibinfo {author} {\bibfnamefont {R.}~\bibnamefont
  {\ifmmode~\check{Z}\else \v{Z}\fi{}itko}}, \bibinfo {author} {\bibfnamefont
  {J.~S.}\ \bibnamefont {Lim}}, \bibinfo {author} {\bibfnamefont
  {R.}~\bibnamefont {L\'opez}}, \ and\ \bibinfo {author} {\bibfnamefont
  {R.}~\bibnamefont {Aguado}},\ }\href@noop {} {\bibfield  {journal} {\bibinfo
  {journal} {Phys. Rev. B}\ }\textbf {\bibinfo {volume} {91}},\ \bibinfo
  {pages} {045441} (\bibinfo {year} {2015})}\BibitemShut {NoStop}%
\bibitem [{\citenamefont {Futterer}\ \emph {et~al.}(2013)\citenamefont
  {Futterer}, \citenamefont {Swiebodzinski}, \citenamefont {Governale},\ and\
  \citenamefont {K{\"o}nig}}]{Futterer13}%
  \BibitemOpen
  \bibfield  {author} {\bibinfo {author} {\bibfnamefont {D.}~\bibnamefont
  {Futterer}}, \bibinfo {author} {\bibfnamefont {J.}~\bibnamefont
  {Swiebodzinski}}, \bibinfo {author} {\bibfnamefont {M.}~\bibnamefont
  {Governale}}, \ and\ \bibinfo {author} {\bibfnamefont {J.}~\bibnamefont
  {K{\"o}nig}},\ }\href@noop {} {\bibfield  {journal} {\bibinfo  {journal}
  {Phys. Rev. B}\ }\textbf {\bibinfo {volume} {87}},\ \bibinfo {pages} {014509}
  (\bibinfo {year} {2013})}\BibitemShut {NoStop}%
\bibitem [{\citenamefont {Rozhkov}\ and\ \citenamefont
  {Arovas}(2000)}]{Rozhkov00}%
  \BibitemOpen
  \bibfield  {author} {\bibinfo {author} {\bibfnamefont {A.~V.}\ \bibnamefont
  {Rozhkov}}\ and\ \bibinfo {author} {\bibfnamefont {D.~P.}\ \bibnamefont
  {Arovas}},\ }\href@noop {} {\bibfield  {journal} {\bibinfo  {journal} {Phys.
  Rev. B}\ }\textbf {\bibinfo {volume} {62}},\ \bibinfo {pages} {6687}
  (\bibinfo {year} {2000})}\BibitemShut {NoStop}%
\bibitem [{\citenamefont {Gavish}\ \emph {et~al.}(2000)\citenamefont {Gavish},
  \citenamefont {Levinson},\ and\ \citenamefont {Imry}}]{Gavish00}%
  \BibitemOpen
  \bibfield  {author} {\bibinfo {author} {\bibfnamefont {U.}~\bibnamefont
  {Gavish}}, \bibinfo {author} {\bibfnamefont {Y.}~\bibnamefont {Levinson}}, \
  and\ \bibinfo {author} {\bibfnamefont {Y.}~\bibnamefont {Imry}},\ }\href@noop
  {} {\bibfield  {journal} {\bibinfo  {journal} {Phys. Rev. B}\ }\textbf
  {\bibinfo {volume} {62}},\ \bibinfo {pages} {R10637} (\bibinfo {year}
  {2000})}\BibitemShut {NoStop}%
\bibitem [{\citenamefont {Shockley}(1938)}]{Shockley38}%
  \BibitemOpen
  \bibfield  {author} {\bibinfo {author} {\bibfnamefont {W.}~\bibnamefont
  {Shockley}},\ }\href@noop {} {\bibfield  {journal} {\bibinfo  {journal} {J.
  Appl. Phys.}\ }\textbf {\bibinfo {volume} {9}},\ \bibinfo {pages} {635}
  (\bibinfo {year} {1938})}\BibitemShut {NoStop}%
\bibitem [{\citenamefont {Pala}\ \emph {et~al.}(2007)\citenamefont {Pala},
  \citenamefont {Governale},\ and\ \citenamefont {K\"onig}}]{Pala07}%
  \BibitemOpen
  \bibfield  {author} {\bibinfo {author} {\bibfnamefont {M.~G.}\ \bibnamefont
  {Pala}}, \bibinfo {author} {\bibfnamefont {M.}~\bibnamefont {Governale}}, \
  and\ \bibinfo {author} {\bibfnamefont {J.}~\bibnamefont {K\"onig}},\
  }\href@noop {} {\bibfield  {journal} {\bibinfo  {journal} {New. J. Phys.}\
  }\textbf {\bibinfo {volume} {9}},\ \bibinfo {pages} {278} (\bibinfo {year}
  {2007})}\BibitemShut {NoStop}%
\bibitem [{\citenamefont {Clerk}\ \emph {et~al.}(2010)\citenamefont {Clerk},
  \citenamefont {Devoret}, \citenamefont {Girvin}, \citenamefont {Marquardt},\
  and\ \citenamefont {Schoelkopf}}]{Clerk10}%
  \BibitemOpen
  \bibfield  {author} {\bibinfo {author} {\bibfnamefont {A.~A.}\ \bibnamefont
  {Clerk}}, \bibinfo {author} {\bibfnamefont {M.~H.}\ \bibnamefont {Devoret}},
  \bibinfo {author} {\bibfnamefont {S.~M.}\ \bibnamefont {Girvin}}, \bibinfo
  {author} {\bibfnamefont {F.}~\bibnamefont {Marquardt}}, \ and\ \bibinfo
  {author} {\bibfnamefont {R.~J.}\ \bibnamefont {Schoelkopf}},\ }\href
  {\doibase 10.1103/RevModPhys.82.1155} {\bibfield  {journal} {\bibinfo
  {journal} {Rev. Mod. Phys.}\ }\textbf {\bibinfo {volume} {82}},\ \bibinfo
  {pages} {1155} (\bibinfo {year} {2010})}\BibitemShut {NoStop}%
\bibitem [{\citenamefont {Emary}\ and\ \citenamefont {Aguado}(2011)}]{Emary11}%
  \BibitemOpen
  \bibfield  {author} {\bibinfo {author} {\bibfnamefont {C.}~\bibnamefont
  {Emary}}\ and\ \bibinfo {author} {\bibfnamefont {R.}~\bibnamefont {Aguado}},\
  }\href@noop {} {\bibfield  {journal} {\bibinfo  {journal} {Phys. Rev. B}\
  }\textbf {\bibinfo {volume} {84}},\ \bibinfo {pages} {085425} (\bibinfo
  {year} {2011})}\BibitemShut {NoStop}%
\bibitem [{Note1()}]{Note1}%
  \BibitemOpen
  \bibinfo {note} {Note that we here consider a symmetrized noise spectrum.
  Therefore both contributions, independently on whether the excitation energy
  of a process is positive or negative, enter the noise spectrum with equal
  weight at positive frequencies.}\BibitemShut {Stop}%
\bibitem [{\citenamefont {Deacon}\ \emph
  {et~al.}(2010{\natexlab{b}})\citenamefont {Deacon}, \citenamefont {Tanaka},
  \citenamefont {Oiwa}, \citenamefont {Sakano}, \citenamefont {Yoshida},
  \citenamefont {Shibata}, \citenamefont {Hirakawa},\ and\ \citenamefont
  {Tarucha}}]{Deacon10b}%
  \BibitemOpen
  \bibfield  {author} {\bibinfo {author} {\bibfnamefont {R.}~\bibnamefont
  {Deacon}}, \bibinfo {author} {\bibfnamefont {Y.}~\bibnamefont {Tanaka}},
  \bibinfo {author} {\bibfnamefont {A.}~\bibnamefont {Oiwa}}, \bibinfo {author}
  {\bibfnamefont {R.}~\bibnamefont {Sakano}}, \bibinfo {author} {\bibfnamefont
  {K.}~\bibnamefont {Yoshida}}, \bibinfo {author} {\bibfnamefont
  {K.}~\bibnamefont {Shibata}}, \bibinfo {author} {\bibfnamefont
  {K.}~\bibnamefont {Hirakawa}}, \ and\ \bibinfo {author} {\bibfnamefont
  {S.}~\bibnamefont {Tarucha}},\ }\href@noop {} {\bibfield  {journal} {\bibinfo
   {journal} {Phys. Rev. B}\ }\textbf {\bibinfo {volume} {81}},\ \bibinfo
  {pages} {121308} (\bibinfo {year} {2010}{\natexlab{b}})}\BibitemShut
  {NoStop}%
\end{thebibliography}

%


\end{document}